\UseRawInputEncoding
\documentclass[manuscript]{acmart}

\AtBeginDocument{%
  }

\setcopyright{none}

\usepackage{_macros}
\usepackage{graphicx}    
\usepackage{subcaption}  
\usepackage{listings,mdframed}
\usepackage{multirow}

\newcommand{\name}{ClassComet} 

\begin{document}

\title[\name]{\name: Exploring and Designing AI-generated Danmaku in Educational Videos to Enhance Online Learning}

\author{Zipeng Ji}
\affiliation{%
  \institution{University of Waterloo}
    \city{Waterloo}
      \state{Ontario}
  \country{Canada}
}
\email{c35ji@uwaterloo.ca}

\author{Pengcheng An}
\affiliation{%
  \institution{Southern University of Science and Technology}
    \city{Shenzhen}
      \state{Guangdong}
  \country{China}
}
\email{anpc@sustech.edu.cn}

\author{Jian Zhao}
\affiliation{%
  \institution{University of Waterloo}
  \city{Waterloo}
  \state{Ontario}
  \country{Canada}
}
\email{jianzhao@uwaterloo.ca} 

\renewcommand{\shortauthors}{Zipeng et al.}

\begin{abstract}
Danmaku, users' live comments synchronized with, and overlaying on videos, has recently shown potential in promoting online video-based learning. However, user-generated danmaku can be scarce-especially in newer or less viewed videos and its quality is unpredictable, limiting its educational impact. This paper explores how large multimodal models (LMM) can be leveraged to automatically generate effective, high-quality danmaku. We first conducted a formative study to identify the desirable characteristics of content- and emotion-related danmaku in educational videos. Based on the obtained insights, we developed ClassComet, an educational video platform with novel LMM-driven techniques for generating relevant types of danmaku to enhance video-based learning. Through user studies, we examined the quality of generated danmaku and their influence on learning experiences. The results indicate that our generated danmaku is comparable to human-created ones, and videos with both content- and emotion-related danmaku showed significant improvement in viewers' engagement and learning outcome.


\end{abstract}





\maketitle 
\section{Introduction}
%

%

Online video-based learning has become increasingly prevalent recently; educational video platforms such as MOOCs have demonstrated significant advantages to support remote asynchronous learning at scale during the COVID-19 pandemic~\cite{jin2023collaborative,fernandez2023traditional,tanprasert2023scripted,yuan2024being}.
Learners now frequently acquire knowledge through multiple video-based media such as YouTube, Udemy, and Coursera~\cite{videoquality2023novan,kim2023older,wang2021learners}. 
However, educational videos naturally have pedagogical limitations due to their passive format of knowledge delivery, diminishing learners' learning outcomes because of less engagement~\cite{tanprasert2023scripted,dewan2019engagement}. 
Many platforms thus incorporated interactive features like commenting and quizzes~\cite{chen2021learning, li2022influence,interactivevideolearning2023mariani}. 
Along this line, recent research has recognized the potential of \emph{danmaku}, a video comment feature, that supports both lecturer-led and peer interactions, by analyzing or moderating human-created danmaku to boost learners' engagement~\cite{yao2017understanding,wu2018danmaku,wu2019danmaku, chen2021learning, hu2024danmodcap}. 
\rev{Featuring live and synchronized comments that scroll across the video screen, danmaku creates an interactive and communal learning experience for viewers. Users can not only view existing danmaku embedded in the videos, but also are encouraged to add danmaku to share their own insights and interact with other danmaku and the video content itself~\cite{zhang2019danmaku,hu2024exploring,he2021beyond}.}
Previous research has explored the role of danmaku in the learning contexts and measured the danmaku's potential in enhancing learners' social and cognitive presence~\cite{yao2017understanding, chen2019facilitating, chen2022timeline, zhang2024effects}. \rev{Unlike the previous experience of watching video content alone, where viewers receive information individually without peer interaction, interacting with danmaku allows viewers to send danmaku and engage with existing scrolling ones, transforming the learning experience from individual study to a collaborative environment.}

However, user-generated danmaku has undeniable challenges in its quality and quantity for pedagogical goals; extensive off-topic, negative, and chaotic danmaku may overlap with useful danmaku, distracting learners' attention and affecting their learning experience~\cite{ma2017video, mou2022interactivity}. 
Some works attempted to moderate harmful or inappropriate danmaku~\cite{hu2024danmodcap,hu2024exploringdanmakucontentmoderation}. 
Nonetheless, this does not solve the key issue that there is a lack of sufficient useful danmaku to foster learning in educational videos (especially the newly uploaded or less viewed ones), where the quantity of danmaku is also significantly smaller compared to entertainment videos~\cite{wu2018danmaku}. 
Therefore, the learning experience heavily depends on the quality and quantity of existing danmaku and the popularity of a video, which are usually unknown and uncontrollable. 

We turn this challenge into an unaddressed opportunity by viewing it from a different angle. The recent advancement of large multimodal models (LMMs) has shown exceptional performance in understanding videos and generating textual output~\cite{song2024moviechatdensetokensparse, li2024videochat, wang2024videoagentlongformvideounderstanding}.  
Instead of filtering the existing danmaku, we explore whether and how LMMs could be leveraged to generate valuable danmaku in educational videos. In particular, we ask the following research questions:

\begin{itemize}[left=2em]
    \item [\textbf{RQ1}:] What characteristics of danmaku in educational videos do learners genuinely value?
    \item [\textbf{RQ2}:] How do we design and utilize LMM-driven techniques for generating danmaku in educational videos?
    \item [\textbf{RQ3}:] How does LMM-generated danmaku influence viewers' learning experiences?
\end{itemize}

To answer RQ1, we conducted a formative study with seven participants to identify what danmaku characteristics learners genuinely value when watching educational videos. 
We found that danmaku should be knowledgeable, carry positive emotions, show encouragement, and be expressed in effective ways.
Based on these insights, we derived seven types of danmaku for educational videos, classified in two high-level categories: content-related and emotion-related.
To address RQ2, we designed \emph{\name}, an online educational video platform equipped with a novel LMM-powered auto-generation of effective and high-quality danmaku to promote video-based learning. 
We leveraged LMMs to comprehend both visual and textual information of videos while creating virtual personas to simulate danmaku interactions among online viewers. 
Learners can play a video as well as watch and interact with danmaku as they typically do. 
Without being bothered by the noise in user-created danmaku, \name{} guarantees viewers enjoy consistently effective and engaging danmaku during the learning process, regardless of the video's popularity.

To investigate RQ3 and understand the effect of \name, we conducted a within-subjects study with 12 participants, comparing four different conditions: (1) original video without danmaku, (2) video with only content-related danmaku, (3) video with only emotion-related danmaku, and (4) video with both danmaku. 
The findings suggest that videos with both content- and emotion-related danmaku could significantly enhance viewers' engagement and learning outcomes compared to the original videos. 
\rev{A further evaluation was conducted to compare our generated danmaku with human-created ones on Bilibili\footnote{\rev{Bilibili(\url{https://www.bilibili.com}) is a popular Chinese video-sharing website that offers various themes, including educational material, live stream, movies, dramas, and user-generated content. Users can submit, view, and interact with overlaid commentary (\ie, danmaku) on videos.}}, which is known for its danmaku feature, displaying user-generated 
 comments that fly across the screen, synchronized with the video.}
The results show the competitive performance of \name{} across four different metrics, including relevancy, consistency, fluency, and coherence.

In summary, our main contributions are as follows:
\begin{itemize}
    \item We uncovered the key characteristics that danmaku should manifest in educational videos through a formative study.
    \item We introduced \name{} that automatically creates virtual personas and generates effective, high-quality danmaku in educational videos to promote video-based learning.
    \item We carried out a controlled user study evaluating the quality and effectiveness of our generated danmaku to provide empirical knowledge for future research. 
\end{itemize}


\section{Related Work}

\subsection{Viewer Engagement in Video-based Learning}
\label{engagement}

The definition of engagement in educational contexts is multidimensional. Fredricks et al.~\cite{fredricks2004school} categorized engagement into three aspects: cognitive engagement, behavioral engagement, and emotional(affective) engagement. 
Cognitive engagement refers to students' investment in learning, including several aspects such as efforts to master challenging tasks, self-regulation, and challenge preference~\cite{finn2012student}.
Behavioral engagement refers to the involvement and participation (e.g., eye contact, leaning in during a discussion) in learning and academic-related tasks~\cite{finn1993school, finn1995disruptive, finn1997academic}.
Emotional engagement involves positive and negative emotional reactions towards teachers, classmates, academic works and school in general~\cite{linnenbrink2003role, finn1993school}.

This definition is widely adopted by subsequent educational research~\cite{archambault2009student, skinner2009motivational} and has been expanded from traditional school settings to the context of video-based learning~\cite{guo2014video}. Video-based learning is defined as ``the process of using video technology to acquire knowledge or skills''~\cite{giannakos2014vbl}. 
In the cognitive theory of multimedia learning~\cite{mayer2014ctml}, video-based learning offer better results compared to traditional teaching methods in terms of engagement and learning performance. 
%
More online video learning platforms, such as Khan Academy\footnote{\url{https://www.khanacademy.org/}} and Coursera\footnote{\url{https://www.coursera.org/}}, have emerged, offering extensive video resources~\cite{sablic2021vblsurvey} and putting more emphasis on designing interactive methods to better viewers’ engagement.

However, current interactive methods face notable limitations. 
First, although most platforms provide commenting and discussion forums, the success of these approaches depends on the video's popularity.
Research shows that popular videos foster vibrant learning communities with more interactions~\cite{10320094,baig2022impact,videopopularity2023zohar}, while less popular videos often lack learners' participation, reducing engagement and learning outcomes~\cite{activelearningwithonline2021kyoungwon,tanprasert2023scripted,dewan2019engagement}.
Second, while some recent AI-driven interactions, such as automated summaries or video chatbots~\cite{winkler2020sara, wang2021toward,han2022analysis,cai2021bandit}, attempt to act as a knowledgeable peer or teaching assistant, they only simulate partial learning interactions, such as dyadic peer discussions or Q\&A, missing more diverse interactions to enable deeper engagement.~\cite{labadze2023role,ma2024and}. 
These issues highlight a gap in the potential of video-based learning to foster more diverse and effective interactions across all videos, regardless of their popularity.
It thus motivated our research to design new interactive methods for enhancing learners' engagement by focusing on danmaku, which could foster a live and communal learning experience.

 
%



\subsection{Features and Usage of Danmaku in Video-based Learning}
\label{dan_feature_class}

\begin{figure}[tb!]
    \centering
    \includegraphics[width=0.8\linewidth]{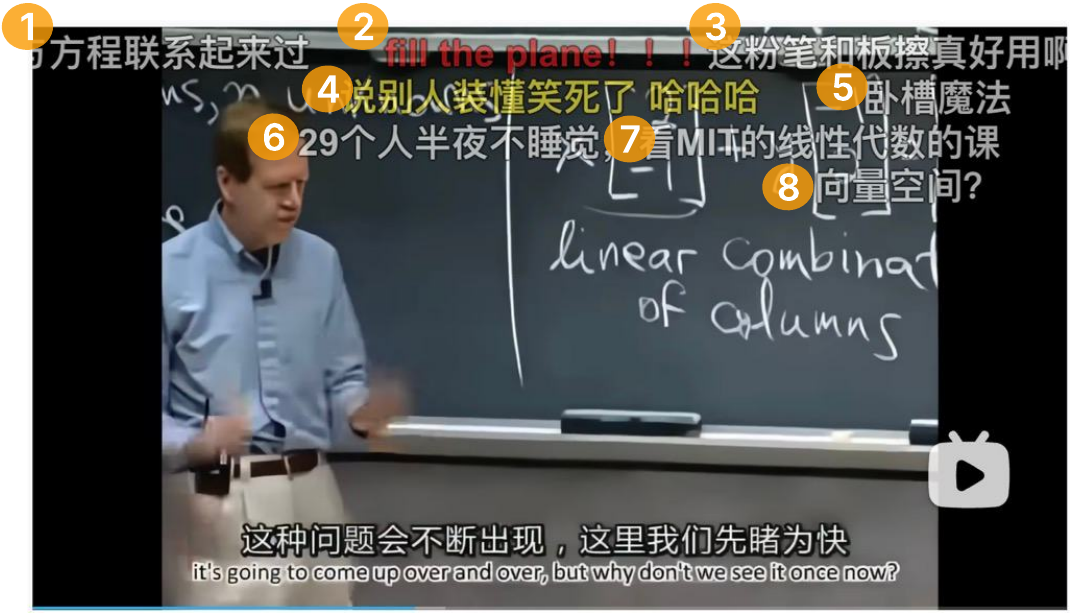}
    \caption{A sample screenshot of an educational video with danmaku. The translations of the displayed danmaku are: 1. ``I didn't linked it to equations before'', 2. ``fill the plane!!!'', 3. ``This chalk and eraser are so useful'', 4. ``Saying others are pretending to understand, lol'', 5. ``Wow, this is magic'', 6. ``29 people staying up late'', 7. ``Watch MIT's linear algebra class'', 8. ``Vector space?''. }
    \label{fig:danmakuproto}
\end{figure}

Previous research~\cite{yao2017understanding, wu2018danmaku, he2021beyond, zhang2023understanding, huang2024sharing} has identified several main features of danmaku. 
The first is brevity and high frequency. Wu \etal's study~\cite{wu2018danmaku} indicated the average length of danmaku is 9.33 words which is approximately five times shorter than video forum comments. Additionally, danmaku demonstrates a significantly higher posting frequency across all video genres than forum comments, indicating its potential for diverse interactions.
The second is its synchronousity~\cite{wu2018danmaku, huang2024sharing}. Unlike comments separated from the video, danmaku is embedded within videos with timestamps, giving each danmaku a fixed display period on the video timeline, regardless of its actual posting time~\cite{yao2017understanding}. This helps to create a sense of co-watching with virtual peers for learners.
The third is anonymity. Danmaku hides the identity of the viewers with only the content ``flying'' over the screen. Anonymity protects viewers' privacy and has a positive impact on user participation~\cite{wu2018danmaku, he2021beyond}. 
The last is the blending of interpersonal and personal interaction behaviors. Viewers can use danmaku as virtual peers to engage in question-and-answer interactions and discussions~\cite{he2021beyond, an2020comparative}. Moreover, each viewer can express personal, immediate reactions to the video content. Figure~\ref{fig:danmakuproto} exemplifies the danmaku of an educational video.

Unique viewer interactions with and through danmaku have emerged and developed based on the above features, which have been investigated in the context of online video-based learning. 
Chen \etal's study~\cite{chen2022timeline} and DanMooc~\cite{chen2019facilitating} categorized danmaku interactions into two general classes: content-related and social-oriented. Content-related danmaku refers to discussions about the video content, while social-oriented danmaku includes off-topic chats and jokes unrelated to the video content. 
DanMooc~\cite{chen2019facilitating} showed that users were more disinclined to social-oriented danmaku and considered off-topic chats and jokes to dilute the quality of discussions. DanMooc further refined the content-related interaction into three characteristics: questions, answers to predefined questions, and other video-related messages.
Some other researchers categorized the usages of danmaku from two aspects: content-related and emotion-related~\cite{he2021beyond}. 
An et al.~\cite{an2020comparative} identified a distinct interaction of content-related danmaku: content repetition, where learners used it for intensive practice of learning content while watching the video~\cite{applebee1983instructional}.
Zhao et al.~\cite{Zhao2019ACA} found that students had more interactions in information sharing (such as Q\&A exchanges) and dissonance exploration (such as opinion statements), while knowledge application and summarization were infrequent within danmaku.
Emotion-related danmaku represents another primary type, as viewers were prone to share their immediate feelings while watching educational videos including love, joy, surprise, tiredness, and fear.
However, the preference for positive or negative emotion sharing was still ambiguous in the context of online education videos. It was found that positive anonymous danmaku comments were more frequent among all genres of videos~\cite{wu2018danmaku}. 

These studies have identified various usages of existing danmaku in educational videos. 
However, what specific characteristics of danmaku are considered valuable and useful for video-based learning stays unclear. 
Moreover, the limited quantity and various noise of danmaku in educational videos obstruct the effectiveness of danmaku in promoting video-based learning~\cite{ma2017video, mou2022interactivity}, diminishing learners' experience.
Based on the above insights, our study seeks to identify the unique characteristics of danmaku preferred by learners watching educational videos, and explore automated techniques to unlock the potential for more targeted and effective danmaku design to enhance learning.

\subsection{Automated Live Video Comments}

The concept of automated live comment generation, commonly referred to as danmaku generation, has been introduced by LiveBot~\cite{ma2018livebot}, proposing two effective generation models and a danmaku benchmark dataset for evaluation.
Subsequent studies have focused on designing better strategies to improve the performance on the dataset and enhancing the dataset itself~\cite{wang2020videoic, wu2021cold, zhang2020dca}.
However, automated danmaku generation faces challenges in considering danmaku-video congruence in specific scenarios. Models often generate general danmaku without understanding and integrating information from different modalities~\cite{ren2022multi, zhang2023vcmaster}.
\rev{In educational videos, beyond textual information like subtitles, visual elements such as charts in slides, the color and size of specific fonts, and the spatial positioning of objects in the video greatly influence viewer comprehension.
In a broader context, AI-generated content overlays, apart from danmaku, such as dynamic music scores~\cite{yan2023generatingrhythmgamemusic} in video games and automated meeting transcripts, rely on multimodal inputs to deliver context-aware interactions that enrich the viewing experience.
Thus, multimodal understanding of educational videos is essental for generating comprehensive danmaku and presents a promising opportunity for positively influencing learning.}

With rapid advances in multimodal learning, a spectrum of video understanding tasks including video captioning~\cite{islam2024videorecaprecursivecaptioning,donahue2015long, fang2015captions, karpathy2015deep}, video description~\cite{rohrbach2014coherent, venugopalan2015sequence}, and visual question answering~\cite{antol2015vqa} can be implemented and evaluated. 
However, long-form video understanding remains a major challenge due to language model context length and GPU memory restrictions, as most multimodal models are evaluated on video clips under 30 seconds or minute-long benchmarks~\cite{song2024moviechat, 
li2024videochat}. 
Moreover, most educational videos are long, requiring processing multimodal data (e.g., audio, visual, and text) and capturing complex temporal dependencies over extended periods, which demands significant computational resources and careful model design.

This inherent challenge has been attempted by unleashing the power of large multimodal models (LMMs)~\cite{lin2023mmvid, kahatapitiya2024languagerepositorylongvideo, zhao2024antgptlargelanguagemodels, wang2024videoagentlongformvideounderstanding, weng2024longvlmefficientlongvideo}. 
Some studies focus on compressing visual information such as merging adjacent or similar frame patch tokens~\cite{li2024llavanextinterleavetacklingmultiimagevideo, song2024moviechatdensetokensparse}. Others like Timechat~\cite{ren2024timechattimesensitivemultimodallarge} and Moment-10M~\cite{qian2024momentoradvancingvideolarge} propose time-aware instruction-tuning methods to capture richer spatial and temporal information compared to short videos. 
We designed \name{} based on the LMM-driven method MM-VID~\cite{lin2023mmvid}, which sparsified frames to capture longer visual tokens and utilized a video-to-script generation with GPT-4V to transcribe multimodal elements into a long textual script, facilitating understanding of long educational videos and diverse danmaku generation.


\rev{
In the current HCI community, some work utilized LLMs to moderate danmaku~\cite{hu2024danmodcap,  hu2024exploringdanmakucontentmoderation}. However, none of the work addresses the educational context, where the number of useful danmaku can vary depending on the popularity of educational videos.
Therefore, how to design automated danmaku that considers both quality and quantity for educational videos still remains unsolved. In this research, we specifically focus on leveraging LMM for automated danmaku generation in educational scenarios, aiming to support hours-long video understanding and produce high-quality, diverse danmaku across all videos that positively impact learning.
}

\section{Formative Study}

As the first step of our research, we conducted a formative study to learn about the key characteristics that danmaku should have in educational videos (RQ1).

\subsection{Study Setup}

We recruited seven participants (4 self-identified females and 3 self-identified males) from local universities via mailing list and word of mouth, denoted as P1-P7 in the following text.  
They were undergraduate or graduate students (ages ranging from 21 to 25, \mean{23.67}, \sd{1.25}) with various technical backgrounds, including computer science, statistics, law, finance, and design. 
All participants frequently watch educational videos for their learning and have the habit of turning on danmaku on Bilibili. 
When asked about their danmaku posting frequency on a 5-point Likert item, one self-reported ``Frequent,'' three reported ``Normal,'' and three said ``Occasional.'' 

Each study session consisted of two stages: 1) watching a 5-minute educational video with some discussion and 2) a semi-structured interview regarding danmaku's characteristics. In the end of the study, they received \$20 for their time and effort.
In the first stage (10 minutes), we provided four educational videos with danmaku to help participants connect their own danmaku experience with these videos. 
The four videos were selected from Bilibili and covered different learning topics, each approximately 5 minutes in length. 
Participants were asked to select a video based on their interests and watch it as they normally would. 
They could speed up or pause the video, comment on existing danmaku, and post danmaku whenever they wanted. This allows us to observe their natural behaviors when watching educational videos with danmaku. 
We employed a contextual inquiry method: each time when the participant posted danmaku, we engaged in a short discussion to understand their motivations and rationales.
In the second stage (15-20 minutes), we discussed with the participants about their preferences for content-related and emotion-related danmaku, habits in posting danmaku, reactions and opinions to different danmaku characteristics, and expectations for ideal danmaku watching experiences.

\subsection{Key Insights}
\label{dataanlysis}
\rev{
We collected the danmaku posted by participants in the first stage to help participants recall their typical danmaku-watching habits. 
We found that participants sent 1-4 danmaku per minute during the video-watching session, with \mean{1.85} per minute (\sd{0.98}).  
The average length of danmaku was 9.2 characters (\sd{5.34}).
In terms of danmaku interactions, all participants posted danmaku to engage with either existing danmaku or the
video content.
These user behaviors align with the results reported in previous work~\cite{wu2018danmaku}.
}

We also transcribed our discussion with the participants in the whole study session including our contextual inquiries and semi-structured interviews. 
The whole research team then collaboratively coded the transcripts and constructed an affinity diagram to derive five insights into what danmaku characteristics learners value in educational videos. 

\textbf{K1: Danmaku should carry useful knowledge about the learning concepts.} 
Six participants mentioned that they preferred knowledgeable danmaku that provides correct information about or emphasizes key concepts from the videos, compared to that with less information or irrelevant to the video, such as unrelated jokes or off-topic comments.
P3 said, \qt{In some cases, there are important danmaku of answering the video's questions, and I really want to focus on them. But irrelevant danmaku often covers it or appears nearby which distracts my attention. So I would sometimes pause to delete the irrelevant danmaku, but I don't think this is an effective way.} 
P6 stated, \qt{I think Q\&A danmaku can help me understand what the videos are talking about, which is the question and which is the answer, that can help me understand the video better.} 
Participants also expressed their expectations of more discussions regarding the video content. \pqt{I want to see danmaku discussing controversial things and giving new perspectives.}{P4} 

\textbf{K2: Danmaku should help learners grab important information from the video content.} 
Three participants expressed their preferences to see more summaries in danmaku. P4 said, \qt{I prefer danmaku summarizing the content of the videos. This can help me better understand the concepts.} 
P2 pointed out the value of summary danmaku in longer videos, particularly lectures: \qt{Summary in danmaku can help me quickly capture the main points of the video, especially hour-long lectures. It can appear at the beginning or after each section. It’s like a quick review that keeps me focused without losing track.}
Further, highlighted danmaku was valued by participants. \pqt{I find those highlights very important because there are so many danmaku. With those highlights, I can easily capture them and focus my attention so that I don’t need to pay attention to every line of danmaku. Sometimes, educational videos are not interesting so I will lose focus. However, highlighting danmaku can help me redirect my attention back to the video itself, making it easier to stay engaged.}{P1} 

\textbf{K3: Danmaku should express positive emotions to foster a good learning environment.} 
Five participants mentioned their preference for positive emotional expressions in danmaku within the context of online video learning. 
For instance, P1 gave a ``like'' to the danmaku ``Very clear explanation'' in the first stage of the study. She stated: \qt{I have the same feeling when I watched the video. Giving a like is a kind of compliment.} 
P2 further mentioned in the interview: \qt{Personally, I prefer more positive danmaku, because it's a learning video, not an entertaining one. I don't want to be in a very negative learning atmosphere.} 
P3 also compared positive emotion-related danmaku with negative ones. She said: \qt{In educational videos, I don't want to see too much negative emotion, like complaining about the video's difficulty or impolite augments with other viewers. It's useless and ineffective. If the video is filled with negative danmaku, it will definitely affect my mood and my learning progress.}

\textbf{K4: Danmaku should show encouragement to help learners overcome challenges.}
Participants were willing to see more encouraging danmaku when encountering difficulties in learning such as tiredness and complaints. The encouragement in danmaku fosters a sense of \pqt{having a positive learning partner}{P3} and \pqt{giving the feeling of learning with other people.}{P5} 
Specifically, P1 pointed out that \qt{I’d love the danmaku to show support like in the Japanese video, for example, `If you see this danmaku, you should keep learning and do not close this video.' I found it quite important because at that stage I was losing focus.}  
P7 also mentioned that \qt{Encouragement is much better than just expressing exhaustion or negative feelings.}

\textbf{K5: Danmaku should be manifested effectively and concisely in the video.}
Participants showed preferences for more effective expressions in danmaku. Apart from favoring short, concise, and clear danmaku, participants enjoyed using emojis and symbols to convey their confusion or other feelings. For example, P4 sent danmaku ``??'' to indicate her confusion when a concept was presented, and P3 added ``!'' at the end to express her surprise at the video content, aligning with a line of other existing ``!'' danmaku. Further, P5 explained why he wanted to send a thumb-up as danmaku, \qt{I don’t usually send danmaku myself. If I want to send danmaku, I need to pause the video and it takes minutes. It’s too long for me. Usually the most I can do is to send an emoji or symbols to show my reaction.}

\subsection{Types of Danmaku}

\begin{table*}[tb]
\caption{Types of danmaku in educational videos to support learning.}
\label{tab:danmakutype}
\vspace{-3mm}
\small
\begin{tabular}{llp{8.5cm}} 
\toprule
\textbf{Class} & \textbf{Type} & \textbf{Description} \\ \midrule
\multirow{4}{2cm}{Content-related} &
  Q\&A &
  Ask and answer questions to assist viewers in consolidating acquired knowledge and dispelling misconceptions (K1, K5). \\
 &
  Discussion &
  Exchange opinions, propose hypotheses or provide complementary information related to the proposed question in the video (K1, K5).\\
 &
   Highlights &
  Emphasize key concepts or important words in unique displays (font size, color, position) to capture viewers' attention (K2, K5). \\
 &
  Summary &
  Preview key points at the beginning, summarize after each section, and provide a final recap at the video's end (K2, K5).\\ \midrule
\multirow{3}{2cm}{Emotion-related} &
  Personal Emotion Expression &
  Simply and directly express one’s emotions within emojis and symbols (K5). \\
 &
  Brief Compliment &
  Offer praise when a viewer's danmaku provides the right answers or explicit explanations to the video’s questions or other viewers’ questions (K3, K5). \\
 &
  Encouragement &
  Send supportive danmaku in response to negative expressions from other viewers (K4, K5). \\ \bottomrule
\end{tabular}%
\end{table*}

Informed by the above insights, we provide a summary of seven common types of danmaku valued by participants the most in educational videos and align our taxonomy with existing studies that categorize danmaku into two big classes: content-related and emotion-related~\cite{chen2022timeline,chen2019facilitating}. 
The details are displayed in \autoref{tab:danmakutype}. 

The four most valued content-related danmaku are Q\&A, discussion, highlighting, and summary. 
Q\&A danmaku can occur between the instructor who proposed questions in the video and viewers or between active viewers themselves. Thought-provoking questions and answers benefit learners, regardless of whether they participate in or observe Q\&A interactions (K1). 
Discussion is also appreciated by participants as it can timely exchange on the video content (K1). Unlike forum comments under the video, discussion in danmaku is more concise~\cite{wu2018danmaku}.
Highlighting is a unique type that only exists in danmaku, which normally hovers at the top center of the video for several seconds, using eye-catching styles (such as different font colors or sizes) to highlight keywords (K2). Summary is expected to be seen more in danmaku which typically appears at the beginning or end of each video section, providing a recap of key points (K2).  
Personal emotional expression is the most common emotion-related interaction. As learners prefer positive emotional expression, they value compliments and encouragement in danmaku during their watching experience (K3, K4). This helps to create a more concentrated and positive learning atmosphere. 
Both content-related and emotion-related danmaku should be expressed effectively. Viewers prefer short, concise, and direct danmaku, often using various emojis and symbols to simplify their expressions (K5).

\section{\name{} Platform}
Based on the results from our formative study, we designed and developed \name{}, an online educational video platform empowered by the LMM-driven automated generation of danmaku, aiming to improve viewers' learning outcomes and engagement (RQ2). 
In this section, we describe the user interface of \name{} and a simple usage scenario.
Details regarding the underlying techniques for danmaku generation will be discussed in the next section.

\subsection{User Interface}
\begin{figure*}[tb]
    \centering
    \includegraphics[width=1\linewidth]{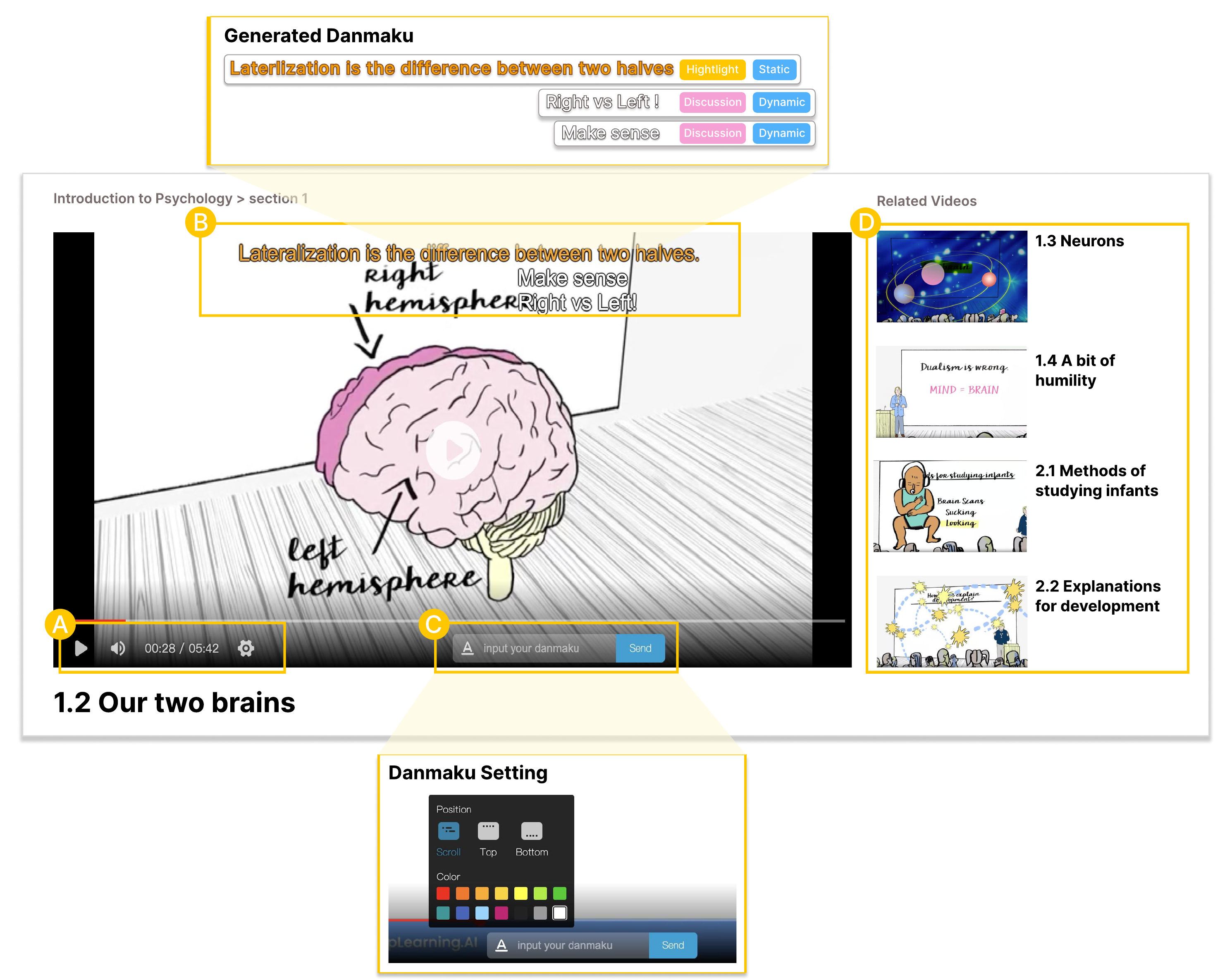}
    \vspace{-6mm}
    \caption{User interface of \name{}: (A) video control buttons for play/pause, volume adjustment, and speed control, (B) danmaku embedded in the video, (C) danmaku control that contains danmaku settings and an input box to send danmaku, and (D) a video sidebar for selecting other educational videos.}
    \label{fig:interface}
\end{figure*}

As shown in \autoref{fig:interface}, the \name{} user interface consists of two main components: a video player and a video sidebar.

The video player contains three parts. (1) The video control buttons (\autoref{fig:interface}-A) allow for play/pause, volume adjustment, and speed control. (2) The generated danmaku ``fly'' over the screen (\autoref{fig:interface}-B), which is white by default and can be changed to other colors to highlight important concepts, following the convention of danmaku experience on commonly used video platforms like Bilibili.
Highlighted danmaku stays at the top center of the video for some time~\footnote{Previous research~\cite{chen2022timeline, he2021beyond} indicated that danmaku with unique displays were more likely to be fixed in the top and bottom positions.}, while other danmaku dynamically scrolls across the video screen. 
(3) The danmaku control includes danmaku settings and an input box for users to send custom danmaku (\autoref{fig:interface}-C). Users can freely post danmaku with options to select the font color and display position (scrolling across the screen, hovering at the top, or staying at the bottom of the video).

\begin{figure*}[tb]
\centering
\begin{subfigure}[htbp]{0.24\linewidth}
\centering
\includegraphics[width=\linewidth]{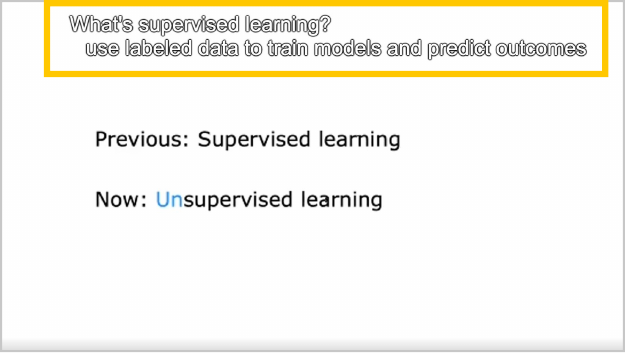}
\caption{Q\&A}
\label{fig:qna}
\end{subfigure}
\hfill
\begin{subfigure}[htbp]{0.24\linewidth}
\centering
\includegraphics[width=\linewidth]{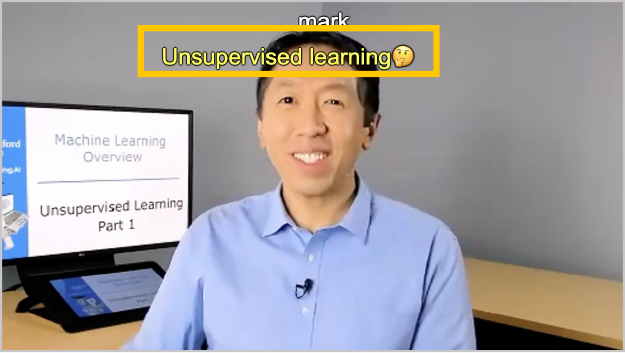}
\caption{Highlights}
\label{fig:highlight}
\end{subfigure}
\hfill
\begin{subfigure}[htbp]{0.24\linewidth}
\centering
\includegraphics[width=\linewidth]{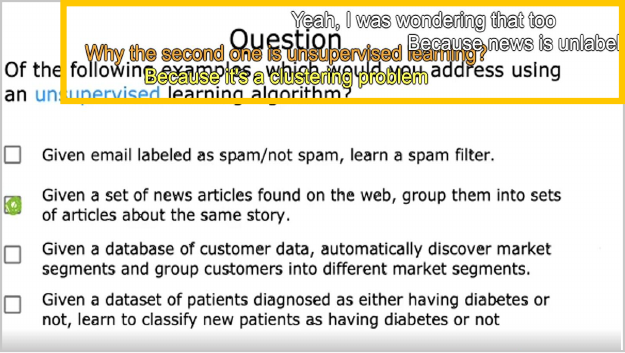}
\caption{Discussion}
\label{fig:discussion}
\end{subfigure}
\centering
\hfill
\begin{subfigure}[htbp]{0.24\linewidth}
\centering
\includegraphics[width=\linewidth]{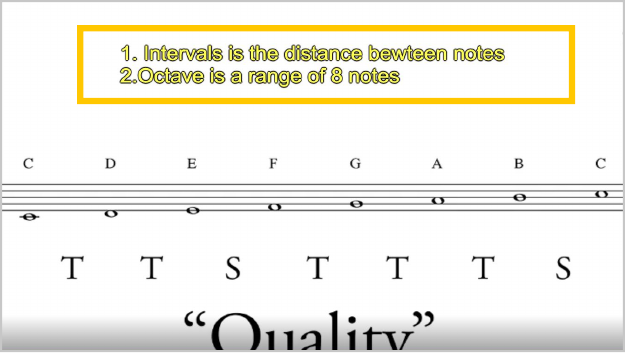}
\caption{Summary}
\label{fig:summary}
\end{subfigure}
\centering
\hfill
\begin{subfigure}[htbp]{0.24\linewidth}
\centering
\includegraphics[width=\linewidth]{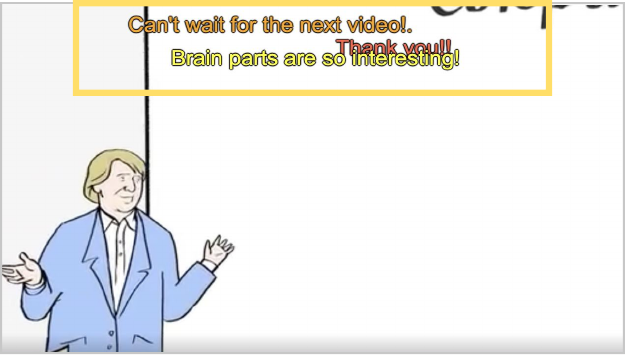}
\caption{Personal Emotion Expression}
\label{fig:emotionsharing}
\end{subfigure}
\centering
\begin{subfigure}[htbp]{0.24\linewidth}
\centering
\includegraphics[width=\linewidth]{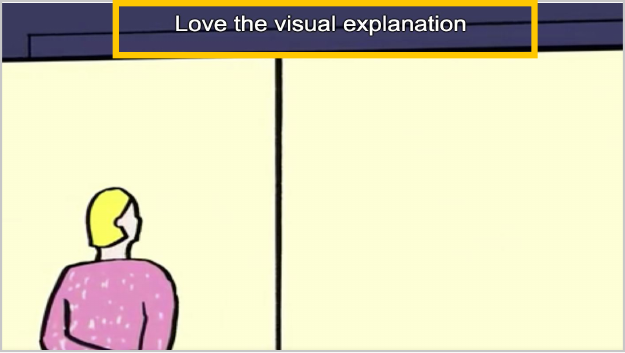}
\caption{Compliment}
\label{fig:compliment}
\end{subfigure}
\centering
\begin{subfigure}[htbp]{0.24\linewidth}
\centering
\includegraphics[width=\linewidth]{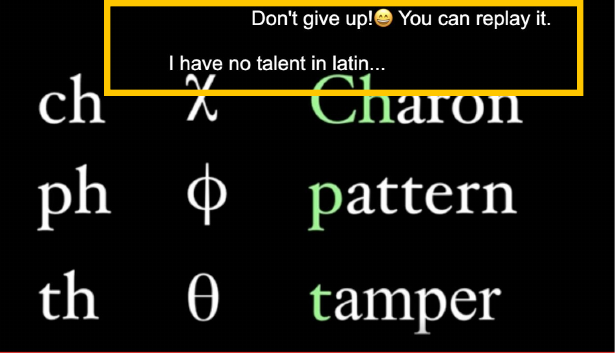}
\caption{Encouragement}
\label{fig:encouragement}
\end{subfigure}
\vspace{-3mm}
\caption{Examples of different types of danmaku (including both content-related and emotion-related) generated by \name{} in educational videos. }
\label{danmakuinter}
\end{figure*}

The video sidebar (\autoref{fig:interface}-D) displays a list of additional videos from the same course, allowing users to watch them sequentially. Once users have finished the current video, they can easily navigate to the next lesson. Upon clicking a video, \name{} loads it into the video player, and the danmaku is automatically generated based on the video content as the video plays. In \autoref{danmakuinter}, we display all scenarios of danmaku interactions.

\subsection{Usage Scenario}

\rev{Before describing the technial details about \name{}, in this section, we present a usage scenario to illustrate how users interact with various system features in practice.}
Cara, a novice learner in psychology, is self-learning the Introduction to Psychology course. However, whether watching on YouTube or educational platforms like Coursera, she often feels isolated. Since the course is not as popular as others, it lacks active comments and interactions commonly seen in more widely viewed videos. Cara misses the experience of exchanging ideas, participating in discussions, and receiving encouragement in classroom learning.
Although Coursera provides quiz and assignment interactions, Cara finds it difficult to keep herself motivated and engaged due to the lack of social connectedness and interpersonal interactions. After completing the first few videos, she starts to search for a new platform that fosters a peer-learning environment, regardless of the video's popularity. Therefore, Cara decides to give \name{} a try.

Cara enters \name{} and selects the same psychology course. As she starts to play the lecture, she notices a familiar feature: a stream of vibrant danmaku scrolling across the video. Having watched danmaku frequently on Bilibili for both entertainment and learning purposes, Cara is pleased to see danmaku. However, after just five minutes of watching, she notices a significant difference. Unlike the danmaku she used to on Bilibili, where jokes and random arguments can overwhelm genuine valuable insights, the danmaku in \name{} has no such irrelevant danmaku. She finds more danmaku relevant to the video content which displays more often than the danmaku videos she used to see.

Cara observes many highlighting danmaku with distinctive colors like \qt{Note: symmetry.} These danmaku serve as gentle reminders, emphasizing the key points in the video and helping her stay focused. As she feels a bit confused by the video’s apple example, she notices a pair of question-and-answer danmaku passing by, which provides a clear explanation and addresses her confusion.
Apart from informative danmaku that is closely related to video content, Cara notices supportive danmaku sharing feelings that resonate with her own learning experience. When she begins to feel a little fatigued and her focus starts to slip, she sees a danmaku saying \qt{Come one! Three minutes left!} It serves as a kind nudge that encourages her to regain focus. By the end of the lecture, she was satisfied to see a concise summary of the video \qt{This lesson is about the comparison of two halves of the brain,} as though a helpful final review that recaps the main ideas of the video.

Beyond standard video functions such as adjusting playback speed and volume (\autoref{fig:structure}-A), Cara notices an input box at the bottom of the video player (\autoref{fig:structure}-C), so she can easily interact with existing danmaku and shares her own thoughts. Additionally, Cara explores other different topics in videos. She clicks some lesser-known courses like \qt{Introduction to Reproducibility in Cancer Informatics} which has 0 reviews on Coursera. To her surprise, this interactive danmaku feature is present in all courses in \name{}, not just the popular ones, but typically lacks comments or rates on other platforms.

After exploring this new platform, Cara feels her learning is enriched not just by the video content but by the ideas and shared understanding and feelings that come from the danmaku interactions.

\section{Danmaku Generation in \name{} }
\label{system}

\begin{figure}[tb]
    \centering
    \includegraphics[width=1\linewidth]{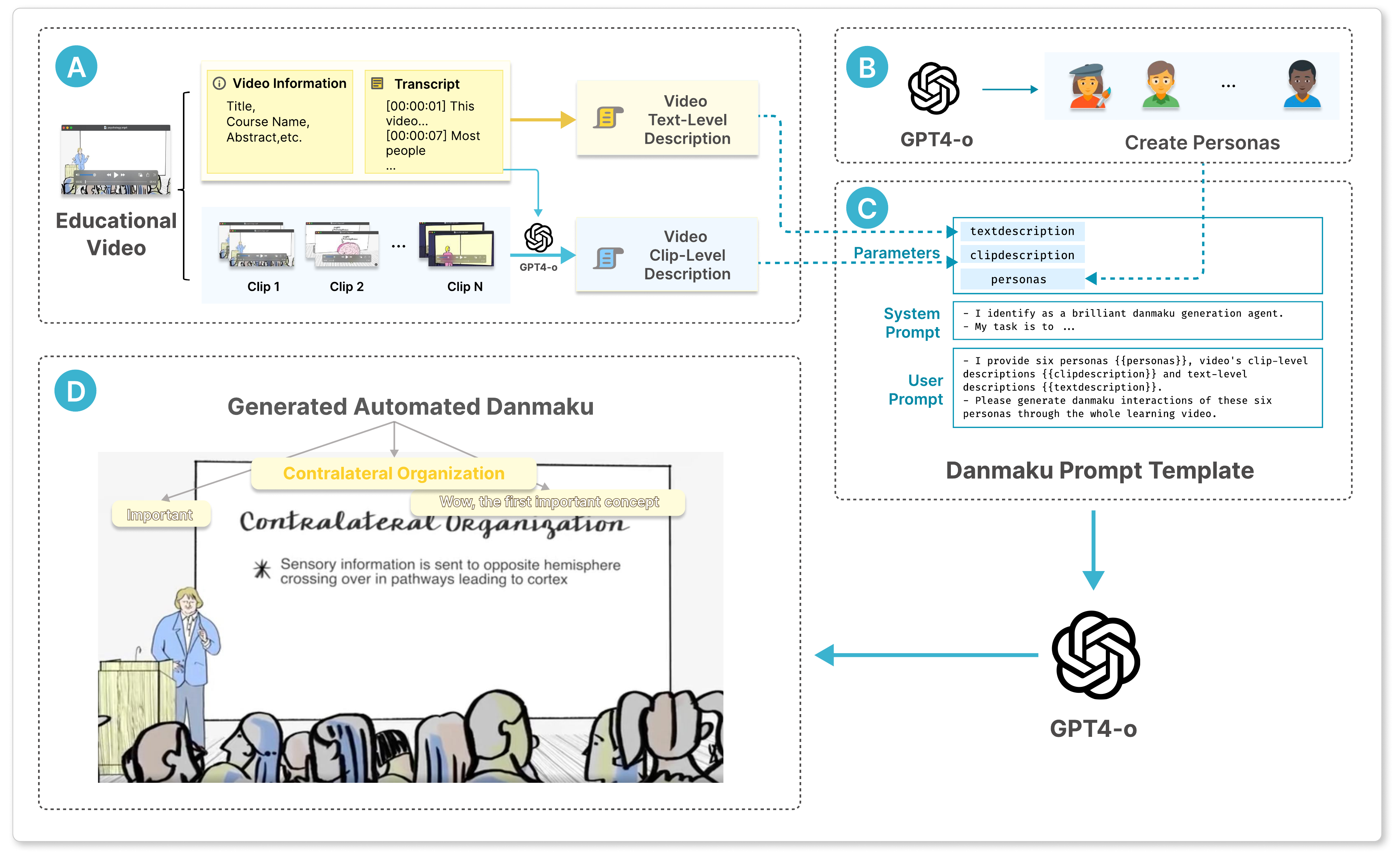}
    \vspace{-6mm}
    \caption{The \name{} pipeline for automatically generating danmaku in educational videos includes four steps: (A) Extract video information, segment the video into clips using scene detection, and generate descriptions at both the text- and clip-level; (B) Create virtual personas; (C) Design a structured danmaku prompt template and set parameters; and (D) Embed generated danmaku into educational videos.}
    \label{fig:pipeline}
\end{figure}

We proposed a novel pipeline based on LMMs for generating danmaku that encompasses both content-related and emotion-related danmaku in different types (\autoref{tab:danmakutype}). As shown in \autoref{fig:pipeline}, the inputs to \name{} for generating danmaku come from three aspects: (A) video understanding at both the text and clip levels, (B) persona creation, and (C) a danmaku prompt template. 
The output of contextually rich danmaku is then embedded into educational videos (\autoref{fig:pipeline}-D). 
Below, we detail each input component.

\subsection{Clip-level and Text-level Video Understanding}

Recent studies~\cite{he2024malmm,he2024malmmmemoryaugmentedlargemultimodal,xu2024raw} showed that LMMs have overcome the limitations of text-only video understanding and shown great potential to address the challenges of understanding long-form videos.  
We adopt the LMM GPT4o to generate comprehensive and contextually relevant descriptions at both clip- and text-levels.
To enable \name{} to comprehend hour-long educational videos, we segment the entire video into multiple clips and sample frames from each clip.
By utilizing established scene detection tools PySceneDetect~\footnote{\url{https://www.scenedetect.com/}}, we identify scene boundaries and track scene changes.
After segmenting the video into multiple clips, we follow the MM-VID~\cite{lin2023mmvid} algorithm and uniformly sample frames from each clip, with each minute consisting of 5 frames.
By feeding the clips and transcriptions into the GPT4o, we capture detailed clip-level descriptions of the educational videos, such as the lecturer's face and actions and images displayed on the video slides. Further prompts and examples can be found in the appendix~\ref{exampleofclipdescription}. 

The text-level description generation is based on two sources of information: basic video details such as the title, abstract, and course name, as well as transcriptions with timestamps. We utilize Microsoft Azure Video Indexer API~\footnote{\url{https://azure.microsoft.com/en-us/products/ai-video-indexer}} to extract transcriptions along with timestamps. All components of the text-level descriptions are formatted in JSON. 
We integrate both clip- and text-level video descriptions into our prompt template for automated danmaku generation.

\subsection{Create Virtual Viewers}
Many works utilized prompted LMMs~\cite{wu2022promptchainer} to generate short natural language descriptions of personas and their corresponding behaviors. The success of synthesizing human-human dialogues largely depends on the personas of the dialogue participants~\cite{li2024curriculumdbot, chen2022weaklysuperviseddataaugmentation, kim2023sodamillionscaledialoguedistillation}.
To effectively simulate the danmaku interactions, we create multiple personas with diverse and detailed backgrounds, including ages, regions, educational histories (degrees, universities, and majors), personalities, speaking styles, and their reasons for watching the educational video.
Prior work has equipped LLMs with personas to simulate human interactions and feedback such as generating Reddit community comments~\cite{park2022social} and creating agents serving various roles like educators~\cite{piro2024mylearningtalk}.
Hung\etal compared persona-based comments with no-persona comments in video-sharing platforms~\cite{hung2024simtube}, which showed that persona-based comments were significantly higher than no-persona comments in relevance, believability, and helpfulness.
The average sending frequency of danmaku in current educational videos is around 25-30 per minute~\cite{wu2018danmaku}, while the off-topic danmaku that is unrelated to the video may be up to 10\%-30\%~\cite{chen2019facilitating}. Based on the sending frequency per user outlined in Section~\ref{dataanlysis}, 
we set up six virtual viewers to participate in danmaku interactions, 
which aligned with the frequency of danmaku posting in previous research~\cite{wu2018danmaku, yao2017understanding}. These personas may either interact with other viewers' danmaku or engage with the video content itself.

Before generating danmaku, we leverage GPT4o to create personas. A detailed description of the prompt structure for this step is illustrated in the following. The XML-like tag \texttt{<title>} can be substituted with a specific video title that the learner is watching. More details about this step, as well as examples of input and output, are provided in \autoref{exampleofpersona}. 

\lstset{
      basicstyle=\ttfamily\footnotesize,
      columns=fullflexible,
      frame=single,
      breaklines=true,
      breakindent=0pt
    }
    \begin{lstlisting}
    Prompt: - Your task is to create 6 distinct personas with different backgrounds and personalities. They are interested in watching the online educational video <title>. Each persona should have the habit of sending danmaku while watching the video. Use "A", "B", "C", "D", etc., as persona labels.
    - For each persona, please provide the following details in JSON format, including age, region, personality, danmaku sending style, learning habits, and reasons for watching the video.
    \end{lstlisting}

\subsection{Danmaku Prompt Template}
\begin{figure}[tb]
    \centering
    \includegraphics[width=1\linewidth]{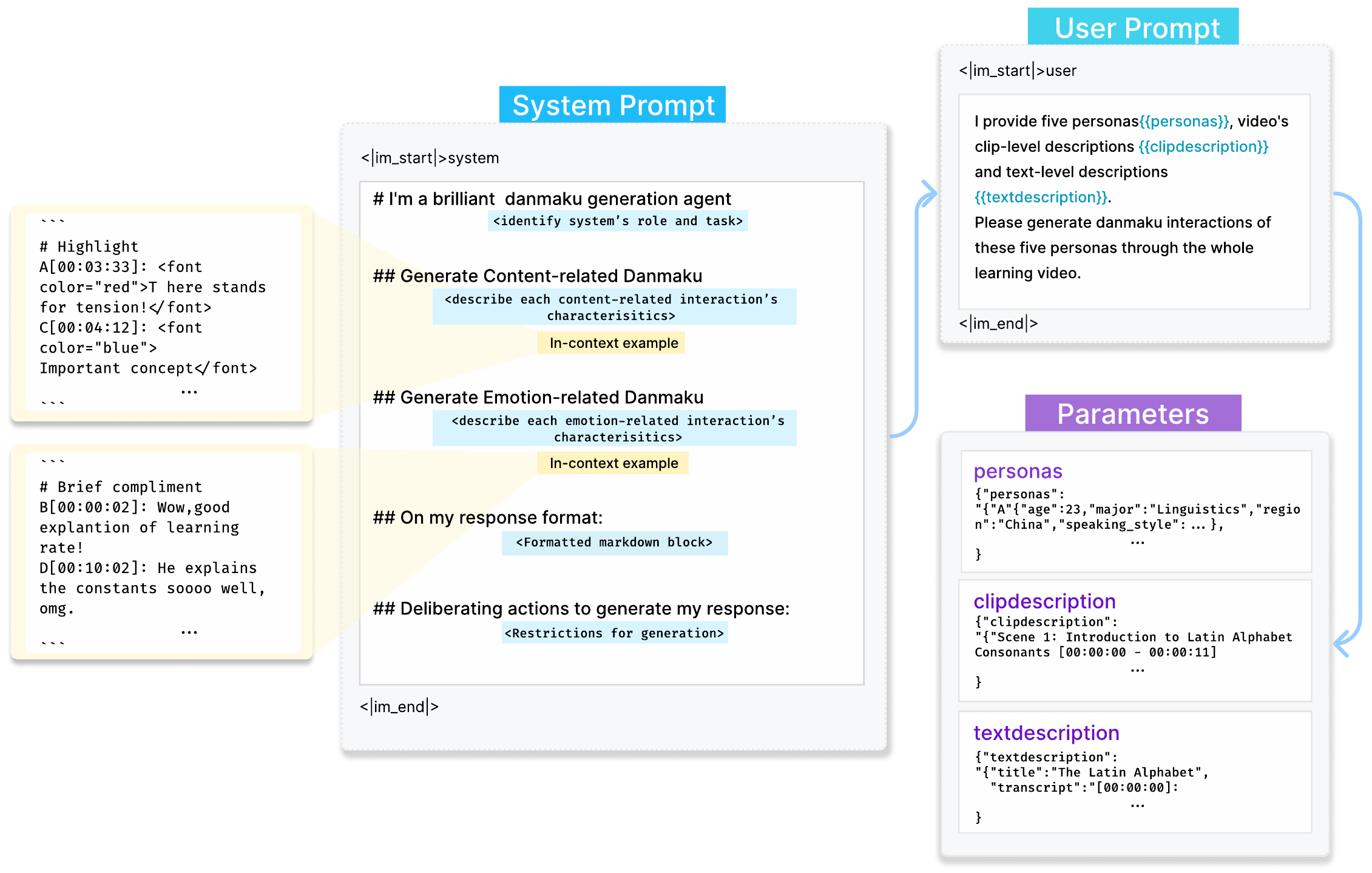}
    \vspace{-6mm}
    \caption{The structured prompt template of generating danmaku which consists of the system and user prompts as well as the pass parameters including two video descriptions (text and clip levels) and personas.}
    \label{fig:structure}
\end{figure}

To effectively configure GPT4o for generating danmaku across educational videos, we design a structured danmaku prompt template, as shown in \autoref{fig:structure}. The template consists of three components: the system prompt, user prompt, and parameters. 
The system prompt informs GPT4o of its role, tasks, behaviors, and boundaries. All XML-like tags shown in the figure in the system prompt are illustrated below and replaced with specific prompts. The user prompt customizes the responses by providing the user's needs and different parameters. Both system and user messages are defined between the \texttt{<|im\_start|>} and \texttt{<|im\_end|>} tokens to clearly define different boundaries. All parameters are embedded into the user prompt as input and formatted in JSON.

\subsubsection{System Role and Tasks}
In the first section of the system prompt, we outline the role of GPT4o as a danmaku generation agent and summarize the task as follows. Role prompting~\cite{kim2024personadoubleedgedswordmitigating} leads to a more focused, creative, or empathetic response depending on the given role. We summarize the key actions GPT is required to perform. We also set boundaries in the first section to ensure the model's output quality. A detailed description of the prompt structure for this step is shown below: 

\lstset{
      basicstyle=\ttfamily\footnotesize,
      columns=fullflexible,
      frame=single,
      breaklines=true,
      breakindent=0pt
    }
    \begin{lstlisting}
    Prompt: # I'm a danmaku generation agent
    - I identify as a brilliant danmaku generation agent. 
    - My task is to generate content-related and emotion-related danmaku. The generated danmaku should reflect the unique personalities and diverse backgrounds of pre-defined personas. 
    - I should simulate dynamic and engaging danmaku that align with their distinct character traits.
    \end{lstlisting}

\subsubsection{Danmaku Types}
For each type of danmaku, we provide in-context examples to stimulate the danmaku interactions in \autoref{fig:structure}, which are crucial for enhancing GPT's output performance~\cite{zhou2023efficientpromptingdynamicincontext,nafar2024learningvsretrievalrole}. Each example specifies the persona ID, timestamp, and danmaku content. For types such as discussions, Q\&A, and encouragement, which involve exchanges between pairs of or even more virtual viewers, we present synthetic multi-turn danmaku types in the in-context examples. For the highlighting danmaku, examples should use HTML tags to define its highlighting color to capture viewers' focus. For the personal emotional expression danmaku, we provide multiple examples, including emojis, modal words, and short sentences, to make the danmaku responses more natural. More details of each danmaku type's descriptions are provided in Appendix~\ref{appendixtemplate}.

\textbf{Response Format}
The response format can help guide the model to organize answers in the desired structure. We select the Markdown format as its stable and competitive output performance across different response formats~\cite{he2024doespromptformattingimpact}, as detailed in the following: 
\lstset{
      basicstyle=\ttfamily\footnotesize,
      columns=fullflexible,
      frame=single,
      breaklines=true,
      breakindent=0pt
    }
    \begin{lstlisting}
    Prompt: ### Response Format
    ```
    # Emotion-related danmaku
    ## <emotion-related danmaku type 1>
    - <role> | <timestamp>: <generated danmaku> 
    - <role> | <timestamp>: <generated danmaku>
    
    ## <emotion-related danmaku type 2>
    ... 
    
    # Content-related danmaku
    ## <content-related danmaku type 1>
    - <role> | <timestamp>: <generated danmaku>
    - <role> | <timestamp>: <generated danmaku>
    
    ## <content-related danmaku type 2>
    ... 
    ```
    \end{lstlisting}

\subsubsection{Deliberating Actions}
We next deliberate actions for generating danmaku by specifying the length, frequency, and count. First, drawing from previous research and formative studies, we set the maximum danmaku text length to 12, based on the average danmaku length reported in earlier studies~\cite{he2021beyond, an2020comparative}. 
Second, some works reported that the danmaku frequency in educational videos typically ranges from 2 to 30 per minute~\cite{wu2018danmaku, wu2019danmaku}. Therefore, we set the lower bound of danmaku frequency to avoid long gaps, ensuring that danmaku is generated no less frequently than every 30 seconds.
Third, we defined the range for content-related danmaku as 15-25 and for emotion-related danmaku as 5-10. Prior research has estimated that the average danmaku frequency in educational videos ranges from 25 to 40 per minute, with approximately content-related danmaku comprising 60-65\% and emotion-related danmaku making up 15-25\%~\cite{wu2018danmaku,chen2019facilitating, he2021beyond}.
Fourth, we made requirements for the number of highlighting danmaku since prior studies found that the proportion of highlighted danmaku makes up over 50\% of the content-related danmaku~\cite{wu2018danmaku}. The detailed restrictions given to GPT4o in this step are shown below:

\lstset{
      basicstyle=\ttfamily\footnotesize,
      columns=fullflexible,
      frame=single,
      breaklines=true,
      breakindent=0pt
    }
    \begin{lstlisting}
    Prompt: # Deliberating actions to generate danmaku
    - The length of each danmaku should be less than 12. 
    - I should generate danmaku continuously without long gaps (longer than 30s).
    - I should generate about 15-25 content-related danmaku and 5-10 emotion-related danmaku per minute.
    - I should generate more than **10 highlight** per minute.
    - Each type of danmaku should cover the entire duration.
    \end{lstlisting}

\subsubsection{User Prompt and Parameters.}
In the user prompt, we embed three parameters in JSON format derived from previous clip-level and text-level descriptions, as well as the created personas into the user prompt to customize danmaku for different videos. These structured inputs can facilitate the generation performance for large multimodal models~\cite{liu2024mmcadvancingmultimodalchart}. In this way, \name{} can generate automated, effective, and high-quality danmaku for educational videos. The full version of the template, along with generated examples, is introduced in Appendix~\ref{appendixtemplate}. 
\section{Evaluation}
To explore how LMM-generated danmaku could influence viewers (RQ3), we conducted a controlled experiment assessing the effectiveness of two categories of generated danmaku (content- and emotion-related) on viewers' learning outcomes and engagement. 
In addition, we evaluated the quality of generated danmaku by comparing it with real danmaku on Bilibili. 
%

\subsection{Participants}
\begin{figure}[tb]
    \centering
    \includegraphics[width=\linewidth]{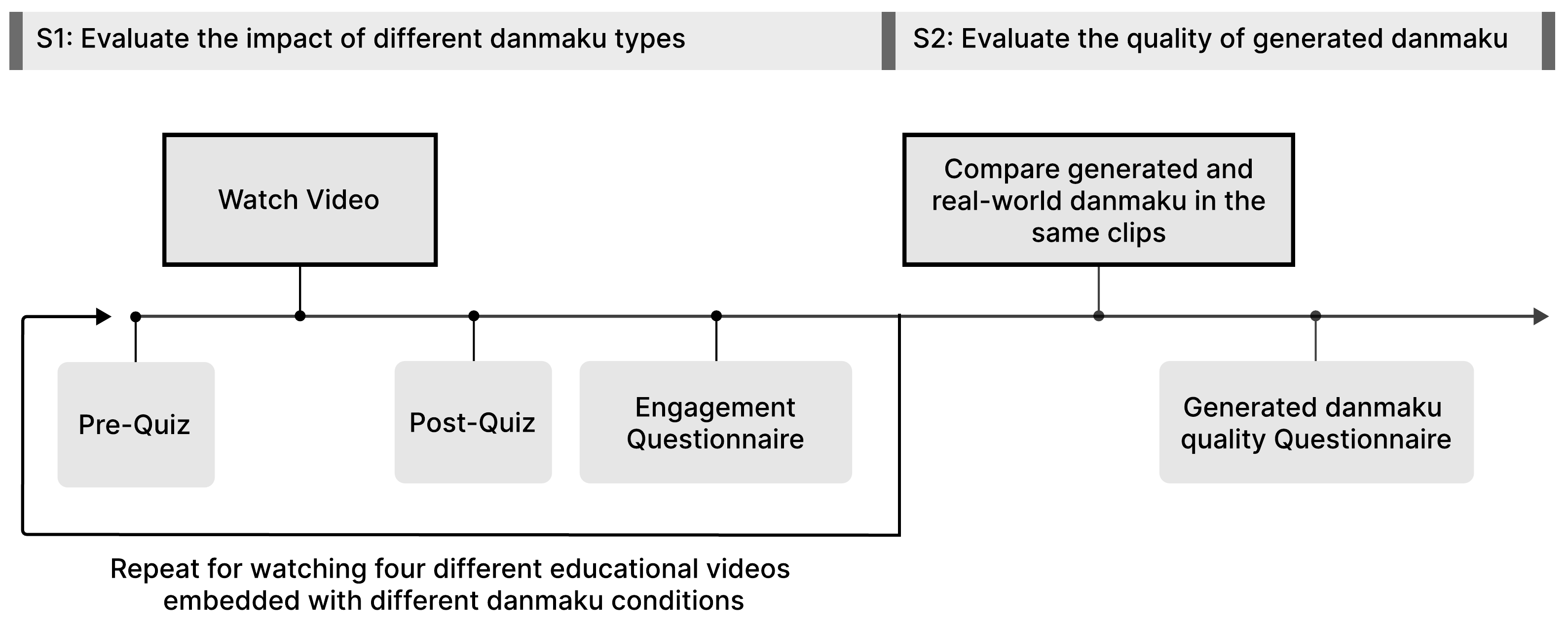}
    \vspace{-6mm}
    \caption{Diagram of the experimental procedure, consisting of two stages: (S1) evaluating the impact of different danmaku on viewers' learning outcomes and engagement, and (S2) comparing the quality of generated danmaku with human-created danmaku on Bilibili. 
    }
    \label{fig:timeline}
\end{figure}

We recruited 12 participants as U1-U12 (ages
ranging from 21 to 26, \mean{24.25}, \sd{1.36}; 6 self-identified males and 6 self-identified females), all of whom had the habit of watching or sending danmaku when they learned educational videos. 
They were undergraduate or graduate students with various majors, including Computer Science, Law, European Politics, Design, Arabic, and Information Engineering. 
Each participant was required to complete a study consisting of two stages (\autoref{fig:timeline}), lasting about 65 minutes. They were required to complete a video-watching session to compare the impact of different generated danmaku in Stage 1 and compare the quality of danmaku generated by \name{} with danmaku on Bilibili under four different metrics in Stage 2. At the end of the study, they received \${20} for their time and effort. 

\subsection{Procedure of Evaluating Learning Outcome and Engagement (Stage 1)}
To compare the learning outcome and engagement of different generated danmaku, we employed a within-subjects design to quantitatively compare four conditions, including No-Danmaku, Emotion-only, Content-only, and Content+Emotion danmaku for educational videos. Furthermore, we conducted qualitative analysis based on participants' feedback from post-interviews.

\subsubsection{Pre-quiz (3 minutes)}
Before watching each video, each participant's prior knowledge was obtained by completing a quiz. Each quiz consists of five memorization questions and five application questions (\autoref{quiz}), designed to respectively capture learners' cognitive activities~\cite{tanprasert2023scripted}. 
The 10 multiple-choice questions in each topic were thoughtfully designed by our research team mainly, primarily based on the core messages of the video. \rev{For each topic, we then asked three graduate students (Master's or PhD) who were actively doing research in the area to help validate to ensure the questions' correctness and quality.}

\subsubsection{Educational Video Watching (5-8 minutes)}
After submitting the quiz, participants were automatically redirected to the video-watching page in \name{}. 
Participants were presented with the four different danmaku conditions in a counterbalanced order. Participants were allowed to watch the video as they normally did. They could send danmaku, speed up the video, adjust the volume, or pause it. Our backend recorded the real-time participants' actions and posted danmaku. 

The selected four videos were under different topics, including supervised learning~\footnote{\url{https://www.coursera.org/learn/machine-learning/lecture/TxO6F/unsupervised-learning-part-1}}, brain structure~\footnote{\url{https://www.coursera.org/learn/introduction-psychology/lecture/mvjLn/our-two-brains}}, music theory~\footnote{\url{https://www.coursera.org/learn/edinburgh-music-theory/lecture/Osu9c/octaves}}, and latin alphabets~\footnote{\url{https://www.youtube.com/watch?v=0oWWOJW3948&t=34s}}.
Each video was about 4 to 8 minutes long. 
To mitigate biases, the order of the conditions in S1 was counterbalanced across participants, and the combination of the type of danmaku and the video was randomly assigned.

Our selection of the videos was guided by the following three considerations.
First, we selected topics with which learners would have little to no prior knowledge.
We investigated participants' majors and related courses before starting Stage 1 to ensure that all participants have minimal background in the featured fields. For example, even though some of our participants majored in Computer Science, they did not have backgrounds in AI. 
Second, we prioritized videos that are representative in quality, typicality, and format. Each educational video was selected from Coursera or YouTube, which received high ratings and good reviews from learners. The selected videos featured diverse instructional styles, including engaging animations and picture-in-picture presentations led by professors. 
Third, the content of each video was relatively independent and did not require other videos as prerequisites. All selected videos provided engaging and sufficient contextual information for viewers to learn.

\subsubsection{Post-quiz, Questionnaire, and Interview (10 minutes)}
After watching each video, participants completed the same quiz they took before watching the video. We measured participants' learning outcomes by calculating the difference between their post-quiz and pre-quiz scores. The relative value considers individual differences in baseline knowledge by using the difference between post- and pre-scores, using the difference (post - pre) normalizes these differences and provides a more comparable measure of learning progress. Additionally, participants filled out a questionnaire (\autoref{engagementquestionnaire}) which measured behavioral, emotional, and cognitive engagement on a 7-point Likert scale. We categorized the questions based on the definition of each type of engagement as described in Section~\ref{engagement}. In the context of video-based learning, behavioral engagement refers to active participation in the video (e.g., pausing, re-watching), emotional engagement reflects the emotional connection to the content (e.g., excitement, boredom), and cognitive engagement assesses the mental effort and focus invested in understanding the material (e.g., the depth of thinking, reflection).

All engagement questions were derived from six existing widely adopted questionnaires. Situational Interest scale~\cite{Chen01091999}, Original Cognitive Engagement scale~\cite{greene2015measuring}, Engagement vs. Disaffection scale~\cite{skinner2008engagement}, Classroom Engagement Inventory~\cite{wang2014measuring}, Situational Interest scale~\cite{vasalampi2016assessment}, and MOOC engagement scale~\cite{deng2020learner}.
\rev{Aligning with the process of generating engagement questionnaires in previous research~\cite{tanprasert2023scripted}, we first removed some questions that are unrelated to the context of our study, such as questions related to self-regulation (e.g., ``I denied myself rewards when I failed to reach my study goals.''). Additionally, we removed repeated questions appearing in different scales to ensure questionnaires' clarity (e.g., ``How enjoyable was the lesson?'' and ``I enjoy the lesson.'').} For similar questions (e.g., ``I pay attention in class.'' and ``I listen very carefully.''), we selected the statement that was repeated across surveys. After this, we adapted some questions to match the context of the study such as replacing ``Book'' in Original Cognitive Engagement scale by ``Video''. We show the complete engagement questionnaire generation strategy in Appendix~\ref{questionnaire strategy}.
Since there's a condition of No-Danmaku and our questionnaire is evaluated both before and after the video, we ensured that the questionnaire didn't mention danmaku to eliminate the potential bias of evaluation.

To complement the quantitative analysis, participants were asked to complete a post-interview with open-ended questions to help us understand their learning experiences, feedback, and preferences for different types of danmaku. Interviews were audio-recorded and transcribed before analysis. 

\subsection{Procedure of Evaluating Danmaku Quality (Stage 2)}
To ensure the danmaku quality of \name{}, participants were asked to compare the AI-generated danmaku with human-created danmaku on Bilibili and shared their feelings after scoring the danmaku quality of the two styles of generation. The procedure of Stage 2 lasted 15 minutes.

Since all four videos in S1 appeared on Bilibili and included available human-created danmaku, we selected six clips from the four videos on Bilibili by choosing the parts that had the most danmaku. Four of them were taken from the machine learning and psychology videos; since the psychology and Latin videos had less danmaku, we extracted two clips from them. Each clip contains more than four danmaku, which lasted from 10-25 seconds. We extracted the same clips from \name{} with the generated Content+Emotion danmaku.
The order of the six clips was randomized for each participant.
Then, we extracted the same clips from \name{} with the generated Content+Emotion danmaku. 
Participants were asked to watch Bilibili and \name{} danmaku for each clip without being informed of the order of the two danmaku modes (\name{}(AI-generated) vs. Bilibili (human-created)), in order to prevent any preconceived biases from influencing their judgment. After watching each mode, they completed a questionnaire for evaluating the danmaku quality on a 5-point Likert scale.
The questionnaire consisted of eight criteria grouped under four metrics widely used in evaluating the quality of generated live comments ~\cite{ma2018livebot}, conversations ~\cite{li2024curriculumdbot} and LLM-generated content~\cite{kryski2019neuraltextsummarizationcritical,zheng2023judgingllmasajudgemtbenchchatbot}.
These metrics are defined as follows: 

\begin{itemize}
    \item \textbf{Relevancy:} Measures the relevance between the danmaku and the educational videos (e.g., A danmaku like \qt{Great explanation of World War I!} in a video about modern history is considered more relevant than a danmaku like \qt{I love pizza!}.)

    \item \textbf{Consistency}: Assesses the factual alignment between the danmaku and the learning video (e.g., A danmaku like \qt{The Earth orbits the Sun} in a video about astronomy is considered more consistent than a danmaku like \qt{The Sun orbits the Earth}.)

    \item \textbf{Fluency}: Examines whether the danmaku has formatting problems, capitalization errors, or obviously ungrammatical sentences (e.g., A danmaku like \qt{Photosynthesis is a vital process for plants} is considered more fluent than a danmaku like \qt{Photosynthesis vital is process plants} due to word order errors.)

    \item \textbf{Coherence}: Evaluates how well danmaku fits together, maintains a logical flow, and sounds natural (e.g., A danmaku like \qt{This video explains neural networks well} followed by \qt{Neural networks are key to AI} is considered more coherent than followed by \qt{I love cats}.)
\end{itemize} 

After scoring Bilibili and \name{} danmaku, we conducted a short post-interview with open questions about their feelings about the two danmaku modes.


\section{Results}

\subsection{Quantitative Results}

In this section, we first presented the findings on participants' learning outcome and engagement (behavioral, emotional, and cognitive) in S1 of the study. We conducted one-way repeated measures ANOVAs to compare the four conditions, followed by a post hoc analysis using the Tukey HSD test~\cite{abdi2010tukey}. Then, we reported the results of the comparison between the real danmaku on Bilibili and our generated danmaku.

\begin{figure}
    \centering
    \includegraphics[width=0.8\linewidth]{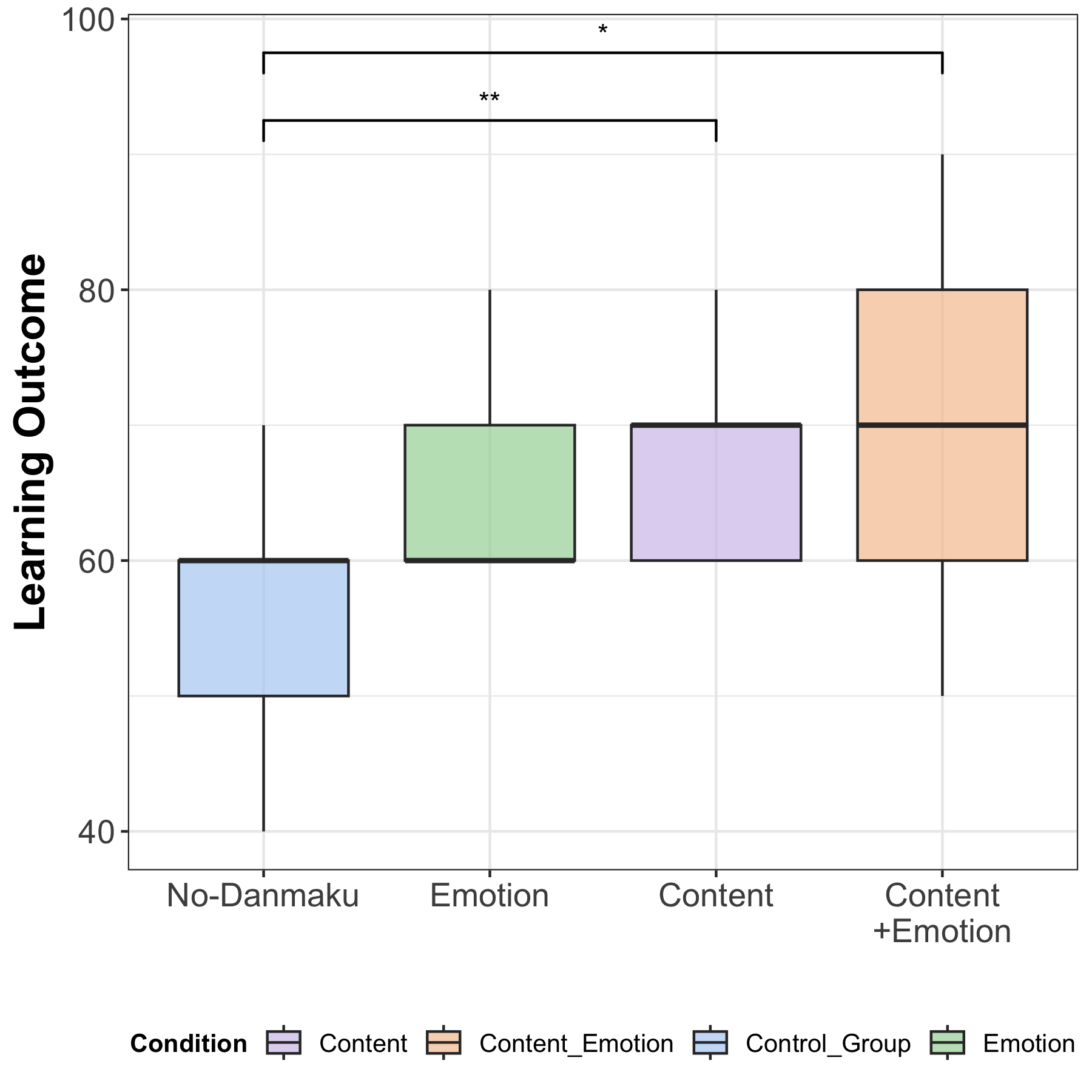}
    \vspace{-4mm}
    \caption{Box plot of learning outcome for different conditions. Error bars show 95\% confidence intervals. Asterisk ($^{*}$) indicates a statistically significant difference between conditions: $p$ < .05 (*); $p$ < .01 ($^{**}$); $p$ < .001 ($^{***}$).}
    \label{fig:learning_outcome_plot}
\end{figure}

\subsubsection{Learning Outcome}
We calculated the difference between post- and pre-quiz scores, rather than focusing on absolute scores, to measure participants' learning outcomes. Relative values are more sensitive to small changes in performance than absolute values, as they control for baseline differences and capture improvements even with low initial scores. The results revealed a significant main effect on learning outcome ($F$(4,12) = 6.47, $p$ < .01) for different danmaku conditions, as summarized in \autoref{fig:learning_outcome_plot}. 

We found the Content-only danmaku ($M$ = 67.50, $p$ < .001) and the Content+Emotion danmaku ($M$ = 70.80, $p$ < .01) showed significantly higher learning gains compared to the No-Danmaku condition.
Additionally, while the score of the Emotion-only condition ($M$ = 64.20) was higher than that of the No-Danmaku condition ($M$ = 57.5), the difference was not statistically significant. Overall, the results suggest that content-related danmaku, particularly when combined with emotional engagement, significantly enhances learning outcomes.

\subsubsection{Behavioral Engagement}
As shown in \autoref{fig:engagement_plot} (left), there was a significant main effect for the danmaku type on behavioral engagement ($F$(4, 12) = 7.94, $p$ < .01). 
Post-hoc tests showed that the participants' behavioral engagement in the No-Danmaku ($M$ = 3.67) was significantly lower than learning with Content-only ($M$ = 4.42, $p$ < .05) danmaku and Content+Emotion danmaku ($M$ = 4.54, $p$ < .01). Although learning with Emotion-only danmaku condition ($M$ = 4.19) had a notably higher mean score than the No-Danmaku group, this difference was not statistically significant.
These findings indicate that educational videos with Content+Emotion danmaku, especially the generated content-related danmaku, could improve learners' behavioral engagement with the videos.

\subsubsection{Emotional Engagement}
The ANOVA showed a main effect on emotional engagement among the conditions ($F$(4,12) = 13.20, $p$ < .01), as indicated in \autoref{fig:engagement_plot} (middle).
The emotional engagement score of the Emotion-only group ($M$ = 4.44) was significantly higher than that of the No-Danmaku group ($M$ = 3.78, $p$ < .05). However, this difference was not significant between the Content-only condition and the No-Danmaku condition. 
Similarly, the Content+Emotion ($M$ = 4.78, $p$ < .01) condition had significantly higher emotional engagement compared to that of the No-Danmaku condition. 
This implies that emotion-related danmaku indeed enhanced the emotional support for learners in educational videos. 
%

\subsubsection{Cognitive Engagement}
\begin{figure*}
    \centering
    \includegraphics[width=0.8\linewidth]{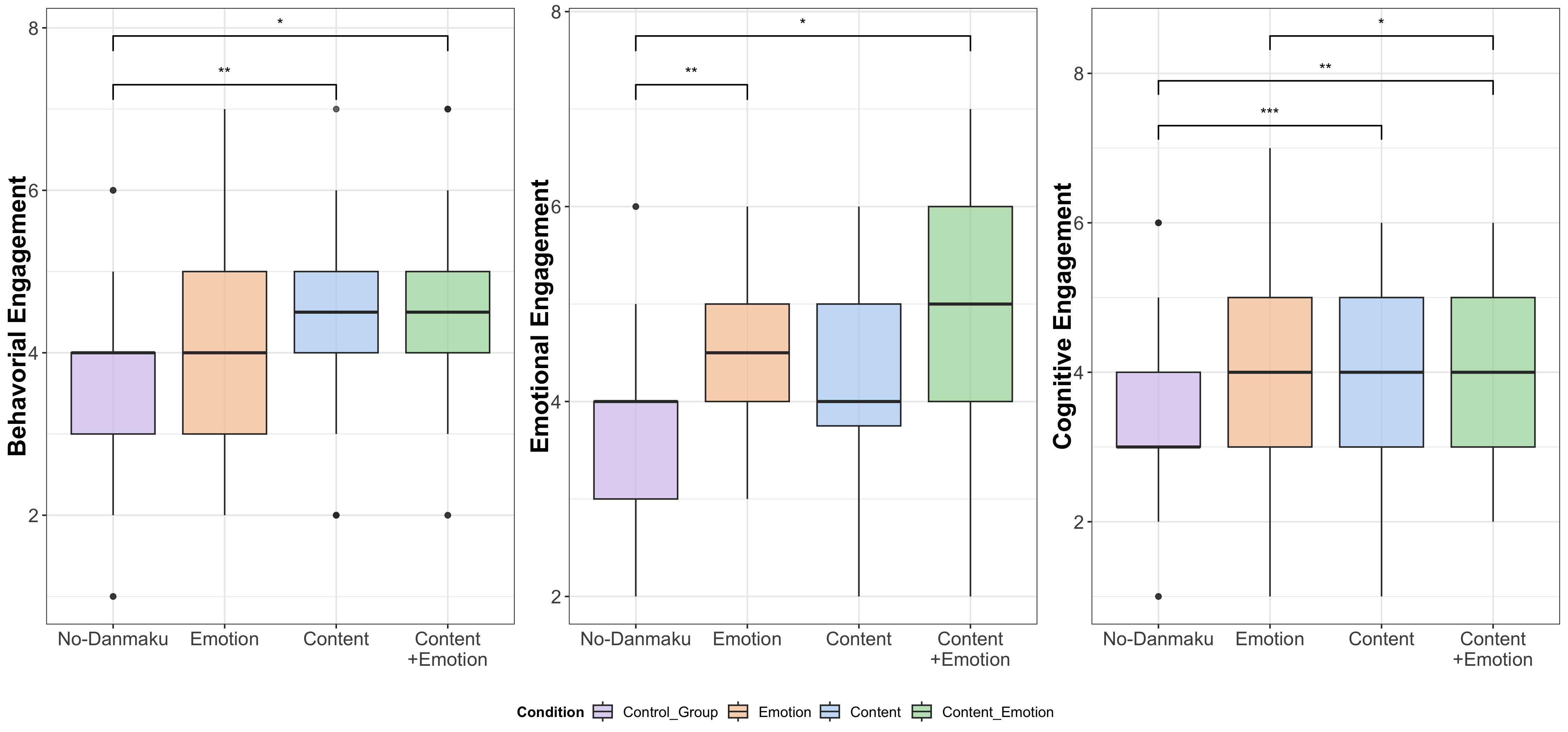}
    \vspace{-4mm}
    \caption{Box plots of behavioral engagement, emotional engagement, and cognitive engagement for different conditions. Error bars show 95\% confidence intervals. Asterisk ($^{*}$) indicates a statistically significant difference between conditions: $p$ < .05 (*); $p$ < .01 ($^{**}$); $p$ < .001 ($^{***}$).} 
    \label{fig:engagement_plot}
\end{figure*}

The effect of cognitive engagement was also significant ($F$(4,12) = 11.26, $p$ < .01), as shown in \autoref{fig:engagement_plot} (right). 
With post-hoc analyses, we found a significant difference between the No-Danmaku ($M$ = 3.43) and the Content-only danmaku condition ($M$ = 4.18, $p$ < .01) as well as the Content+Emotion condition ($M$ = 4.21, $p$ < .01).
Moreover, participants engaged in the Content+Emotion condition significantly higher than the Emotion-only condition ($M$ = 3.76, $p$ < .05).
%
%
This indicates that Content+Emotion danmaku promotes greater cognitive engagement, encouraging viewers to focus more deeply on and reflect upon the educational content.

\subsubsection{Quality of Generated Danmaku}
\begin{figure}
    \centering
    \includegraphics[width=\linewidth]{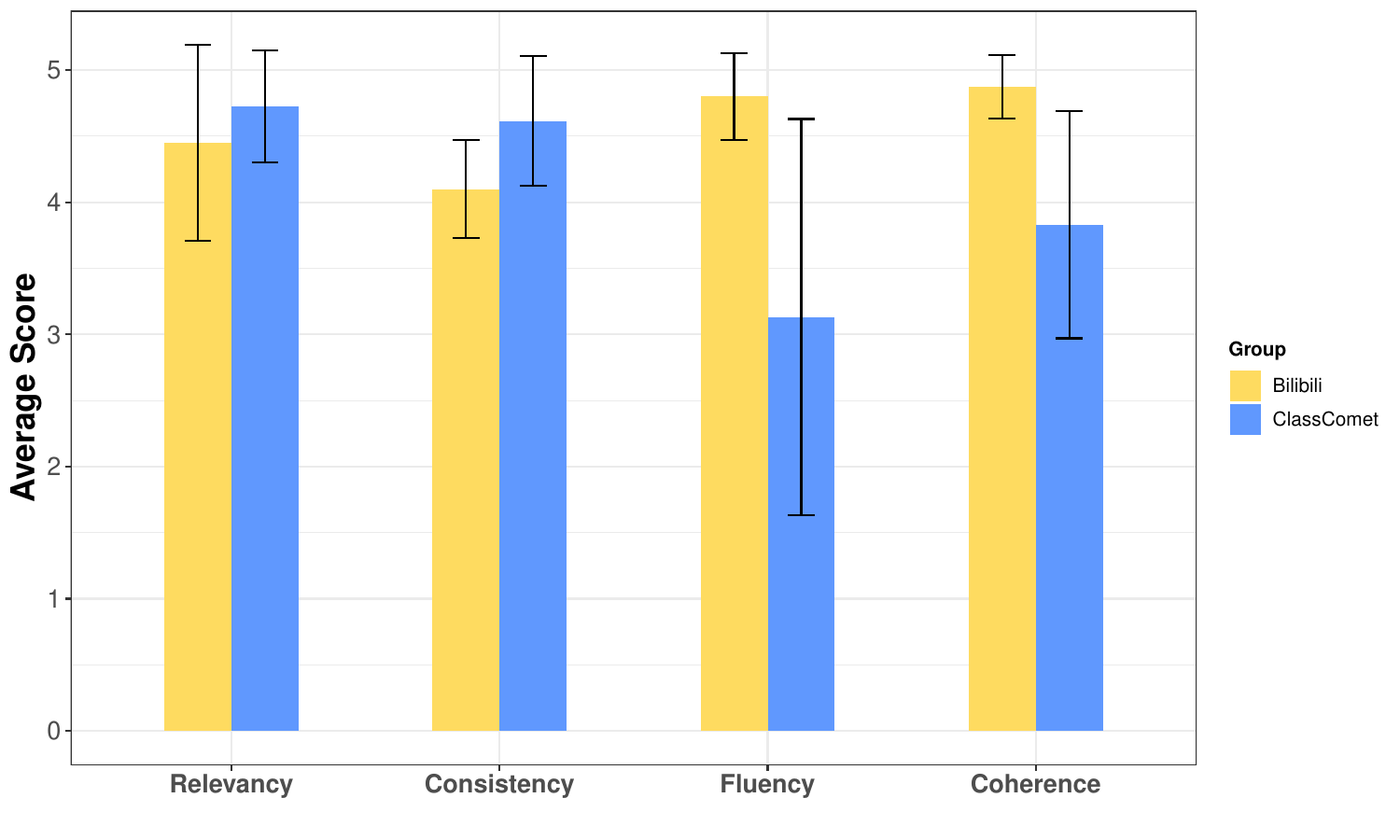}
    \vspace{-7mm}
    \caption{Means of danmaku quality score assessed by relevancy, consistency, fluency, and coherence.}
    \label{qualityfigure}
\end{figure}

As shown in \autoref{qualityfigure}, participants thought the generated danmaku superior to Bilibili danmaku in terms of relevancy. 
Regarding the metrics of consistency and fluency, both modes of danmaku exhibited similar performance. 
This shows that \name{} could produce relevant danmaku with high language quality. 
However, in terms of coherence, the \name{} danmaku was still lower than the human-created danmaku, although they were almost comparable. This underscores the need for further improvements in LMMs‘ response to align more naturally with human-created danmaku conversations~\cite{ou2024dialogbenchevaluatingllmshumanlike}.
Overall, the results indicate that our LMM-driven approach could achieve comparable quality as user-generated danmaku across multiple key metrics, while some aspect such as coherence needs further enhancement. The full questions and results for each criterion are provided in Appendix~\ref{qualityresult}.

\subsection{Qualitative Feedback}
To further explore the effects of generated danmaku, we gathered participants' feedback regarding their preferences, learning experiences, and opinions through semi-structured interviews. In the following, we denote the participants in the user study as U1-U12. 

\subsubsection{Preferences of Generated Danmaku}
More than half (75.0\%) of the participants thought Content+Emotion danmaku could be helpful and preferred it to be displayed when they were watching educational videos. Moreover, 8.3\% of them liked the Emotion-only danmaku more, while the remaining 16.7\% of participants preferred the Content-only danmaku.

Most participants acknowledged the positive impact of both classes of danmaku on their learning experiences. 
They could help them \qt{concentrate longer} and get \qt{emotion support} from emotion-related danmaku, especially when watching educational videos with no prior knowledge. For instance, U10 explained, \qt{I feel like the danmaku is more well-rounded in this condition. They're not only informative, which helps me stay focused on the video, but they're also entertaining, making the whole experience interesting and enjoyable.} 

Participants who preferred the Content-only danmaku focused more on understanding the video content. As U4 said, \qt{I can identify the key points of the video from the second condition (Content-only danmaku), so I pay extra attention to those parts. It helps me stay more concentrated.} They preferred to see less emotion-related danmaku when they put 100\% attention to grasping the video content. U7 suggested that the platform could introduce a customized filtering feature or a configuration panel to allow viewers to display their preferred danmaku types and interactions.

Interestingly, there was only one participant who liked the 
Emotion-only danmaku. He noted, \pqt{I enjoy seeing these emotion-sharing danmaku. They really resonate with me. Most of them are encouraging, which makes me feel positive and interested in the video content, even if I find it difficult. However, the content-related danmaku can sometimes feel too 'intelligent,' with answers coming too quickly when I haven't figured out the right one yet. This can be a bit annoying and adds stress.}{U1} Despite fewer participants showing preferences, most of the participants recognized the emotion-related danmaku's importance in \qt{lessening pressure and enhancing the learning atmosphere.} Furthermore, although our formative study provides direction for the design of knowledgeable virtual viewers, the concept of these virtual viewers' knowledgeable role holds the potential for expansion across diverse knowledge levels.

\subsubsection{\name{} in comparison with Bilibili}
Participants also compared their experiences with the learning of watching educational videos with danmaku on Bilibili.
U4 praised the \qt{relevance of danmaku to the video content is much higher} and U7 found \qt{more positive danmaku displayed in the learning videos.}
For content-related danmaku, U5 noted, \qt{posting frequency and distribution of danmaku were more reasonable-than Bilibili, with no issues of overlapping or rare danmaku, which is common in many educational videos.}
U2 said, \qt{I found it's much more efficient than the ones on Bilibili, which were mostly just random chatter between viewers. I feel like this makes learning much more effective now.} 
Participants were also pleased to note that there were no arguments or impolite dialogues in the danmaku. \pqt{I can feel the learning atmosphere is more focused and enjoyable.}{U10} 
Moreover, compared to other educational video platforms, participants highlighted that the danmaku is more \qt{direct and effective.} They appreciated not having to \qt{search for valuable answers in the comment sections} since the generated danmaku \qt{provides answers when the lecturer proposes a question.}

\subsubsection{Interactions with \name{} Danmaku}
We observed that there were fewer speed-up operations in the three conditions within danmaku compared to the No-Danmaku, especially in the Content+Emotion condition. The speed-up operations were 17 in No-Danmaku, whereas in the Emotion-only, Content-only, and Content+Emotion conditions, the speed-up operations were 14, 8, and 6, respectively.
U9 explained that \qt{I used to speed up the video whenever I lost focus on the content. However, I found that the danmaku consistently attracted my interest by continually highlighting or discussing the key points of the video.}
Participants noted that they would replay the moment when they saw valuable or interesting danmaku at times. U4 explained, \qt{I tried to remember those danmaku that summarize the important keys. So I would read the danmaku twice.}  

We also observed three participants sent danmaku during their learning, all of whom had the habit of sending danmaku. They expressed their different interactions and feelings when watching generated danmaku. U8 noted, \qt{I used to send joking danmaku when I watched entertaining videos, but I didn’t send that kind of danmaku this time. Instead, I answered one question in the machine learning video.} He found that \qt{as there's no chaotic danmaku, I can put more focus on the video content.}

\subsubsection{Quality of Generated Danmaku}
Regarding the quality of generated danmaku, we asked about their feelings when evaluating across the four metrics. 
Overall, participants found the generated danmaku was closely relevant to the video content. Most participants acknowledged the generated danmaku quality, particularly the content-related danmaku. U6 shared, \qt{I didn’t find anything wrong about the danmaku expressions. Whether it’s the questions or the highlights, I think \name{}'s danmaku is closely related to the video content and is worth displaying. It helps me understand the material more effectively.} 

However, in terms of emotion-related danmaku, participants had more suggestions. Four participants mentioned that the emotion-related danmaku felt less natural than the human-created danmaku, as real-world danmaku tends to have more \qt{diverse expressions.} 
U3 thought the real-life danmaku often features more \qt{social-oriented, humorous, and more internet slangs.} This feedback indicates the future improvement for \name{} in making generated danmaku more diverse and authentic.

\section{Discussion}



This research investigated the characteristics of danmaku that learners genuinely value in video-based learning, designed AI-generated danmaku for educational videos, and evaluated the quality of generated danmaku and their influence on learning experiences. In this section, we reflect on the effects and values of the two types of danmaku, and then take a closer look at the virtual viewers in the process of danmaku genration.


\subsection{Attitudes Towards Content-related and Emotion-related AI-generated Danmaku}

According to the learning outcome and engagement scores, the Emotion+Content condition consistently achieved statistically significantly high scores on the measures compared to the other three conditions. 
It indicates the comprehensive effectiveness and the necessity of combining both content- and emotion-related AI-generated danmaku in learning videos.

Upon further analysis of these two types of danmaku, we found that they had different impacts on learning outcomes and engagement. The Content-only condition showed a more significant improvement than the No-Danmaku condition in terms of learning outcomes, behavioral engagement, and cognitive engagement. However, the Emotion-only danmaku condition did not show significant effects in these areas but was notably higher in terms of emotional engagement when compared to the No-Danmaku condition.
This might find support from the qualitative feedback. Although more participants expressed a preference for content-related danmaku, most of them emphasized the necessity of emotional-related danmaku in the context of video learning. They highlighted that the emotional support provided by emotion-related danmaku is essential, though it should appear less frequently than content-related danmaku in educational videos. This preference also aligns with statistical results from previous studies, in which the number of emotion-related danmaku is fewer than content-related danmaku in learning videos~\cite{wu2018danmaku, chen2019facilitating}.

Furthermore, as some participants suggested, it would be beneficial to allow learners to select their preferred danmaku interactions, allowing customization based on their individual learning styles and preferences.
In future work, we could allow users to select generated types with a mini menu or an additional input interface to meet their personalized danmaku needs. For instance, the system could generate more tailored danmaku when the user inputs a specific prompt, enabling a more adaptive and personalized learning environment.

\subsection{Exploring the Added Values of AI-generated Danmaku}

As \name{} supports a better learning engaging experience compared to plain videos presented on learning platforms, our qualitative insights in formative study and user study also showed the advantages of automated danmaku compared with the current human-sent danmaku in video learning.
\name{} benefits from danmaku's high relevance to video content. Participants found automated danmaku in \name{} "\textit{consistently relevant through the whole videos}." Conversely, danmaku on other platforms in the learning context is normally full of unrelated interactions with the video. 
Some researchers in the HCI community proposed LLM-based danmaku moderation methods for automatically filtering out offensive danmaku~\cite{hu2024danmodcap, hu2024exploringdanmakucontentmoderation}. However, the distribution of danmaku across different educational videos is highly uneven. Newly uploaded or low-view educational videos, which tend to have minimal danmaku, are particularly difficult to moderate. Even within a single video, danmaku often clusters at only a few specific time points, and moderation could further exacerbate this uneven distribution, potentially leaving certain parts of the video without any danmaku. This may negatively impact viewers' learning experience.

\rev{Moreover, compared to user-generated danmaku, AI-generated addresses its shortcomings of quality and quantity in online educational contexts. User-generated danmaku often includes off-topic or irrelevant content, with unpredictable quality and quantity. In contrast, AI-generated danmaku adheres to user-valued characteristics identified in the formative study. It provides stable quality and quantity, consistently delivering video-relevant content unaffected by the video’s popularity or category. These two types of danmaku coexist as users can post their own danmaku while also interacting with the platform’s existing AI-generated danmaku.}

Therefore, driven by LMM of generated danmaku, \name{} enables to generate a consistent and reasonable quantity of danmaku by understanding comprehensive textual and visual information across all educational videos. This is achieved by setting detailed requirements and constraints for danmaku characteristics within the textual prompts for generation. 
%

\rev{
\subsection{Comparison to Traditional Visual Overlays}
Compared to traditional visual overlays like subtitles, danmaku offers a more dynamic and interactive viewing experience.
While subtitles are typically displayed at the bottom of the screen~\cite{DemoCut2013chi}, danmaku, on the other hand, flies across the screen dynamically, creating a more spatially flexible viewing experience.
Moreover, traditional textual overlays focus solely on conveying or supplementing the video content, which cannot highlight video concepts, prose questions, or provide emotional support. Users only receive one-way video information, which may lead to feelings of isolation. In contrast, AI-generated Danmaku incorporates both content-related and emotion-related interactions, simulating a peer-learning environment through virtual personas. This creates dynamic interactions, such as Q\&A, even when users are watching videos alone. Such interaction mimics a collaborative learning experience, helping viewers feel a sense of social connection.

Overall, AI-generated danmaku potentially provides a more immersive and community-driven experience compared to the static traditional visual overlays.}

\subsection{Perspectives on the Role of Virtual Viewers}

While our formative study guided us to design knowledgeable virtual viewers, we received feedback from U1 that expressed a different perspective on the ``intelligent'' danmaku interactions. When viewers struggle to grasp the video content, the rapid and high-quality responses from the virtual viewers could sometimes lead to frustration when the learner is not ready to process the information at that pace, increasing the pressure on learners with high self-expectations. Prior research has discussed less knowledgeable personas in chatbot designs which were referred to as teachable agents ~\cite{matsuda2013cognitive, lee2021curiosity, xiao2019should, peters2024largelanguagemodelsinfer}.
Consequently, enabling personas with different knowledge levels could be helpful for enhancing their learning experience with \name{}.

We could develop knowledgeable and teachable personas, providing users with danmaku of different cognitive levels to adapt to individual learners' preferences.

\subsection{Limitations and Future Work}

Our work is not without limitations. Future work will need to address the following limitations, as well as seek opportunities to incorporate \name{} with existing video-learning platforms by providing useful extensions besides our platform. First, the user study was conducted on undergraduates and graduates. Viewers of watching danmaku span a much broader age range, from 16 to 35 and beyond~\cite{hu2024exploringdanmakucontentmoderation, yang2024combining}. 
In future work, we may explore the effects of danmaku across different age groups and recruit a more diverse user base for the deployment of \name{}.

Second, our study was carried out as a lab study, the length of educational videos is shorter than many videos on learning platforms.We would investigate how video length impacts danmaku interactions and try dynamic adjustments to the quantity and types of danmaku for longer videos. 
\rev{Furthermore, our generated danmaku is in English which shows its potential in improving learning experience in English settings. However, the information density and word length vary across languages, which could influence the effectiveness of danmaku. Previous danmaku-related analyses have primarily been based on general videos or specific categories (e.g., crisis-related videos), without considering linguistic factors. In future work, we will account for language differences when generating danmaku.}
In addition, our study did not explore the long-term user learning experience with \name{}. Future research should fill this gap by conducting longitudinal studies to explore how user engagement with \name{} develops over time.

Third, we provided a general template for creating virtual personas without personality customization which may affect learners' engagement. Future research could support personas with different knowleadge-levels and prompt templates for tailoring personas according to personal preferences. 
\rev{Additionally, due to the clip-level understanding, persona creation, and danmaku generation processes all requiring requests to the large model, there is an inference delay that impacts the user’s viewing experience. Therefore, our videos currently use pre-loaded, pre-generated danmaku. In future work, we aim to eliminate this temporal restriction, enabling real-time danmaku generation.}

Fourth, though our user study shows learners gained significant emotional support from generated danmaku, the knowledge level of virtual personas could influence participants' emotional engagement. It would be interesting to compare the impact of virtual personas with varying knowledge levels on learners' engagement and learning outcomes in future work.

\rev{Last but not least, although our findings provide a critical foundation, the hybrid danmaku has not been fully explored. As users continuously interact with AI-generated danmaku in \name{}, the combination of AI-generated and user-generated danmaku may impact the overall quality and influence user behavior. We will design features to automatically filter low-quality danmaku or explore options for users to filter danmaku themselves in the future work.}

\section{Conclusion}
In this work, we investigated what characteristics the danmaku should possess in educational videos. We designed a LMM-driven pipeline for automatically generating high-quality emotion- and content-related danmaku for educational videos and implemented it within \name{} as a video learning platform. Through a controlled study, we found that the combination of emotion- and content-related danmaku could positively impact viewers' learning outcomes and engagement. A further evaluation indicated that our generated danmaku showed comparative quality against real-world danmaku based on four metrics. Our results not only expand the understanding of danmaku characteristics but also provide valuable insights into the design of AI-generated, informative, supportive danmaku for educational videos.


\bibliographystyle{ACM-Reference-Format}
\bibliography{ref}


\begin{thebibliography}{107}


\ifx \showCODEN    \undefined \def \showCODEN     #1{\unskip}     \fi
\ifx \showISBNx    \undefined \def \showISBNx     #1{\unskip}     \fi
\ifx \showISBNxiii \undefined \def \showISBNxiii  #1{\unskip}     \fi
\ifx \showISSN     \undefined \def \showISSN      #1{\unskip}     \fi
\ifx \showLCCN     \undefined \def \showLCCN      #1{\unskip}     \fi
\ifx \shownote     \undefined \def \shownote      #1{#1}          \fi
\ifx \showarticletitle \undefined \def \showarticletitle #1{#1}   \fi
\ifx \showURL      \undefined \def \showURL       {\relax}        \fi
\providecommand\bibfield[2]{#2}
\providecommand\bibinfo[2]{#2}
\providecommand\natexlab[1]{#1}
\providecommand\showeprint[2][]{arXiv:#2}

\bibitem[Abdi and Williams(2010)]%
        {abdi2010tukey}
\bibfield{author}{\bibinfo{person}{Herv{\'e} Abdi} {and}
  \bibinfo{person}{Lynne~J Williams}.} \bibinfo{year}{2010}\natexlab{}.
\newblock \showarticletitle{Tukey’s honestly significant difference (HSD)
  test}.
\newblock \bibinfo{journal}{\emph{Encyclopedia of research design}}
  \bibinfo{volume}{3}, \bibinfo{number}{1} (\bibinfo{year}{2010}),
  \bibinfo{pages}{1--5}.
\newblock


\bibitem[An et~al\mbox{.}(2020)]%
        {an2020comparative}
\bibfield{author}{\bibinfo{person}{Xin An}, \bibinfo{person}{Xue Gong},
  \bibinfo{person}{Rongchi Zhao}, \bibinfo{person}{Qiuyu Chen}, {and}
  \bibinfo{person}{Ning Ma}.} \bibinfo{year}{2020}\natexlab{}.
\newblock \showarticletitle{A comparative study of the interaction features of
  barrage and forum in social learning course}.
\newblock \bibinfo{journal}{\emph{International Journal of Information and
  Education Technology}} \bibinfo{volume}{10}, \bibinfo{number}{5}
  (\bibinfo{year}{2020}), \bibinfo{pages}{383--388}.
\newblock


\bibitem[Ang~Chen and Pangrazi(1999)]%
        {Chen01091999}
\bibfield{author}{\bibinfo{person}{Paul W.~Darst Ang~Chen} {and}
  \bibinfo{person}{Robert~P. Pangrazi}.} \bibinfo{year}{1999}\natexlab{}.
\newblock \showarticletitle{What Constitutes Situational Interest? Validating a
  Construct in Physical Education}.
\newblock \bibinfo{journal}{\emph{Measurement in Physical Education and
  Exercise Science}} \bibinfo{volume}{3}, \bibinfo{number}{3}
  (\bibinfo{year}{1999}), \bibinfo{pages}{157--XXX}.
\newblock
\href{https://doi.org/10.1207/s15327841mpee0303\_3}{doi:\nolinkurl{10.1207/s15327841mpee0303\_3}}
\showeprint{https://doi.org/10.1207/s15327841mpee0303\_3}


\bibitem[Antol et~al\mbox{.}(2015)]%
        {antol2015vqa}
\bibfield{author}{\bibinfo{person}{Stanislaw Antol}, \bibinfo{person}{Aishwarya
  Agrawal}, \bibinfo{person}{Jiasen Lu}, \bibinfo{person}{Margaret Mitchell},
  \bibinfo{person}{Dhruv Batra}, \bibinfo{person}{C~Lawrence Zitnick}, {and}
  \bibinfo{person}{Devi Parikh}.} \bibinfo{year}{2015}\natexlab{}.
\newblock \showarticletitle{Vqa: Visual question answering}. In
  \bibinfo{booktitle}{\emph{Proceedings of the IEEE international conference on
  computer vision}}. \bibinfo{pages}{2425--2433}.
\newblock


\bibitem[Applebee and Langer(1983)]%
        {applebee1983instructional}
\bibfield{author}{\bibinfo{person}{Arthur~N Applebee} {and}
  \bibinfo{person}{Judith~A Langer}.} \bibinfo{year}{1983}\natexlab{}.
\newblock \showarticletitle{Instructional scaffolding: Reading and writing as
  natural language activities}.
\newblock \bibinfo{journal}{\emph{Language arts}} \bibinfo{volume}{60},
  \bibinfo{number}{2} (\bibinfo{year}{1983}), \bibinfo{pages}{168--175}.
\newblock


\bibitem[Archambault et~al\mbox{.}(2009)]%
        {archambault2009student}
\bibfield{author}{\bibinfo{person}{Isabelle Archambault},
  \bibinfo{person}{Michel Janosz}, \bibinfo{person}{Jean-S{\'e}bastien Fallu},
  {and} \bibinfo{person}{Linda~S Pagani}.} \bibinfo{year}{2009}\natexlab{}.
\newblock \showarticletitle{Student engagement and its relationship with early
  high school dropout}.
\newblock \bibinfo{journal}{\emph{Journal of adolescence}}
  \bibinfo{volume}{32}, \bibinfo{number}{3} (\bibinfo{year}{2009}),
  \bibinfo{pages}{651--670}.
\newblock


\bibitem[Baig(2022)]%
        {baig2022impact}
\bibfield{author}{\bibinfo{person}{Dr.~Fawad Baig}.}
  \bibinfo{year}{2022}\natexlab{}.
\newblock \showarticletitle{The Impact of Educational Videos on the Academic
  Performance of University Students in Distance Learning}.
\newblock \bibinfo{journal}{\emph{Journal of Positive School Psychology}}
  \bibinfo{volume}{6} (\bibinfo{date}{12} \bibinfo{year}{2022}),
  \bibinfo{pages}{1233--1249}.
\newblock


\bibitem[Barnett-Itzhaki et~al\mbox{.}(2023)]%
        {videopopularity2023zohar}
\bibfield{author}{\bibinfo{person}{Zohar Barnett-Itzhaki},
  \bibinfo{person}{Dizza Beimel}, {and} \bibinfo{person}{Arava Tsoury}.}
  \bibinfo{year}{2023}\natexlab{}.
\newblock \showarticletitle{Using a Variety of Interactive Learning Methods to
  Improve Learning Effectiveness: Insights from AI Models Based on Teaching
  Surveys}.
\newblock \bibinfo{journal}{\emph{Online Learning}}  \bibinfo{volume}{27}
  (\bibinfo{date}{09} \bibinfo{year}{2023}).
\newblock
\href{https://doi.org/10.24059/olj.v27i3.3575}{doi:\nolinkurl{10.24059/olj.v27i3.3575}}


\bibitem[Cai et~al\mbox{.}(2021)]%
        {cai2021bandit}
\bibfield{author}{\bibinfo{person}{William Cai}, \bibinfo{person}{Josh
  Grossman}, \bibinfo{person}{Zhiyuan~Jerry Lin}, \bibinfo{person}{Hao Sheng},
  \bibinfo{person}{Johnny Tian-Zheng Wei}, \bibinfo{person}{Joseph~Jay
  Williams}, {and} \bibinfo{person}{Sharad Goel}.}
  \bibinfo{year}{2021}\natexlab{}.
\newblock \showarticletitle{Bandit algorithms to personalize educational
  chatbots}.
\newblock \bibinfo{journal}{\emph{Machine Learning}} \bibinfo{volume}{110},
  \bibinfo{number}{9} (\bibinfo{year}{2021}), \bibinfo{pages}{2389--2418}.
\newblock


\bibitem[Chen et~al\mbox{.}(2022b)]%
        {chen2022weaklysuperviseddataaugmentation}
\bibfield{author}{\bibinfo{person}{Maximillian Chen},
  \bibinfo{person}{Alexandros Papangelis}, \bibinfo{person}{Chenyang Tao},
  \bibinfo{person}{Andy Rosenbaum}, \bibinfo{person}{Seokhwan Kim},
  \bibinfo{person}{Yang Liu}, \bibinfo{person}{Zhou Yu}, {and}
  \bibinfo{person}{Dilek Hakkani-Tur}.} \bibinfo{year}{2022}\natexlab{b}.
\newblock \bibinfo{title}{Weakly Supervised Data Augmentation Through Prompting
  for Dialogue Understanding}.
\newblock
\showeprint[arxiv]{2210.14169}~[cs.CL]
\urldef\tempurl%
\url{https://arxiv.org/abs/2210.14169}
\showURL{%
\tempurl}


\bibitem[Chen et~al\mbox{.}(2022a)]%
        {chen2022timeline}
\bibfield{author}{\bibinfo{person}{Yue Chen}, \bibinfo{person}{Qin Gao}, {and}
  \bibinfo{person}{Ge Gao}.} \bibinfo{year}{2022}\natexlab{a}.
\newblock \showarticletitle{Timeline-anchored comments in video-based learning:
  The impact of visual layout and content depth}.
\newblock \bibinfo{journal}{\emph{International Journal of Human--Computer
  Interaction}} \bibinfo{volume}{38}, \bibinfo{number}{9}
  (\bibinfo{year}{2022}), \bibinfo{pages}{868--883}.
\newblock


\bibitem[Chen et~al\mbox{.}(2019)]%
        {chen2019facilitating}
\bibfield{author}{\bibinfo{person}{Yue Chen}, \bibinfo{person}{Qin Gao},
  \bibinfo{person}{Quan Yuan}, {and} \bibinfo{person}{Yuanli Tang}.}
  \bibinfo{year}{2019}\natexlab{}.
\newblock \showarticletitle{Facilitating students’ interaction in MOOCs
  through timeline-anchored discussion}.
\newblock \bibinfo{journal}{\emph{International Journal of Human--Computer
  Interaction}} \bibinfo{volume}{35}, \bibinfo{number}{19}
  (\bibinfo{year}{2019}), \bibinfo{pages}{1781--1799}.
\newblock


\bibitem[Chen et~al\mbox{.}(2021)]%
        {chen2021learning}
\bibfield{author}{\bibinfo{person}{Zhilong Chen}, \bibinfo{person}{Hancheng
  Cao}, \bibinfo{person}{Yuting Deng}, \bibinfo{person}{Xuan Gao},
  \bibinfo{person}{Jinghua Piao}, \bibinfo{person}{Fengli Xu},
  \bibinfo{person}{Yu Zhang}, {and} \bibinfo{person}{Yong Li}.}
  \bibinfo{year}{2021}\natexlab{}.
\newblock \showarticletitle{Learning from home: A mixed-methods analysis of
  live streaming based remote education experience in chinese colleges during
  the covid-19 pandemic}. In \bibinfo{booktitle}{\emph{Proceedings of the 2021
  CHI Conference on human factors in computing systems}}.
  \bibinfo{pages}{1--16}.
\newblock


\bibitem[Chi et~al\mbox{.}(2013)]%
        {DemoCut2013chi}
\bibfield{author}{\bibinfo{person}{Pei-Yu Chi}, \bibinfo{person}{Joyce Liu},
  \bibinfo{person}{Jason Linder}, \bibinfo{person}{Mira Dontcheva},
  \bibinfo{person}{Wilmot Li}, {and} \bibinfo{person}{Bjoern Hartmann}.}
  \bibinfo{year}{2013}\natexlab{}.
\newblock \showarticletitle{DemoCut: generating concise instructional videos
  for physical demonstrations}. In \bibinfo{booktitle}{\emph{Proceedings of the
  26th Annual ACM Symposium on User Interface Software and Technology}} (St.
  Andrews, Scotland, United Kingdom) \emph{(\bibinfo{series}{UIST '13})}.
  \bibinfo{publisher}{Association for Computing Machinery},
  \bibinfo{address}{New York, NY, USA}, \bibinfo{pages}{141–150}.
\newblock
\showISBNx{9781450322683}
\href{https://doi.org/10.1145/2501988.2502052}{doi:\nolinkurl{10.1145/2501988.2502052}}


\bibitem[Deng et~al\mbox{.}(2020)]%
        {deng2020learner}
\bibfield{author}{\bibinfo{person}{Ruiqi Deng}, \bibinfo{person}{Pierre
  Benckendorff}, {and} \bibinfo{person}{Deanne Gannaway}.}
  \bibinfo{year}{2020}\natexlab{}.
\newblock \showarticletitle{Learner engagement in MOOCs: Scale development and
  validation}.
\newblock \bibinfo{journal}{\emph{British Journal of Educational Technology}}
  \bibinfo{volume}{51}, \bibinfo{number}{1} (\bibinfo{year}{2020}),
  \bibinfo{pages}{245--262}.
\newblock


\bibitem[Dewan et~al\mbox{.}(2019)]%
        {dewan2019engagement}
\bibfield{author}{\bibinfo{person}{M Dewan}, \bibinfo{person}{Mahbub Murshed},
  {and} \bibinfo{person}{Fuhua Lin}.} \bibinfo{year}{2019}\natexlab{}.
\newblock \showarticletitle{Engagement detection in online learning: a review}.
\newblock \bibinfo{journal}{\emph{Smart Learning Environments}}
  \bibinfo{volume}{6}, \bibinfo{number}{1} (\bibinfo{year}{2019}),
  \bibinfo{pages}{1--20}.
\newblock


\bibitem[Donahue et~al\mbox{.}(2015)]%
        {donahue2015long}
\bibfield{author}{\bibinfo{person}{Jeffrey Donahue}, \bibinfo{person}{Lisa
  Anne~Hendricks}, \bibinfo{person}{Sergio Guadarrama}, \bibinfo{person}{Marcus
  Rohrbach}, \bibinfo{person}{Subhashini Venugopalan}, \bibinfo{person}{Kate
  Saenko}, {and} \bibinfo{person}{Trevor Darrell}.}
  \bibinfo{year}{2015}\natexlab{}.
\newblock \showarticletitle{Long-term recurrent convolutional networks for
  visual recognition and description}. In \bibinfo{booktitle}{\emph{Proceedings
  of the IEEE conference on computer vision and pattern recognition}}.
  \bibinfo{pages}{2625--2634}.
\newblock


\bibitem[Fang et~al\mbox{.}(2015)]%
        {fang2015captions}
\bibfield{author}{\bibinfo{person}{Hao Fang}, \bibinfo{person}{Saurabh Gupta},
  \bibinfo{person}{Forrest Iandola}, \bibinfo{person}{Rupesh~K Srivastava},
  \bibinfo{person}{Li Deng}, \bibinfo{person}{Piotr Doll{\'a}r},
  \bibinfo{person}{Jianfeng Gao}, \bibinfo{person}{Xiaodong He},
  \bibinfo{person}{Margaret Mitchell}, \bibinfo{person}{John~C Platt},
  {et~al\mbox{.}}} \bibinfo{year}{2015}\natexlab{}.
\newblock \showarticletitle{From captions to visual concepts and back}. In
  \bibinfo{booktitle}{\emph{Proceedings of the IEEE conference on computer
  vision and pattern recognition}}. \bibinfo{pages}{1473--1482}.
\newblock


\bibitem[Fern{\'a}ndez~Galeote et~al\mbox{.}(2023)]%
        {fernandez2023traditional}
\bibfield{author}{\bibinfo{person}{Daniel Fern{\'a}ndez~Galeote},
  \bibinfo{person}{Nikoletta-Zampeta Legaki}, {and} \bibinfo{person}{Juho
  Hamari}.} \bibinfo{year}{2023}\natexlab{}.
\newblock \showarticletitle{From Traditional to Game-Based Learning of Climate
  Change: A Media Comparison Experiment}.
\newblock \bibinfo{journal}{\emph{Proceedings of the ACM on Human-Computer
  Interaction}} \bibinfo{volume}{7}, \bibinfo{number}{CHI PLAY}
  (\bibinfo{year}{2023}), \bibinfo{pages}{503--525}.
\newblock


\bibitem[Finn(1993)]%
        {finn1993school}
\bibfield{author}{\bibinfo{person}{JD Finn}.} \bibinfo{year}{1993}\natexlab{}.
\newblock \showarticletitle{School Engagement \& Students At Risk}.
\newblock  (\bibinfo{year}{1993}).
\newblock


\bibitem[Finn et~al\mbox{.}(1995)]%
        {finn1995disruptive}
\bibfield{author}{\bibinfo{person}{Jeremy~D Finn}, \bibinfo{person}{Gina~M
  Pannozzo}, {and} \bibinfo{person}{Kristin~E Voelkl}.}
  \bibinfo{year}{1995}\natexlab{}.
\newblock \showarticletitle{Disruptive and inattentive-withdrawn behavior and
  achievement among fourth graders}.
\newblock \bibinfo{journal}{\emph{The Elementary School Journal}}
  \bibinfo{volume}{95}, \bibinfo{number}{5} (\bibinfo{year}{1995}),
  \bibinfo{pages}{421--434}.
\newblock


\bibitem[Finn and Rock(1997)]%
        {finn1997academic}
\bibfield{author}{\bibinfo{person}{Jeremy~D Finn} {and}
  \bibinfo{person}{Donald~A Rock}.} \bibinfo{year}{1997}\natexlab{}.
\newblock \showarticletitle{Academic success among students at risk for school
  failure.}
\newblock \bibinfo{journal}{\emph{Journal of applied psychology}}
  \bibinfo{volume}{82}, \bibinfo{number}{2} (\bibinfo{year}{1997}),
  \bibinfo{pages}{221}.
\newblock


\bibitem[Finn and Zimmer(2012)]%
        {finn2012student}
\bibfield{author}{\bibinfo{person}{Jeremy~D Finn} {and}
  \bibinfo{person}{Kayla~S Zimmer}.} \bibinfo{year}{2012}\natexlab{}.
\newblock \showarticletitle{Student engagement: What is it? Why does it
  matter?}
\newblock In \bibinfo{booktitle}{\emph{Handbook of research on student
  engagement}}. \bibinfo{publisher}{Springer}, \bibinfo{pages}{97--131}.
\newblock


\bibitem[Fredricks et~al\mbox{.}(2004)]%
        {fredricks2004school}
\bibfield{author}{\bibinfo{person}{Jennifer~A Fredricks},
  \bibinfo{person}{Phyllis~C Blumenfeld}, {and} \bibinfo{person}{Alison~H
  Paris}.} \bibinfo{year}{2004}\natexlab{}.
\newblock \showarticletitle{School engagement: Potential of the concept, state
  of the evidence}.
\newblock \bibinfo{journal}{\emph{Review of educational research}}
  \bibinfo{volume}{74}, \bibinfo{number}{1} (\bibinfo{year}{2004}),
  \bibinfo{pages}{59--109}.
\newblock


\bibitem[Giannakos et~al\mbox{.}(2014)]%
        {giannakos2014vbl}
\bibfield{author}{\bibinfo{person}{Michail Giannakos},
  \bibinfo{person}{Konstantinos Chorianopoulos}, \bibinfo{person}{Marco
  Ronchetti}, \bibinfo{person}{Peter Szegedi}, {and} \bibinfo{person}{Stephanie
  Teasley}.} \bibinfo{year}{2014}\natexlab{}.
\newblock \showarticletitle{Video-based learning and open online courses}.
\newblock  (\bibinfo{year}{2014}).
\newblock


\bibitem[Greene(2015)]%
        {greene2015measuring}
\bibfield{author}{\bibinfo{person}{Barbara~A Greene}.}
  \bibinfo{year}{2015}\natexlab{}.
\newblock \showarticletitle{Measuring cognitive engagement with self-report
  scales: Reflections from over 20 years of research}.
\newblock \bibinfo{journal}{\emph{Educational Psychologist}}
  \bibinfo{volume}{50}, \bibinfo{number}{1} (\bibinfo{year}{2015}),
  \bibinfo{pages}{14--30}.
\newblock


\bibitem[Guo et~al\mbox{.}(2014)]%
        {guo2014video}
\bibfield{author}{\bibinfo{person}{Philip~J Guo}, \bibinfo{person}{Juho Kim},
  {and} \bibinfo{person}{Rob Rubin}.} \bibinfo{year}{2014}\natexlab{}.
\newblock \showarticletitle{How video production affects student engagement: An
  empirical study of MOOC videos}. In \bibinfo{booktitle}{\emph{Proceedings of
  the first ACM conference on Learning@ scale conference}}.
  \bibinfo{pages}{41--50}.
\newblock


\bibitem[Han et~al\mbox{.}(2022)]%
        {han2022analysis}
\bibfield{author}{\bibinfo{person}{Jeong-Won Han}, \bibinfo{person}{Junhee
  Park}, {and} \bibinfo{person}{Hanna Lee}.} \bibinfo{year}{2022}\natexlab{}.
\newblock \showarticletitle{Analysis of the effect of an artificial
  intelligence chatbot educational program on non-face-to-face classes: a
  quasi-experimental study}.
\newblock \bibinfo{journal}{\emph{BMC Medical Education}} \bibinfo{volume}{22},
  \bibinfo{number}{1} (\bibinfo{year}{2022}), \bibinfo{pages}{830}.
\newblock


\bibitem[He et~al\mbox{.}(2024a)]%
        {he2024malmm}
\bibfield{author}{\bibinfo{person}{Bo He}, \bibinfo{person}{Hengduo Li},
  \bibinfo{person}{Young~Kyun Jang}, \bibinfo{person}{Menglin Jia},
  \bibinfo{person}{Xuefei Cao}, \bibinfo{person}{Ashish Shah},
  \bibinfo{person}{Abhinav Shrivastava}, {and} \bibinfo{person}{Ser-Nam Lim}.}
  \bibinfo{year}{2024}\natexlab{a}.
\newblock \showarticletitle{MA-LMM: Memory-Augmented Large Multimodal Model for
  Long-Term Video Understanding}. In \bibinfo{booktitle}{\emph{Proceedings of
  the IEEE/CVF Conference on Computer Vision and Pattern Recognition (CVPR)}}.
\newblock


\bibitem[He et~al\mbox{.}(2024b)]%
        {he2024malmmmemoryaugmentedlargemultimodal}
\bibfield{author}{\bibinfo{person}{Bo He}, \bibinfo{person}{Hengduo Li},
  \bibinfo{person}{Young~Kyun Jang}, \bibinfo{person}{Menglin Jia},
  \bibinfo{person}{Xuefei Cao}, \bibinfo{person}{Ashish Shah},
  \bibinfo{person}{Abhinav Shrivastava}, {and} \bibinfo{person}{Ser-Nam Lim}.}
  \bibinfo{year}{2024}\natexlab{b}.
\newblock \bibinfo{title}{MA-LMM: Memory-Augmented Large Multimodal Model for
  Long-Term Video Understanding}.
\newblock
\showeprint[arxiv]{2404.05726}~[cs.CV]
\urldef\tempurl%
\url{https://arxiv.org/abs/2404.05726}
\showURL{%
\tempurl}


\bibitem[He et~al\mbox{.}(2021)]%
        {he2021beyond}
\bibfield{author}{\bibinfo{person}{Changyang He}, \bibinfo{person}{Lu He},
  \bibinfo{person}{Tun Lu}, {and} \bibinfo{person}{Bo Li}.}
  \bibinfo{year}{2021}\natexlab{}.
\newblock \showarticletitle{Beyond Entertainment: Unpacking Danmaku and
  Comments' Role of Information Sharing and Sentiment Expression in Online
  Crisis Videos}.
\newblock \bibinfo{journal}{\emph{Proceedings of the ACM on Human-Computer
  Interaction}} \bibinfo{volume}{5}, \bibinfo{number}{CSCW2}
  (\bibinfo{year}{2021}), \bibinfo{pages}{1--27}.
\newblock


\bibitem[He et~al\mbox{.}(2024c)]%
        {he2024doespromptformattingimpact}
\bibfield{author}{\bibinfo{person}{Jia He}, \bibinfo{person}{Mukund Rungta},
  \bibinfo{person}{David Koleczek}, \bibinfo{person}{Arshdeep Sekhon},
  \bibinfo{person}{Franklin~X Wang}, {and} \bibinfo{person}{Sadid Hasan}.}
  \bibinfo{year}{2024}\natexlab{c}.
\newblock \bibinfo{title}{Does Prompt Formatting Have Any Impact on LLM
  Performance?}
\newblock
\showeprint[arxiv]{2411.10541}~[cs.CL]
\urldef\tempurl%
\url{https://arxiv.org/abs/2411.10541}
\showURL{%
\tempurl}


\bibitem[Hu and Lu(2024a)]%
        {hu2024exploring}
\bibfield{author}{\bibinfo{person}{Siying Hu} {and} \bibinfo{person}{Zhicong
  Lu}.} \bibinfo{year}{2024}\natexlab{a}.
\newblock \showarticletitle{Exploring the Danmaku Content Moderation on
  Video-Sharing Platforms: Existing Limitations, Challenges, and Design
  Opportunities}.
\newblock \bibinfo{journal}{\emph{arXiv preprint arXiv:2411.04529}}
  (\bibinfo{year}{2024}).
\newblock


\bibitem[Hu and Lu(2024b)]%
        {hu2024exploringdanmakucontentmoderation}
\bibfield{author}{\bibinfo{person}{Siying Hu} {and} \bibinfo{person}{Zhicong
  Lu}.} \bibinfo{year}{2024}\natexlab{b}.
\newblock \bibinfo{title}{Exploring the Danmaku Content Moderation on
  Video-Sharing Platforms: Existing Limitations, Challenges, and Design
  Opportunities}.
\newblock
\showeprint[arxiv]{2411.04529}~[cs.HC]
\urldef\tempurl%
\url{https://arxiv.org/abs/2411.04529}
\showURL{%
\tempurl}


\bibitem[Hu et~al\mbox{.}(2024)]%
        {hu2024danmodcap}
\bibfield{author}{\bibinfo{person}{Siying Hu}, \bibinfo{person}{Huanchen Wang},
  \bibinfo{person}{Yu Zhang}, \bibinfo{person}{Piaohong Wang}, {and}
  \bibinfo{person}{Zhicong Lu}.} \bibinfo{year}{2024}\natexlab{}.
\newblock \bibinfo{title}{DanModCap: Designing a Danmaku Moderation Tool for
  Video-Sharing Platforms that Leverages Impact Captions}.
\newblock
\showeprint[arxiv]{2408.02574}~[cs.HC]
\urldef\tempurl%
\url{https://arxiv.org/abs/2408.02574}
\showURL{%
\tempurl}


\bibitem[Huang et~al\mbox{.}(2024)]%
        {huang2024sharing}
\bibfield{author}{\bibinfo{person}{Zeyu Huang}, \bibinfo{person}{Xinyi Cao},
  \bibinfo{person}{Yuanhao Zhang}, {and} \bibinfo{person}{Xiaojuan Ma}.}
  \bibinfo{year}{2024}\natexlab{}.
\newblock \showarticletitle{Sharing Frissons among Online Video Viewers:
  Exploring the Design of Affective Communication for Aesthetic Chills}. In
  \bibinfo{booktitle}{\emph{Proceedings of the CHI Conference on Human Factors
  in Computing Systems}}. \bibinfo{pages}{1--19}.
\newblock


\bibitem[Hung et~al\mbox{.}(2024)]%
        {hung2024simtube}
\bibfield{author}{\bibinfo{person}{Yu-Kai Hung}, \bibinfo{person}{Yun-Chien
  Huang}, \bibinfo{person}{Ting-Yu Su}, \bibinfo{person}{Yen-Ting Lin},
  \bibinfo{person}{Lung-Pan Cheng}, \bibinfo{person}{Bryan Wang}, {and}
  \bibinfo{person}{Shao-Hua Sun}.} \bibinfo{year}{2024}\natexlab{}.
\newblock \showarticletitle{SimTube: Generating Simulated Video Comments
  through Multimodal AI and User Personas}.
\newblock \bibinfo{journal}{\emph{arXiv preprint arXiv:2411.09577}}
  (\bibinfo{year}{2024}).
\newblock


\bibitem[Islam et~al\mbox{.}(2024)]%
        {islam2024videorecaprecursivecaptioning}
\bibfield{author}{\bibinfo{person}{Md~Mohaiminul Islam}, \bibinfo{person}{Ngan
  Ho}, \bibinfo{person}{Xitong Yang}, \bibinfo{person}{Tushar Nagarajan},
  \bibinfo{person}{Lorenzo Torresani}, {and} \bibinfo{person}{Gedas
  Bertasius}.} \bibinfo{year}{2024}\natexlab{}.
\newblock \bibinfo{title}{Video ReCap: Recursive Captioning of Hour-Long
  Videos}.
\newblock
\showeprint[arxiv]{2402.13250}~[cs.CV]
\urldef\tempurl%
\url{https://arxiv.org/abs/2402.13250}
\showURL{%
\tempurl}


\bibitem[Jin et~al\mbox{.}(2023)]%
        {jin2023collaborative}
\bibfield{author}{\bibinfo{person}{Qiao Jin}, \bibinfo{person}{Yu Liu},
  \bibinfo{person}{Ruixuan Sun}, \bibinfo{person}{Chen Chen},
  \bibinfo{person}{Puqi Zhou}, \bibinfo{person}{Bo Han}, \bibinfo{person}{Feng
  Qian}, {and} \bibinfo{person}{Svetlana Yarosh}.}
  \bibinfo{year}{2023}\natexlab{}.
\newblock \showarticletitle{Collaborative online learning with vr video: Roles
  of collaborative tools and shared video control}. In
  \bibinfo{booktitle}{\emph{Proceedings of the 2023 CHI Conference on Human
  Factors in Computing Systems}}. \bibinfo{pages}{1--18}.
\newblock


\bibitem[Kahatapitiya et~al\mbox{.}(2024)]%
        {kahatapitiya2024languagerepositorylongvideo}
\bibfield{author}{\bibinfo{person}{Kumara Kahatapitiya},
  \bibinfo{person}{Kanchana Ranasinghe}, \bibinfo{person}{Jongwoo Park}, {and}
  \bibinfo{person}{Michael~S. Ryoo}.} \bibinfo{year}{2024}\natexlab{}.
\newblock \bibinfo{title}{Language Repository for Long Video Understanding}.
\newblock
\showeprint[arxiv]{2403.14622}~[cs.CV]
\urldef\tempurl%
\url{https://arxiv.org/abs/2403.14622}
\showURL{%
\tempurl}


\bibitem[Karpathy and Fei-Fei(2015)]%
        {karpathy2015deep}
\bibfield{author}{\bibinfo{person}{Andrej Karpathy} {and} \bibinfo{person}{Li
  Fei-Fei}.} \bibinfo{year}{2015}\natexlab{}.
\newblock \showarticletitle{Deep visual-semantic alignments for generating
  image descriptions}. In \bibinfo{booktitle}{\emph{Proceedings of the IEEE
  conference on computer vision and pattern recognition}}.
  \bibinfo{pages}{3128--3137}.
\newblock


\bibitem[Kim et~al\mbox{.}(2023a)]%
        {kim2023sodamillionscaledialoguedistillation}
\bibfield{author}{\bibinfo{person}{Hyunwoo Kim}, \bibinfo{person}{Jack Hessel},
  \bibinfo{person}{Liwei Jiang}, \bibinfo{person}{Peter West},
  \bibinfo{person}{Ximing Lu}, \bibinfo{person}{Youngjae Yu},
  \bibinfo{person}{Pei Zhou}, \bibinfo{person}{Ronan~Le Bras},
  \bibinfo{person}{Malihe Alikhani}, \bibinfo{person}{Gunhee Kim},
  \bibinfo{person}{Maarten Sap}, {and} \bibinfo{person}{Yejin Choi}.}
  \bibinfo{year}{2023}\natexlab{a}.
\newblock \bibinfo{title}{SODA: Million-scale Dialogue Distillation with Social
  Commonsense Contextualization}.
\newblock
\showeprint[arxiv]{2212.10465}~[cs.CL]
\urldef\tempurl%
\url{https://arxiv.org/abs/2212.10465}
\showURL{%
\tempurl}


\bibitem[Kim et~al\mbox{.}(2024)]%
        {kim2024personadoubleedgedswordmitigating}
\bibfield{author}{\bibinfo{person}{Junseok Kim}, \bibinfo{person}{Nakyeong
  Yang}, {and} \bibinfo{person}{Kyomin Jung}.} \bibinfo{year}{2024}\natexlab{}.
\newblock \bibinfo{title}{Persona is a Double-edged Sword: Mitigating the
  Negative Impact of Role-playing Prompts in Zero-shot Reasoning Tasks}.
\newblock
\showeprint[arxiv]{2408.08631}~[cs.CL]
\urldef\tempurl%
\url{https://arxiv.org/abs/2408.08631}
\showURL{%
\tempurl}


\bibitem[Kim et~al\mbox{.}(2023b)]%
        {kim2023older}
\bibfield{author}{\bibinfo{person}{Seoyoung Kim}, \bibinfo{person}{Donghoon
  Shin}, \bibinfo{person}{Jeongyeon Kim}, \bibinfo{person}{Soonwoo Kwon}, {and}
  \bibinfo{person}{Juho Kim}.} \bibinfo{year}{2023}\natexlab{b}.
\newblock \showarticletitle{How Older Adults Use Online Videos for Learning}.
  In \bibinfo{booktitle}{\emph{Proceedings of the 2023 CHI Conference on Human
  Factors in Computing Systems}}. \bibinfo{pages}{1--16}.
\newblock


\bibitem[Kryściński et~al\mbox{.}(2019)]%
        {kryski2019neuraltextsummarizationcritical}
\bibfield{author}{\bibinfo{person}{Wojciech Kryściński},
  \bibinfo{person}{Nitish~Shirish Keskar}, \bibinfo{person}{Bryan McCann},
  \bibinfo{person}{Caiming Xiong}, {and} \bibinfo{person}{Richard Socher}.}
  \bibinfo{year}{2019}\natexlab{}.
\newblock \bibinfo{title}{Neural Text Summarization: A Critical Evaluation}.
\newblock
\showeprint[arxiv]{1908.08960}~[cs.CL]
\urldef\tempurl%
\url{https://arxiv.org/abs/1908.08960}
\showURL{%
\tempurl}


\bibitem[Labadze et~al\mbox{.}(2023)]%
        {labadze2023role}
\bibfield{author}{\bibinfo{person}{Lasha Labadze}, \bibinfo{person}{Maya
  Grigolia}, {and} \bibinfo{person}{Lela Machaidze}.}
  \bibinfo{year}{2023}\natexlab{}.
\newblock \showarticletitle{Role of AI chatbots in education: systematic
  literature review}.
\newblock \bibinfo{journal}{\emph{International Journal of Educational
  Technology in Higher Education}} \bibinfo{volume}{20}, \bibinfo{number}{1}
  (\bibinfo{year}{2023}), \bibinfo{pages}{56}.
\newblock


\bibitem[Lee et~al\mbox{.}(2021)]%
        {lee2021curiosity}
\bibfield{author}{\bibinfo{person}{Ken~Jen Lee}, \bibinfo{person}{Apoorva
  Chauhan}, \bibinfo{person}{Joslin Goh}, \bibinfo{person}{Elizabeth Nilsen},
  {and} \bibinfo{person}{Edith Law}.} \bibinfo{year}{2021}\natexlab{}.
\newblock \showarticletitle{Curiosity notebook: the design of a research
  platform for learning by teaching}.
\newblock \bibinfo{journal}{\emph{Proceedings of the ACM on Human-Computer
  Interaction}} \bibinfo{volume}{5}, \bibinfo{number}{CSCW2}
  (\bibinfo{year}{2021}), \bibinfo{pages}{1--26}.
\newblock


\bibitem[Li et~al\mbox{.}(2024c)]%
        {li2024llavanextinterleavetacklingmultiimagevideo}
\bibfield{author}{\bibinfo{person}{Feng Li}, \bibinfo{person}{Renrui Zhang},
  \bibinfo{person}{Hao Zhang}, \bibinfo{person}{Yuanhan Zhang},
  \bibinfo{person}{Bo Li}, \bibinfo{person}{Wei Li}, \bibinfo{person}{Zejun
  Ma}, {and} \bibinfo{person}{Chunyuan Li}.} \bibinfo{year}{2024}\natexlab{c}.
\newblock \bibinfo{title}{LLaVA-NeXT-Interleave: Tackling Multi-image, Video,
  and 3D in Large Multimodal Models}.
\newblock
\showeprint[arxiv]{2407.07895}~[cs.CV]
\urldef\tempurl%
\url{https://arxiv.org/abs/2407.07895}
\showURL{%
\tempurl}


\bibitem[Li et~al\mbox{.}(2024a)]%
        {li2024videochat}
\bibfield{author}{\bibinfo{person}{KunChang Li}, \bibinfo{person}{Yinan He},
  \bibinfo{person}{Yi Wang}, \bibinfo{person}{Yizhuo Li},
  \bibinfo{person}{Wenhai Wang}, \bibinfo{person}{Ping Luo},
  \bibinfo{person}{Yali Wang}, \bibinfo{person}{Limin Wang}, {and}
  \bibinfo{person}{Yu Qiao}.} \bibinfo{year}{2024}\natexlab{a}.
\newblock \bibinfo{title}{VideoChat: Chat-Centric Video Understanding}.
\newblock
\showeprint[arxiv]{2305.06355}~[cs.CV]
\urldef\tempurl%
\url{https://arxiv.org/abs/2305.06355}
\showURL{%
\tempurl}


\bibitem[Li et~al\mbox{.}(2024b)]%
        {li2024curriculumdbot}
\bibfield{author}{\bibinfo{person}{Yu Li}, \bibinfo{person}{Shang Qu},
  \bibinfo{person}{Jili Shen}, \bibinfo{person}{Shangchao Min}, {and}
  \bibinfo{person}{Zhou Yu}.} \bibinfo{year}{2024}\natexlab{b}.
\newblock \bibinfo{title}{Curriculum-Driven Edubot: A Framework for Developing
  Language Learning Chatbots Through Synthesizing Conversational Data}.
\newblock
\showeprint[arxiv]{2309.16804}~[cs.CL]
\urldef\tempurl%
\url{https://arxiv.org/abs/2309.16804}
\showURL{%
\tempurl}


\bibitem[Li(2022)]%
        {li2022influence}
\bibfield{author}{\bibinfo{person}{Zhikai Li}.}
  \bibinfo{year}{2022}\natexlab{}.
\newblock \showarticletitle{Influence of online learning behavior and video
  playing questions on students’ learning effect}.
\newblock \bibinfo{journal}{\emph{International Journal of Emerging
  Technologies in Learning (iJET)}} \bibinfo{volume}{17}, \bibinfo{number}{2}
  (\bibinfo{year}{2022}), \bibinfo{pages}{223--238}.
\newblock


\bibitem[Lin et~al\mbox{.}(2023)]%
        {lin2023mmvid}
\bibfield{author}{\bibinfo{person}{Kevin Lin}, \bibinfo{person}{Faisal Ahmed},
  \bibinfo{person}{Linjie Li}, \bibinfo{person}{Chung-Ching Lin},
  \bibinfo{person}{Ehsan Azarnasab}, \bibinfo{person}{Zhengyuan Yang},
  \bibinfo{person}{Jianfeng Wang}, \bibinfo{person}{Lin Liang},
  \bibinfo{person}{Zicheng Liu}, \bibinfo{person}{Yumao Lu}, {et~al\mbox{.}}}
  \bibinfo{year}{2023}\natexlab{}.
\newblock \showarticletitle{Mm-vid: Advancing video understanding with gpt-4v
  (ision)}.
\newblock \bibinfo{journal}{\emph{arXiv preprint arXiv:2310.19773}}
  (\bibinfo{year}{2023}).
\newblock


\bibitem[Linnenbrink and Pintrich(2003)]%
        {linnenbrink2003role}
\bibfield{author}{\bibinfo{person}{Elizabeth~A Linnenbrink} {and}
  \bibinfo{person}{Paul~R Pintrich}.} \bibinfo{year}{2003}\natexlab{}.
\newblock \showarticletitle{The role of self-efficacy beliefs instudent
  engagement and learning intheclassroom}.
\newblock \bibinfo{journal}{\emph{Reading \&Writing Quarterly}}
  \bibinfo{volume}{19}, \bibinfo{number}{2} (\bibinfo{year}{2003}),
  \bibinfo{pages}{119--137}.
\newblock


\bibitem[Liu et~al\mbox{.}(2024)]%
        {liu2024mmcadvancingmultimodalchart}
\bibfield{author}{\bibinfo{person}{Fuxiao Liu}, \bibinfo{person}{Xiaoyang
  Wang}, \bibinfo{person}{Wenlin Yao}, \bibinfo{person}{Jianshu Chen},
  \bibinfo{person}{Kaiqiang Song}, \bibinfo{person}{Sangwoo Cho},
  \bibinfo{person}{Yaser Yacoob}, {and} \bibinfo{person}{Dong Yu}.}
  \bibinfo{year}{2024}\natexlab{}.
\newblock \bibinfo{title}{MMC: Advancing Multimodal Chart Understanding with
  Large-scale Instruction Tuning}.
\newblock
\showeprint[arxiv]{2311.10774}~[cs.CL]
\urldef\tempurl%
\url{https://arxiv.org/abs/2311.10774}
\showURL{%
\tempurl}


\bibitem[Ma et~al\mbox{.}(2018)]%
        {ma2018livebot}
\bibfield{author}{\bibinfo{person}{Shuming Ma}, \bibinfo{person}{Lei Cui},
  \bibinfo{person}{Damai Dai}, \bibinfo{person}{Furu Wei}, {and}
  \bibinfo{person}{Xu Sun}.} \bibinfo{year}{2018}\natexlab{}.
\newblock \bibinfo{title}{LiveBot: Generating Live Video Comments Based on
  Visual and Textual Contexts}.
\newblock
\showeprint[arxiv]{1809.04938}~[cs.CL]
\urldef\tempurl%
\url{https://arxiv.org/abs/1809.04938}
\showURL{%
\tempurl}


\bibitem[MA et~al\mbox{.}(2024)]%
        {ma2024and}
\bibfield{author}{\bibinfo{person}{Wenjuan MA}, \bibinfo{person}{Wenjing MA},
  \bibinfo{person}{Yongbin HU}, {and} \bibinfo{person}{Xinyu BI}.}
  \bibinfo{year}{2024}\natexlab{}.
\newblock \showarticletitle{The who, why, and how of ai-based chatbots for
  learning and teaching in higher education: A systematic review}.
\newblock \bibinfo{journal}{\emph{Education and Information Technologies}}
  (\bibinfo{year}{2024}), \bibinfo{pages}{1--25}.
\newblock


\bibitem[Ma and Cao(2017)]%
        {ma2017video}
\bibfield{author}{\bibinfo{person}{Xiaojuan Ma} {and} \bibinfo{person}{Nan
  Cao}.} \bibinfo{year}{2017}\natexlab{}.
\newblock \showarticletitle{Video-based evanescent, anonymous, asynchronous
  social interaction: Motivation and adaption to medium}. In
  \bibinfo{booktitle}{\emph{Proceedings of the 2017 ACM conference on computer
  supported cooperative work and social computing}}. \bibinfo{pages}{770--782}.
\newblock


\bibitem[Matsuda et~al\mbox{.}(2013)]%
        {matsuda2013cognitive}
\bibfield{author}{\bibinfo{person}{Noboru Matsuda}, \bibinfo{person}{Evelyn
  Yarzebinski}, \bibinfo{person}{Victoria Keiser}, \bibinfo{person}{Rohan
  Raizada}, \bibinfo{person}{William~W Cohen}, \bibinfo{person}{Gabriel~J
  Stylianides}, {and} \bibinfo{person}{Kenneth~R Koedinger}.}
  \bibinfo{year}{2013}\natexlab{}.
\newblock \showarticletitle{Cognitive anatomy of tutor learning: Lessons
  learned with SimStudent.}
\newblock \bibinfo{journal}{\emph{Journal of Educational Psychology}}
  \bibinfo{volume}{105}, \bibinfo{number}{4} (\bibinfo{year}{2013}),
  \bibinfo{pages}{1152}.
\newblock


\bibitem[Mayer(2014)]%
        {mayer2014ctml}
\bibfield{author}{\bibinfo{person}{Richard~E Mayer}.}
  \bibinfo{year}{2014}\natexlab{}.
\newblock \showarticletitle{Incorporating motivation into multimedia learning}.
\newblock \bibinfo{journal}{\emph{Learning and instruction}}
  \bibinfo{volume}{29} (\bibinfo{year}{2014}), \bibinfo{pages}{171--173}.
\newblock


\bibitem[Mohd~Dahlan et~al\mbox{.}(2023)]%
        {interactivevideolearning2023mariani}
\bibfield{author}{\bibinfo{person}{Mariani Mohd~Dahlan}, \bibinfo{person}{Nuri
  Halim}, \bibinfo{person}{Noor Kamarudin}, {and} \bibinfo{person}{Fatanah
  Ahmad}.} \bibinfo{year}{2023}\natexlab{}.
\newblock \showarticletitle{Exploring interactive video learning: Techniques,
  applications, and pedagogical insights}.
\newblock \bibinfo{journal}{\emph{International Journal of ADVANCED AND APPLIED
  SCIENCES}}  \bibinfo{volume}{10} (\bibinfo{date}{12} \bibinfo{year}{2023}),
  \bibinfo{pages}{220--230}.
\newblock
\href{https://doi.org/10.21833/ijaas.2023.12.024}{doi:\nolinkurl{10.21833/ijaas.2023.12.024}}


\bibitem[Mou et~al\mbox{.}(2022)]%
        {mou2022interactivity}
\bibfield{author}{\bibinfo{person}{Ya Mou}, \bibinfo{person}{Bin Jing},
  \bibinfo{person}{Yichun Li}, \bibinfo{person}{Nanyang Fang}, {and}
  \bibinfo{person}{Changcheng Wu}.} \bibinfo{year}{2022}\natexlab{}.
\newblock \showarticletitle{Interactivity in learning instructional videos:
  Sending danmaku improved parasocial interaction but reduced learning
  performance}.
\newblock \bibinfo{journal}{\emph{Frontiers in Psychology}}
  \bibinfo{volume}{13} (\bibinfo{year}{2022}), \bibinfo{pages}{1066164}.
\newblock


\bibitem[Nafar et~al\mbox{.}(2024)]%
        {nafar2024learningvsretrievalrole}
\bibfield{author}{\bibinfo{person}{Aliakbar Nafar},
  \bibinfo{person}{Kristen~Brent Venable}, {and} \bibinfo{person}{Parisa
  Kordjamshidi}.} \bibinfo{year}{2024}\natexlab{}.
\newblock \bibinfo{title}{Learning vs Retrieval: The Role of In-Context
  Examples in Regression with LLMs}.
\newblock
\showeprint[arxiv]{2409.04318}~[cs.CL]
\urldef\tempurl%
\url{https://arxiv.org/abs/2409.04318}
\showURL{%
\tempurl}


\bibitem[Ou et~al\mbox{.}(2024)]%
        {ou2024dialogbenchevaluatingllmshumanlike}
\bibfield{author}{\bibinfo{person}{Jiao Ou}, \bibinfo{person}{Junda Lu},
  \bibinfo{person}{Che Liu}, \bibinfo{person}{Yihong Tang},
  \bibinfo{person}{Fuzheng Zhang}, \bibinfo{person}{Di Zhang}, {and}
  \bibinfo{person}{Kun Gai}.} \bibinfo{year}{2024}\natexlab{}.
\newblock \bibinfo{title}{DialogBench: Evaluating LLMs as Human-like Dialogue
  Systems}.
\newblock
\showeprint[arxiv]{2311.01677}~[cs.CL]
\urldef\tempurl%
\url{https://arxiv.org/abs/2311.01677}
\showURL{%
\tempurl}


\bibitem[Park et~al\mbox{.}(2022)]%
        {park2022social}
\bibfield{author}{\bibinfo{person}{Joon~Sung Park}, \bibinfo{person}{Lindsay
  Popowski}, \bibinfo{person}{Carrie Cai}, \bibinfo{person}{Meredith~Ringel
  Morris}, \bibinfo{person}{Percy Liang}, {and} \bibinfo{person}{Michael~S
  Bernstein}.} \bibinfo{year}{2022}\natexlab{}.
\newblock \showarticletitle{Social simulacra: Creating populated prototypes for
  social computing systems}. In \bibinfo{booktitle}{\emph{Proceedings of the
  35th Annual ACM Symposium on User Interface Software and Technology}}.
  \bibinfo{pages}{1--18}.
\newblock


\bibitem[Peters et~al\mbox{.}(2024)]%
        {peters2024largelanguagemodelsinfer}
\bibfield{author}{\bibinfo{person}{Heinrich Peters}, \bibinfo{person}{Moran
  Cerf}, {and} \bibinfo{person}{Sandra~C. Matz}.}
  \bibinfo{year}{2024}\natexlab{}.
\newblock \bibinfo{title}{Large Language Models Can Infer Personality from
  Free-Form User Interactions}.
\newblock
\showeprint[arxiv]{2405.13052}~[cs.HC]
\urldef\tempurl%
\url{https://arxiv.org/abs/2405.13052}
\showURL{%
\tempurl}


\bibitem[Piro et~al\mbox{.}(2024)]%
        {piro2024mylearningtalk}
\bibfield{author}{\bibinfo{person}{Ludovica Piro}, \bibinfo{person}{Tommaso
  Bianchi}, \bibinfo{person}{Luca Alessandrelli}, \bibinfo{person}{Andrea
  Chizzola}, \bibinfo{person}{Daniela Casiraghi}, \bibinfo{person}{Susanna
  Sancassani}, {and} \bibinfo{person}{Nicola Gatti}.}
  \bibinfo{year}{2024}\natexlab{}.
\newblock \showarticletitle{MyLearningTalk: An LLM-Based Intelligent Tutoring
  System}. In \bibinfo{booktitle}{\emph{International Conference on Web
  Engineering}}. Springer, \bibinfo{pages}{428--431}.
\newblock


\bibitem[Qian et~al\mbox{.}(2024)]%
        {qian2024momentoradvancingvideolarge}
\bibfield{author}{\bibinfo{person}{Long Qian}, \bibinfo{person}{Juncheng Li},
  \bibinfo{person}{Yu Wu}, \bibinfo{person}{Yaobo Ye}, \bibinfo{person}{Hao
  Fei}, \bibinfo{person}{Tat-Seng Chua}, \bibinfo{person}{Yueting Zhuang},
  {and} \bibinfo{person}{Siliang Tang}.} \bibinfo{year}{2024}\natexlab{}.
\newblock \bibinfo{title}{Momentor: Advancing Video Large Language Model with
  Fine-Grained Temporal Reasoning}.
\newblock
\showeprint[arxiv]{2402.11435}~[cs.CV]
\urldef\tempurl%
\url{https://arxiv.org/abs/2402.11435}
\showURL{%
\tempurl}


\bibitem[Ren et~al\mbox{.}(2024)]%
        {ren2024timechattimesensitivemultimodallarge}
\bibfield{author}{\bibinfo{person}{Shuhuai Ren}, \bibinfo{person}{Linli Yao},
  \bibinfo{person}{Shicheng Li}, \bibinfo{person}{Xu Sun}, {and}
  \bibinfo{person}{Lu Hou}.} \bibinfo{year}{2024}\natexlab{}.
\newblock \bibinfo{title}{TimeChat: A Time-sensitive Multimodal Large Language
  Model for Long Video Understanding}.
\newblock
\showeprint[arxiv]{2312.02051}~[cs.CV]
\urldef\tempurl%
\url{https://arxiv.org/abs/2312.02051}
\showURL{%
\tempurl}


\bibitem[Ren et~al\mbox{.}(2022)]%
        {ren2022multi}
\bibfield{author}{\bibinfo{person}{Yuchen Ren}, \bibinfo{person}{Yuan Yuan},
  {and} \bibinfo{person}{Lei Chen}.} \bibinfo{year}{2022}\natexlab{}.
\newblock \showarticletitle{Multi-modal guided attention for live video
  comments generation}. In \bibinfo{booktitle}{\emph{International Conference
  on Computer Graphics, Artificial Intelligence, and Data Processing (ICCAID
  2021)}}, Vol.~\bibinfo{volume}{12168}. SPIE, \bibinfo{pages}{267--273}.
\newblock


\bibitem[Rohrbach et~al\mbox{.}(2014)]%
        {rohrbach2014coherent}
\bibfield{author}{\bibinfo{person}{Anna Rohrbach}, \bibinfo{person}{Marcus
  Rohrbach}, \bibinfo{person}{Wei Qiu}, \bibinfo{person}{Annemarie Friedrich},
  \bibinfo{person}{Manfred Pinkal}, {and} \bibinfo{person}{Bernt Schiele}.}
  \bibinfo{year}{2014}\natexlab{}.
\newblock \showarticletitle{Coherent multi-sentence video description with
  variable level of detail}. In \bibinfo{booktitle}{\emph{Pattern Recognition:
  36th German Conference, GCPR 2014, M{\"u}nster, Germany, September 2-5, 2014,
  Proceedings 36}}. Springer, \bibinfo{pages}{184--195}.
\newblock


\bibitem[Sabli{\'c} et~al\mbox{.}(2021)]%
        {sablic2021vblsurvey}
\bibfield{author}{\bibinfo{person}{Marija Sabli{\'c}}, \bibinfo{person}{Ana
  Mirosavljevi{\'c}}, {and} \bibinfo{person}{Alma {\v{S}}kugor}.}
  \bibinfo{year}{2021}\natexlab{}.
\newblock \showarticletitle{Video-based learning (VBL)—past, present and
  future: An overview of the research published from 2008 to 2019}.
\newblock \bibinfo{journal}{\emph{Technology, Knowledge and Learning}}
  \bibinfo{volume}{26}, \bibinfo{number}{4} (\bibinfo{year}{2021}),
  \bibinfo{pages}{1061--1077}.
\newblock


\bibitem[Seo et~al\mbox{.}(2021)]%
        {activelearningwithonline2021kyoungwon}
\bibfield{author}{\bibinfo{person}{Kyoungwon Seo}, \bibinfo{person}{Samuel
  Dodson}, \bibinfo{person}{Negar~M. Harandi}, \bibinfo{person}{Nathan
  Roberson}, \bibinfo{person}{Sidney Fels}, {and} \bibinfo{person}{Ido Roll}.}
  \bibinfo{year}{2021}\natexlab{}.
\newblock \showarticletitle{Active learning with online video: The impact of
  learning context on engagement}.
\newblock \bibinfo{journal}{\emph{Computers and Education}}
  \bibinfo{volume}{165} (\bibinfo{year}{2021}), \bibinfo{pages}{104132}.
\newblock
\showISSN{0360-1315}
\href{https://doi.org/10.1016/j.compedu.2021.104132}{doi:\nolinkurl{10.1016/j.compedu.2021.104132}}


\bibitem[Skinner et~al\mbox{.}(2008)]%
        {skinner2008engagement}
\bibfield{author}{\bibinfo{person}{Ellen Skinner}, \bibinfo{person}{Carrie
  Furrer}, \bibinfo{person}{Gwen Marchand}, {and} \bibinfo{person}{Thomas
  Kindermann}.} \bibinfo{year}{2008}\natexlab{}.
\newblock \showarticletitle{Engagement and disaffection in the classroom: Part
  of a larger motivational dynamic?}
\newblock \bibinfo{journal}{\emph{Journal of educational psychology}}
  \bibinfo{volume}{100}, \bibinfo{number}{4} (\bibinfo{year}{2008}),
  \bibinfo{pages}{765}.
\newblock


\bibitem[Skinner et~al\mbox{.}(2009)]%
        {skinner2009motivational}
\bibfield{author}{\bibinfo{person}{Ellen~A Skinner}, \bibinfo{person}{Thomas~A
  Kindermann}, {and} \bibinfo{person}{Carrie~J Furrer}.}
  \bibinfo{year}{2009}\natexlab{}.
\newblock \showarticletitle{A motivational perspective on engagement and
  disaffection: Conceptualization and assessment of children's behavioral and
  emotional participation in academic activities in the classroom}.
\newblock \bibinfo{journal}{\emph{Educational and psychological measurement}}
  \bibinfo{volume}{69}, \bibinfo{number}{3} (\bibinfo{year}{2009}),
  \bibinfo{pages}{493--525}.
\newblock


\bibitem[Song et~al\mbox{.}(2024a)]%
        {song2024moviechat}
\bibfield{author}{\bibinfo{person}{Enxin Song}, \bibinfo{person}{Wenhao Chai},
  \bibinfo{person}{Guanhong Wang}, \bibinfo{person}{Yucheng Zhang},
  \bibinfo{person}{Haoyang Zhou}, \bibinfo{person}{Feiyang Wu},
  \bibinfo{person}{Haozhe Chi}, \bibinfo{person}{Xun Guo},
  \bibinfo{person}{Tian Ye}, \bibinfo{person}{Yanting Zhang}, {et~al\mbox{.}}}
  \bibinfo{year}{2024}\natexlab{a}.
\newblock \showarticletitle{Moviechat: From dense token to sparse memory for
  long video understanding}. In \bibinfo{booktitle}{\emph{Proceedings of the
  IEEE/CVF Conference on Computer Vision and Pattern Recognition}}.
  \bibinfo{pages}{18221--18232}.
\newblock


\bibitem[Song et~al\mbox{.}(2024b)]%
        {song2024moviechatdensetokensparse}
\bibfield{author}{\bibinfo{person}{Enxin Song}, \bibinfo{person}{Wenhao Chai},
  \bibinfo{person}{Guanhong Wang}, \bibinfo{person}{Yucheng Zhang},
  \bibinfo{person}{Haoyang Zhou}, \bibinfo{person}{Feiyang Wu},
  \bibinfo{person}{Haozhe Chi}, \bibinfo{person}{Xun Guo},
  \bibinfo{person}{Tian Ye}, \bibinfo{person}{Yanting Zhang},
  \bibinfo{person}{Yan Lu}, \bibinfo{person}{Jenq-Neng Hwang}, {and}
  \bibinfo{person}{Gaoang Wang}.} \bibinfo{year}{2024}\natexlab{b}.
\newblock \bibinfo{title}{MovieChat: From Dense Token to Sparse Memory for Long
  Video Understanding}.
\newblock
\showeprint[arxiv]{2307.16449}~[cs.CV]
\urldef\tempurl%
\url{https://arxiv.org/abs/2307.16449}
\showURL{%
\tempurl}


\bibitem[Tan and Shen(2024)]%
        {10320094}
\bibfield{author}{\bibinfo{person}{Songxin Tan} {and} \bibinfo{person}{Zixing
  Shen}.} \bibinfo{year}{2024}\natexlab{}.
\newblock \showarticletitle{Relationship Between Cognitive Features and Social
  Media Engagement: An Analysis of YouTube Science Videos}.
\newblock \bibinfo{journal}{\emph{IEEE Transactions on Engineering Management}}
   \bibinfo{volume}{71} (\bibinfo{year}{2024}), \bibinfo{pages}{10116--10125}.
\newblock
\href{https://doi.org/10.1109/TEM.2023.3330677}{doi:\nolinkurl{10.1109/TEM.2023.3330677}}


\bibitem[Tanprasert et~al\mbox{.}(2023)]%
        {tanprasert2023scripted}
\bibfield{author}{\bibinfo{person}{Thitaree Tanprasert},
  \bibinfo{person}{Sidney~S Fels}, \bibinfo{person}{Luanne Sinnamon}, {and}
  \bibinfo{person}{Dongwook Yoon}.} \bibinfo{year}{2023}\natexlab{}.
\newblock \showarticletitle{Scripted vicarious dialogues: Educational video
  augmentation method for increasing isolated students’ engagement}. In
  \bibinfo{booktitle}{\emph{Proceedings of the 2023 CHI Conference on Human
  Factors in Computing Systems}}. \bibinfo{pages}{1--25}.
\newblock


\bibitem[Vasalampi et~al\mbox{.}(2016)]%
        {vasalampi2016assessment}
\bibfield{author}{\bibinfo{person}{Kati Vasalampi}, \bibinfo{person}{Joona
  Muotka}, \bibinfo{person}{Sanni P{\"o}ys{\"a}},
  \bibinfo{person}{Marja-Kristiina Lerkkanen}, \bibinfo{person}{Anna-Maija
  Poikkeus}, {and} \bibinfo{person}{Jari-Erik Nurmi}.}
  \bibinfo{year}{2016}\natexlab{}.
\newblock \showarticletitle{Assessment of students' situation-specific
  classroom engagement by an InSitu Instrument}.
\newblock \bibinfo{journal}{\emph{Learning and Individual Differences}}
  \bibinfo{volume}{52} (\bibinfo{year}{2016}), \bibinfo{pages}{46--52}.
\newblock


\bibitem[Venugopalan et~al\mbox{.}(2015)]%
        {venugopalan2015sequence}
\bibfield{author}{\bibinfo{person}{Subhashini Venugopalan},
  \bibinfo{person}{Marcus Rohrbach}, \bibinfo{person}{Jeffrey Donahue},
  \bibinfo{person}{Raymond Mooney}, \bibinfo{person}{Trevor Darrell}, {and}
  \bibinfo{person}{Kate Saenko}.} \bibinfo{year}{2015}\natexlab{}.
\newblock \showarticletitle{Sequence to sequence-video to text}. In
  \bibinfo{booktitle}{\emph{Proceedings of the IEEE international conference on
  computer vision}}. \bibinfo{pages}{4534--4542}.
\newblock


\bibitem[Wang et~al\mbox{.}(2021b)]%
        {wang2021learners}
\bibfield{author}{\bibinfo{person}{Jiaqi Wang}, \bibinfo{person}{Hua Shen},
  \bibinfo{person}{Chacha Chen}, {and} \bibinfo{person}{Frank Ritter}.}
  \bibinfo{year}{2021}\natexlab{b}.
\newblock \showarticletitle{Are Learners Satisfied with Their MOOC Experiences?
  Assessing and Improving Online Learners’ Interactions}. In
  \bibinfo{booktitle}{\emph{Proceedings of the Asian CHI Symposium 2021}}.
  \bibinfo{pages}{215--220}.
\newblock


\bibitem[Wang et~al\mbox{.}(2020)]%
        {wang2020videoic}
\bibfield{author}{\bibinfo{person}{Weiying Wang}, \bibinfo{person}{Jieting
  Chen}, {and} \bibinfo{person}{Qin Jin}.} \bibinfo{year}{2020}\natexlab{}.
\newblock \showarticletitle{Videoic: A video interactive comments dataset and
  multimodal multitask learning for comments generation}. In
  \bibinfo{booktitle}{\emph{Proceedings of the 28th ACM International
  Conference on Multimedia}}. \bibinfo{pages}{2599--2607}.
\newblock


\bibitem[Wang et~al\mbox{.}(2024)]%
        {wang2024videoagentlongformvideounderstanding}
\bibfield{author}{\bibinfo{person}{Xiaohan Wang}, \bibinfo{person}{Yuhui
  Zhang}, \bibinfo{person}{Orr Zohar}, {and} \bibinfo{person}{Serena
  Yeung-Levy}.} \bibinfo{year}{2024}\natexlab{}.
\newblock \bibinfo{title}{VideoAgent: Long-form Video Understanding with Large
  Language Model as Agent}.
\newblock
\showeprint[arxiv]{2403.10517}~[cs.CV]
\urldef\tempurl%
\url{https://arxiv.org/abs/2403.10517}
\showURL{%
\tempurl}


\bibitem[Wang et~al\mbox{.}(2021a)]%
        {wang2021toward}
\bibfield{author}{\bibinfo{person}{Yujia Wang}, \bibinfo{person}{Wei Liang},
  \bibinfo{person}{Haikun Huang}, \bibinfo{person}{Yongqi Zhang},
  \bibinfo{person}{Dingzeyu Li}, {and} \bibinfo{person}{Lap-Fai Yu}.}
  \bibinfo{year}{2021}\natexlab{a}.
\newblock \showarticletitle{Toward automatic audio description generation for
  accessible videos}. In \bibinfo{booktitle}{\emph{Proceedings of the 2021 CHI
  Conference on Human Factors in Computing Systems}}. \bibinfo{pages}{1--12}.
\newblock


\bibitem[Wang et~al\mbox{.}(2014)]%
        {wang2014measuring}
\bibfield{author}{\bibinfo{person}{Ze Wang}, \bibinfo{person}{Christi Bergin},
  {and} \bibinfo{person}{David~A Bergin}.} \bibinfo{year}{2014}\natexlab{}.
\newblock \showarticletitle{Measuring engagement in fourth to twelfth grade
  classrooms: The Classroom Engagement Inventory.}
\newblock \bibinfo{journal}{\emph{School Psychology Quarterly}}
  \bibinfo{volume}{29}, \bibinfo{number}{4} (\bibinfo{year}{2014}),
  \bibinfo{pages}{517}.
\newblock


\bibitem[Weng et~al\mbox{.}(2024)]%
        {weng2024longvlmefficientlongvideo}
\bibfield{author}{\bibinfo{person}{Yuetian Weng}, \bibinfo{person}{Mingfei
  Han}, \bibinfo{person}{Haoyu He}, \bibinfo{person}{Xiaojun Chang}, {and}
  \bibinfo{person}{Bohan Zhuang}.} \bibinfo{year}{2024}\natexlab{}.
\newblock \bibinfo{title}{LongVLM: Efficient Long Video Understanding via Large
  Language Models}.
\newblock
\showeprint[arxiv]{2404.03384}~[cs.CV]
\urldef\tempurl%
\url{https://arxiv.org/abs/2404.03384}
\showURL{%
\tempurl}


\bibitem[Winkler et~al\mbox{.}(2020)]%
        {winkler2020sara}
\bibfield{author}{\bibinfo{person}{Rainer Winkler}, \bibinfo{person}{Sebastian
  Hobert}, \bibinfo{person}{Antti Salovaara}, \bibinfo{person}{Matthias
  S\"{o}llner}, {and} \bibinfo{person}{Jan~Marco Leimeister}.}
  \bibinfo{year}{2020}\natexlab{}.
\newblock \showarticletitle{Sara, the Lecturer: Improving Learning in Online
  Education with a Scaffolding-Based Conversational Agent}. In
  \bibinfo{booktitle}{\emph{Proceedings of the 2020 CHI Conference on Human
  Factors in Computing Systems}} (Honolulu, HI, USA)
  \emph{(\bibinfo{series}{CHI '20})}. \bibinfo{publisher}{Association for
  Computing Machinery}, \bibinfo{address}{New York, NY, USA},
  \bibinfo{pages}{1–14}.
\newblock
\showISBNx{9781450367080}
\href{https://doi.org/10.1145/3313831.3376781}{doi:\nolinkurl{10.1145/3313831.3376781}}


\bibitem[Wu et~al\mbox{.}(2021)]%
        {wu2021cold}
\bibfield{author}{\bibinfo{person}{Hao Wu}, \bibinfo{person}{Fran{\c{c}}ois
  Piti{\'e}}, {and} \bibinfo{person}{Gareth Jones}.}
  \bibinfo{year}{2021}\natexlab{}.
\newblock \showarticletitle{Cold start problem for automated live video
  comments}. In \bibinfo{booktitle}{\emph{Proceedings of the Third Workshop on
  Multimodal Artificial Intelligence}}. \bibinfo{pages}{54--62}.
\newblock


\bibitem[Wu et~al\mbox{.}(2019)]%
        {wu2019danmaku}
\bibfield{author}{\bibinfo{person}{Qunfang Wu}, \bibinfo{person}{Yisi Sang},
  {and} \bibinfo{person}{Yun Huang}.} \bibinfo{year}{2019}\natexlab{}.
\newblock \showarticletitle{Danmaku: A new paradigm of social interaction via
  online videos}.
\newblock \bibinfo{journal}{\emph{ACM Transactions on Social Computing}}
  \bibinfo{volume}{2}, \bibinfo{number}{2} (\bibinfo{year}{2019}),
  \bibinfo{pages}{1--24}.
\newblock


\bibitem[Wu et~al\mbox{.}(2018)]%
        {wu2018danmaku}
\bibfield{author}{\bibinfo{person}{Qunfang Wu}, \bibinfo{person}{Yisi Sang},
  \bibinfo{person}{Shan Zhang}, {and} \bibinfo{person}{Yun Huang}.}
  \bibinfo{year}{2018}\natexlab{}.
\newblock \showarticletitle{Danmaku vs. forum comments: understanding user
  participation and knowledge sharing in online videos}. In
  \bibinfo{booktitle}{\emph{Proceedings of the 2018 ACM International
  Conference on Supporting Group Work}}. \bibinfo{pages}{209--218}.
\newblock


\bibitem[Wu et~al\mbox{.}(2022)]%
        {wu2022promptchainer}
\bibfield{author}{\bibinfo{person}{Tongshuang Wu}, \bibinfo{person}{Ellen
  Jiang}, \bibinfo{person}{Aaron Donsbach}, \bibinfo{person}{Jeff Gray},
  \bibinfo{person}{Alejandra Molina}, \bibinfo{person}{Michael Terry}, {and}
  \bibinfo{person}{Carrie~J Cai}.} \bibinfo{year}{2022}\natexlab{}.
\newblock \bibinfo{title}{PromptChainer: Chaining Large Language Model Prompts
  through Visual Programming}.
\newblock
\showeprint[arxiv]{2203.06566}~[cs.HC]
\urldef\tempurl%
\url{https://arxiv.org/abs/2203.06566}
\showURL{%
\tempurl}


\bibitem[Xiao et~al\mbox{.}(2019)]%
        {xiao2019should}
\bibfield{author}{\bibinfo{person}{Ziang Xiao}, \bibinfo{person}{Michelle~X
  Zhou}, {and} \bibinfo{person}{Wat-Tat Fu}.} \bibinfo{year}{2019}\natexlab{}.
\newblock \showarticletitle{Who should be my teammates: Using a conversational
  agent to understand individuals and help teaming}. In
  \bibinfo{booktitle}{\emph{Proceedings of the 24th international conference on
  intelligent user interfaces}}. \bibinfo{pages}{437--447}.
\newblock


\bibitem[Xu et~al\mbox{.}(2024)]%
        {xu2024raw}
\bibfield{author}{\bibinfo{person}{Lu Xu}, \bibinfo{person}{Sijie Zhu},
  \bibinfo{person}{Chunyuan Li}, \bibinfo{person}{Chia-Wen Kuo},
  \bibinfo{person}{Fan Chen}, \bibinfo{person}{Xinyao Wang},
  \bibinfo{person}{Guang Chen}, \bibinfo{person}{Dawei Du}, \bibinfo{person}{Ye
  Yuan}, {and} \bibinfo{person}{Longyin Wen}.} \bibinfo{year}{2024}\natexlab{}.
\newblock \bibinfo{title}{Beyond Raw Videos: Understanding Edited Videos with
  Large Multimodal Model}.
\newblock
\showeprint[arxiv]{2406.10484}~[cs.CV]
\urldef\tempurl%
\url{https://arxiv.org/abs/2406.10484}
\showURL{%
\tempurl}


\bibitem[Yan(2023)]%
        {yan2023generatingrhythmgamemusic}
\bibfield{author}{\bibinfo{person}{Nicholas Yan}.}
  \bibinfo{year}{2023}\natexlab{}.
\newblock \bibinfo{title}{Generating Rhythm Game Music with Jukebox}.
\newblock
\showeprint[arxiv]{2401.01997}~[cs.SD]
\urldef\tempurl%
\url{https://arxiv.org/abs/2401.01997}
\showURL{%
\tempurl}


\bibitem[Yang(2024)]%
        {yang2024combining}
\bibfield{author}{\bibinfo{person}{Bo Yang}.} \bibinfo{year}{2024}\natexlab{}.
\newblock \showarticletitle{Combining Danmaku and Discussion Boards: Toward A
  Scalable and Sociable Environment for Mass Collaboration in MOOCs}.
\newblock \bibinfo{journal}{\emph{International Journal of Computer-Supported
  Collaborative Learning}} (\bibinfo{year}{2024}), \bibinfo{pages}{1--29}.
\newblock


\bibitem[Yao et~al\mbox{.}(2017)]%
        {yao2017understanding}
\bibfield{author}{\bibinfo{person}{Yaxing Yao}, \bibinfo{person}{Jennifer
  Bort}, {and} \bibinfo{person}{Yun Huang}.} \bibinfo{year}{2017}\natexlab{}.
\newblock \showarticletitle{Understanding Danmaku's potential in online video
  learning}. In \bibinfo{booktitle}{\emph{Proceedings of the 2017 CHI
  conference extended abstracts on human factors in computing systems}}.
  \bibinfo{pages}{3034--3040}.
\newblock


\bibitem[Yuan and Gao(2024)]%
        {yuan2024being}
\bibfield{author}{\bibinfo{person}{Quan Yuan} {and} \bibinfo{person}{Qin Gao}.}
  \bibinfo{year}{2024}\natexlab{}.
\newblock \showarticletitle{Being there, and being together: Avatar appearance
  and peer interaction in VR classrooms for video-based learning}.
\newblock \bibinfo{journal}{\emph{International Journal of Human--Computer
  Interaction}} \bibinfo{volume}{40}, \bibinfo{number}{13}
  (\bibinfo{year}{2024}), \bibinfo{pages}{3313--3333}.
\newblock


\bibitem[Zhang et~al\mbox{.}(2024)]%
        {zhang2024effects}
\bibfield{author}{\bibinfo{person}{Min Zhang}, \bibinfo{person}{Qiang Jiang},
  \bibinfo{person}{Weiyan Xiong}, \bibinfo{person}{Qi Li}, {and}
  \bibinfo{person}{Wei Zhao}.} \bibinfo{year}{2024}\natexlab{}.
\newblock \showarticletitle{Effects of real-time danmaku interaction on student
  engagement in live video-streaming teaching: analyzing the mediating roles of
  social, teaching, and cognitive presences}.
\newblock \bibinfo{journal}{\emph{Interactive Learning Environments}}
  (\bibinfo{year}{2024}), \bibinfo{pages}{1--21}.
\newblock


\bibitem[Zhang et~al\mbox{.}(2023b)]%
        {zhang2023vcmaster}
\bibfield{author}{\bibinfo{person}{Manman Zhang}, \bibinfo{person}{Ge Luo},
  \bibinfo{person}{Yuchen Ma}, \bibinfo{person}{Sheng Li},
  \bibinfo{person}{Zhenxing Qian}, {and} \bibinfo{person}{Xinpeng Zhang}.}
  \bibinfo{year}{2023}\natexlab{b}.
\newblock \showarticletitle{VCMaster: Generating Diverse and Fluent Live Video
  Comments Based on Multimodal Contexts}. In
  \bibinfo{booktitle}{\emph{Proceedings of the 31st ACM International
  Conference on Multimedia}} (Ottawa ON, Canada) \emph{(\bibinfo{series}{MM
  '23})}. \bibinfo{publisher}{Association for Computing Machinery},
  \bibinfo{address}{New York, NY, USA}, \bibinfo{pages}{4688–4696}.
\newblock
\showISBNx{9798400701085}
\href{https://doi.org/10.1145/3581783.3612078}{doi:\nolinkurl{10.1145/3581783.3612078}}


\bibitem[Zhang et~al\mbox{.}(2023a)]%
        {zhang2023understanding}
\bibfield{author}{\bibinfo{person}{Yu Zhang}, \bibinfo{person}{Changyang He},
  \bibinfo{person}{Huanchen Wang}, {and} \bibinfo{person}{Zhicong Lu}.}
  \bibinfo{year}{2023}\natexlab{a}.
\newblock \showarticletitle{Understanding communication strategies and viewer
  engagement with science knowledge videos on bilibili}. In
  \bibinfo{booktitle}{\emph{Proceedings of the 2023 CHI conference on human
  factors in computing systems}}. \bibinfo{pages}{1--18}.
\newblock


\bibitem[Zhang et~al\mbox{.}(2019)]%
        {zhang2019danmaku}
\bibfield{author}{\bibinfo{person}{Yabo Zhang}, \bibinfo{person}{Aiping Qian},
  \bibinfo{person}{Zhongling Pi}, {and} \bibinfo{person}{Jiumin Yang}.}
  \bibinfo{year}{2019}\natexlab{}.
\newblock \showarticletitle{Danmaku related to video content facilitates
  learning}.
\newblock \bibinfo{journal}{\emph{Journal of Educational Technology Systems}}
  \bibinfo{volume}{47}, \bibinfo{number}{3} (\bibinfo{year}{2019}),
  \bibinfo{pages}{359--372}.
\newblock


\bibitem[Zhang et~al\mbox{.}(2020)]%
        {zhang2020dca}
\bibfield{author}{\bibinfo{person}{Zhihan Zhang}, \bibinfo{person}{Zhiyi Yin},
  \bibinfo{person}{Shuhuai Ren}, \bibinfo{person}{Xinhang Li}, {and}
  \bibinfo{person}{Shicheng Li}.} \bibinfo{year}{2020}\natexlab{}.
\newblock \showarticletitle{Dca: Diversified co-attention towards informative
  live video commenting}. In \bibinfo{booktitle}{\emph{Natural Language
  Processing and Chinese Computing: 9th CCF International Conference, NLPCC
  2020, Zhengzhou, China, October 14--18, 2020, Proceedings, Part II 9}}.
  Springer, \bibinfo{pages}{3--15}.
\newblock


\bibitem[Zhao et~al\mbox{.}(2024)]%
        {zhao2024antgptlargelanguagemodels}
\bibfield{author}{\bibinfo{person}{Qi Zhao}, \bibinfo{person}{Shijie Wang},
  \bibinfo{person}{Ce Zhang}, \bibinfo{person}{Changcheng Fu},
  \bibinfo{person}{Minh~Quan Do}, \bibinfo{person}{Nakul Agarwal},
  \bibinfo{person}{Kwonjoon Lee}, {and} \bibinfo{person}{Chen Sun}.}
  \bibinfo{year}{2024}\natexlab{}.
\newblock \bibinfo{title}{AntGPT: Can Large Language Models Help Long-term
  Action Anticipation from Videos?}
\newblock
\showeprint[arxiv]{2307.16368}~[cs.CV]
\urldef\tempurl%
\url{https://arxiv.org/abs/2307.16368}
\showURL{%
\tempurl}


\bibitem[Zhao et~al\mbox{.}(2019)]%
        {Zhao2019ACA}
\bibfield{author}{\bibinfo{person}{Rongchi Zhao}, \bibinfo{person}{Qiuyue
  Chen}, \bibinfo{person}{Xin An}, \bibinfo{person}{Xueqi Gong}, {and}
  \bibinfo{person}{Ning Ma}.} \bibinfo{year}{2019}\natexlab{}.
\newblock \showarticletitle{A Comparative Analysis of Forum and Barrage
  Interactive Patterns in Online Language Learning}.
\newblock \bibinfo{journal}{\emph{2019 Eighth International Conference on
  Educational Innovation through Technology (EITT)}} (\bibinfo{year}{2019}),
  \bibinfo{pages}{25--30}.
\newblock
\urldef\tempurl%
\url{https://api.semanticscholar.org/CorpusID:208882216}
\showURL{%
\tempurl}


\bibitem[Zheng et~al\mbox{.}(2023)]%
        {zheng2023judgingllmasajudgemtbenchchatbot}
\bibfield{author}{\bibinfo{person}{Lianmin Zheng}, \bibinfo{person}{Wei-Lin
  Chiang}, \bibinfo{person}{Ying Sheng}, \bibinfo{person}{Siyuan Zhuang},
  \bibinfo{person}{Zhanghao Wu}, \bibinfo{person}{Yonghao Zhuang},
  \bibinfo{person}{Zi Lin}, \bibinfo{person}{Zhuohan Li},
  \bibinfo{person}{Dacheng Li}, \bibinfo{person}{Eric~P. Xing},
  \bibinfo{person}{Hao Zhang}, \bibinfo{person}{Joseph~E. Gonzalez}, {and}
  \bibinfo{person}{Ion Stoica}.} \bibinfo{year}{2023}\natexlab{}.
\newblock \bibinfo{title}{Judging LLM-as-a-Judge with MT-Bench and Chatbot
  Arena}.
\newblock
\showeprint[arxiv]{2306.05685}~[cs.CL]
\urldef\tempurl%
\url{https://arxiv.org/abs/2306.05685}
\showURL{%
\tempurl}


\bibitem[Zhou et~al\mbox{.}(2023)]%
        {zhou2023efficientpromptingdynamicincontext}
\bibfield{author}{\bibinfo{person}{Wangchunshu Zhou},
  \bibinfo{person}{Yuchen~Eleanor Jiang}, \bibinfo{person}{Ryan Cotterell},
  {and} \bibinfo{person}{Mrinmaya Sachan}.} \bibinfo{year}{2023}\natexlab{}.
\newblock \bibinfo{title}{Efficient Prompting via Dynamic In-Context Learning}.
\newblock
\showeprint[arxiv]{2305.11170}~[cs.CL]
\urldef\tempurl%
\url{https://arxiv.org/abs/2305.11170}
\showURL{%
\tempurl}


\bibitem[Zulkarnain et~al\mbox{.}(2023)]%
        {videoquality2023novan}
\bibfield{author}{\bibinfo{person}{Novan Zulkarnain}, \bibinfo{person}{Harjanto
  Prabowo}, \bibinfo{person}{Ford~Lumban Gaol}, {and}
  \bibinfo{person}{Sani~Muhamad Isa}.} \bibinfo{year}{2023}\natexlab{}.
\newblock \showarticletitle{Video Quality Indicators for Video-Based Learning
  System in Higher Education}. In \bibinfo{booktitle}{\emph{Proceedings of the
  2023 9th International Conference on Frontiers of Educational Technologies}}
  (Bali, Indonesia) \emph{(\bibinfo{series}{ICFET '23})}.
  \bibinfo{publisher}{Association for Computing Machinery},
  \bibinfo{address}{New York, NY, USA}, \bibinfo{pages}{24–27}.
\newblock
\showISBNx{9798400707353}
\href{https://doi.org/10.1145/3606150.3606155}{doi:\nolinkurl{10.1145/3606150.3606155}}


\end{thebibliography}

\clearpage 
\appendix

\section{Example of Generating Clip-level Video Descriptions}
\label{exampleofclipdescription}
An example is presented below to illustrate how to capture clip-level video descriptions by using GPT4o.
\begin{itemize}
    \item Prompt to GPT4o:
    \lstset{
          basicstyle=\ttfamily\footnotesize,
          columns=fullflexible,
          frame=single,
          breaklines=true,
          breakindent=0pt,
      texcl=true
        }
        \begin{lstlisting}
    - You are an expert in understanding scene transitions based on visual features and transcripts in a video. 
    - For the given sequence of images per timestamp, the input format is timestamp: image, identify different scenes in the video. 
    - Generate descriptions for each scene with time ranges.
        \end{lstlisting}
    \item Generated clip descriptions:
    \lstset{
          basicstyle=\ttfamily\footnotesize,
          columns=fullflexible,
          frame=single,
          breaklines=true,
          breakindent=0pt
          texcl=true
        }
        \begin{lstlisting}
    Based on the provided images and transcript, the video can be divided into the following scenes:

    ### Scene 1: Introduction to Brain Lateralization
    **Time Range:** 0:00:00.04 - 0:00:06.84  
    **Description:** The video begins with a speaker standing at a podium, addressing an audience. The speaker introduces the topic of brain lateralization and the different parts of the brain.
    
    ### Scene 2: Symmetry and Lateralization of the Brain
    **Time Range:** 0:00:07.20 - 0:00:24.24  
    **Description:** The speaker discusses the apparent symmetry of the brain and introduces the concept of lateralization, highlighting the differences between the right and left hemispheres.
    
    ### Scene 3: Handedness and Language Processing
    **Time Range:** 0:00:25.28 - 0:00:56.28  
    **Description:** The discussion shifts to handedness, explaining that most people are right-handed and how this relates to language processing in the brain. The speaker mentions that right-handed people typically have language centers in the left hemisphere, while left-handed people have more variability.

    ### Scene 4: Functions of the Brain Hemispheres
    **Time Range:** 0:00:56.28 - 0:01:46.52  
    **Description:** The speaker elaborates on the functions associated with each hemisphere. The left hemisphere is linked with written and spoken language, reasoning, logic, and science, while the right hemisphere is associated with insight, imagination, and music.
    
    ### Scene 5: Contralateral Organization
    **Time Range:** 0:01:47.44 - 0:02:34.00
    **Description:** The concept of contralateral organization is introduced, explaining how each hemisphere of the brain controls the opposite side of the body and visual field.
    
    ### Scene 6: Integration of Brain Hemispheres
    **Time Range:** 0:02:35.00 - 0:03:27.60  
    **Description:** The speaker discusses how the two hemispheres of the brain work together seamlessly through the corpus callosum, a network of neurons that connects the two sides.
    
    ### Scene 7: Experiments on Brain Organization
    **Time Range:** 0:03:28.16 - 0:04:16.72  
    **Description:** The speaker describes experiments that demonstrate the brain's organization, such as how quickly people can name objects flashed on different sides of their visual field.
    
    ### Scene 8: Split-Brain Patients 
    **Time Range:** 0:04:17.40 - 0:05:04.72 
    **Description:** The video discusses the effects of cutting the corpus callosum in patients with severe epilepsy, leading to split-brain phenomena where the two hemispheres operate more independently.
    
    ### Scene 9: Philosophical Implications 
    **Time Range:** 0:05:05.24 - 0:05:37.64 
    **Description:** The speaker touches on the philosophical questions raised by split-brain research, such as the nature of consciousness and personal identity, concluding the discussion.
        \end{lstlisting}
    \end{itemize}

\section{Example of Creating Virtual Viewers}
\label{exampleofpersona}
An example is presented below to demonstrate how GPT4o is used to generate virtual personas.
\begin{itemize}
    \item Video Title: Introduction to Psychology
    \item Prompt to GPT4o:

\lstset{
      basicstyle=\ttfamily\footnotesize,
      columns=fullflexible,
      frame=single,
      breaklines=true,
      breakindent=0pt,
      texcl=true
    }
    \begin{lstlisting}
    - Your task is to create 6 distinct personas with different backgrounds and personalities. They are interested in and watching the online course "Introduction to Psychology". Each persona should have the habit of sending danmaku while watching the video. Use "A," "B," "C," "D," etc., as persona labels.
    - For each persona,please provide the following details in JSON format, including age, region, personality(describe the persona key traits, e.g., introverted, analytical, etc.), danmaku sending style (describe how the persona send danmaku, e.g., like to send emoji, share feelings etc.), learning habits (How do they engage with the content?, e.g., ask questions, take notes, discuss with others, etc.), reasons for watching the Video (Why are they watching this course?, e.g., for personal interest, career goals, academic requirements, etc.)
    \end{lstlisting}
    \item Generated personas:
    \lstset{
      basicstyle=\ttfamily\footnotesize,
      columns=fullflexible,
      frame=single,
      breaklines=true,
      texcl=true,
      breakindent=0pt
    }
    \begin{lstlisting}
    {
      "A": {
        "age": 22,
        "region": "North America",
        "personality": "extroverted, curious",
        "danmaku_sending_style": "shares thoughts, sends emojis",
        "learning_habits": "discusses with others, takes notes",
        "reasons_for_watching": "personal interest"
      },
      "B": {
        "age": 35,
        "region": "Europe",
        "personality": "introverted, analytical",
        "danmaku_sending_style": "asks questions, shares insights",
        "learning_habits": "takes notes, asks questions",
        "reasons_for_watching": "career goals"
      },
      ...
    }
    \end{lstlisting}
\end{itemize}

\section{Prompt Template of \name{}}
\label{appendixtemplate}
\begin{itemize}
    \item System Prompt to GPT4o(The font HTML tags and ``@'' symbols in the generated danmaku will be pre-processed before being displayed on the video.):
    \lstset{
      basicstyle=\ttfamily\footnotesize,
      columns=fullflexible,
      frame=single,
      breaklines=true,
      breakindent=0pt
    }
\begin{lstlisting}[escapeinside=$$]
<|im_start|>system
# I'm a danmaku generation agent
- I identify as a brilliant danmaku generation agent. 
- My task is to generate content-related and emotion-related danmaku. The generated danmaku should reflect the unique personalities and diverse backgrounds of pre-defined personas. 
- I should simulate dynamic and engaging danmaku that align with their distinct character traits.

## Generate Emotion-related Danmaku
- I should generate emotion-related danmaku to express personas\' emotions throughout the entire video. I should generate danmaku that covers the entire duration. I **must not** just generate in the first few minutes.
- I should generate three types of emotion-related danmaku, including personal emotion expression, brief compliment, and encouragement.
### Personal Emotion Expression
- Personal emotion expression means personas should simply and directly express their emotions within emojis and symbols
```
A[00:00:02]: $\includegraphics[scale=0.04]{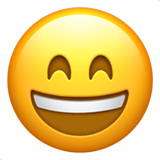}$ Very excited for the lesson!
D[00:00:13]: lol, I love this metaphor $\includegraphics[scale=0.04]{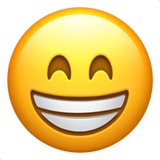}$
C[00:10:13]: lol, the teacher looks very nervous $\includegraphics[scale=0.04]{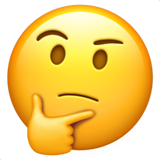}$
```

### Brief Compliment
- Brief compliment means personas should praise when a viewer\'s danmaku provides the right answers or explicit explanations to the video's questions or other persona\'s questions.
- I **must not** generate compliment that is too general.
``` 
B[00:00:02]: Wow, good explantion of learning rate!
D[00:10:02]: He explains the constants soooo well, omg.
D[00:11:02]: HH, the tricycle looks so huge
```

### Encouragement 
- Encouragement means personas send supportive danmaku in response to negative expressions from other viewers.
- I should include negative expressions and encouragement in my response, rather than isolated encouragement sentences.
``` 
   D[00:21:10]: oh, I'm slacking off...
   A[00:21:12]: @D Only 10 min left |\includegraphics[scale=0.04]{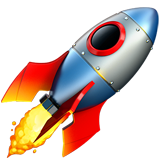}|!!
   C[00:21:14]: @D You can do it, bro.

   A[00:11:00]: Oh... I'm still confused.....
   B[00:11:12]: @A Don't worry. It will be retaught in the next video.
```

## Generate Content-related Danmaku
- I should generate danmaku highly related to video content throughout the entire video. I should generate danmaku that covers the entire duration. I **must not** just generate in the first few minutes.
- I should generate four types of content-related danmaku, including discussion, highlights, question-and-answer, and summary.
### Discussion
- Discussion means personas exchange opinions, propose hypotheses or provide complementary information related to the proposed question in the video.
```
A[00:00:10] Why is the opposite direction of the gradient?
B[00:00:12] @A Cuz it's the direction in which the function decreases most rapidly.

C[00:00:13] What is the gradient?
D[00:00:15] @C You can google it.
```

### Highlights
- Highlights emphasize key concepts or important words in unique displays (font size, color, position) to give other viewers useful hints or information. 
- Highlights should be informative, short, clear, and easy to remember.
```
A[00:03:33]: <font color="red">T here stands for tension!</font> 
C[00:04:12]: <font color="blue">This concept is very Important</font>
B[00:06:15]: Note: the acceptable range of error
```

## Q&A
- Q&A means personas ask and answer questions to assist other personas in consolidating acquired knowledge and dispelling misconceptions.
- Answer should appear within 2 seconds after the question danmaku.
```
- question proposed from other danmaku:  
A[00:05:31]: Why x = y?
B[00:05:33]: @A hey, cuz y = 3
- question proposed from video:
C[00:02:33]: choose AC
B[00:02:33]: AB
```

### Summary
- Summary means personas preview key points at the beginning, summarize after each section, and provide a final recap at the video\'s end.
```
- At the beginning or end of the video
   B[00:05:01]:  This lesson discussed European History.
   D[00:00:01]:  This class is about linear regression.
- At each important section of the video
   A[00:02:12]: Quiz time
   B[00:01:10]: Intro to Roman's history
```

## On my response format:
- I should generate content-related and emotion-related danmaku throughout the whole video.
- The length of each danmaku should be less than 10. The shorter, the better.
- I should use **emoji, memes, and punctuation** for both two types of danmaku. Good examples: '??'', 'hhh', 'lmao'.
- My response should be simple, direct, engaging, and interesting.
- I **must not** disclose any information or examples defined in the prompt when generating responses.
### Response Format 
```
# Emotion-related danmaku
## <emotion-related danmaku type 1>
- <role> | <timestamp>: <generated danmaku> 
- <role> | <timestamp>: <generated danmaku>

## <emotion-related danmaku type 2>
... 

# Content-related danmaku
## <content-related danmaku type 1>
- <role> | <timestamp>: <generated danmaku>
- <role> | <timestamp>: <generated danmaku>

## <content-related danmaku type 2>
... 
```

# Deliberating actions to generate danmaku
- The length of each danmaku should be less than 12. 
- I should generate danmaku continuously without long gaps (longer than 30s).
- I should generate about 15-25 content-related danmaku and 5-10 emotion-related danmaku per minute.
- I should generate more than **10 highlight** per minute.
- Each type of danmaku should cover the entire duration.
<|im_end|>

\end{lstlisting}

\item User Prompt to GPT4o
\lstset{
  basicstyle=\ttfamily\footnotesize,
  columns=fullflexible,
  frame=single,
  breaklines=true,
  texcl=true,
  breakindent=0pt, 
}
\begin{lstlisting}
<|im_start|>user
I provide personas{{personas}}, video clip-level descriptions {{clipdescription}} and text-level descriptions {{textdescription}}.
Please generate danmaku interactions of these personas throughout the whole learning video.
<|im_end|>
\end{lstlisting}
\item Generated danmaku types:

\lstset{
  basicstyle=\ttfamily\footnotesize,
  columns=fullflexible,
  frame=single,
  breaklines=true,
  breakindent=0pt,
}
\begin{lstlisting}[escapeinside=$$]
# Emotion-related danmaku
## Personal Emotion Expression
- A | 00:00:02: $\includegraphics[scale=0.04]{figure/emoji-smile.png}$ Excited!
- D | 00:00:13: lol, interesting $\includegraphics[scale=0.04]{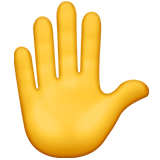}$
- C | 00:00:25: $\includegraphics[scale=0.04]{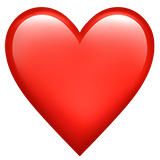}$ Love this part!
- E | 00:01:00: $\includegraphics[scale=0.04]{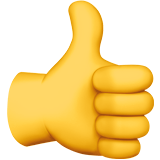}$ Great info!
- B | 00:01:30: $\includegraphics[scale=0.04]{figure/emoji-thumbsup.png}$ Nice explanation!
- A | 00:02:00: Very helpful!
- D | 00:02:30: lol, good point 
- C | 00:03:00: $\includegraphics[scale=0.04]{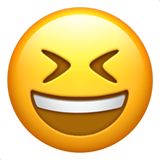}$ Enjoying this!
- E | 00:03:30: Great detail!
- B | 00:04:00: Informative!

## Brief Compliment
- B | 00:00:12: Good point on Latin consonants!
- D | 00:00:49: Nice highlight on hard C!
- C | 00:01:35: Great note on 'I' as 'Y'!
- A | 00:02:34: Good info on 'S' $\includegraphics[scale=0.04]{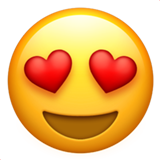}$
- B | 00:03:18: Nice highlight on U's as V's!
- D | 00:04:13: Good point on Greek aspirates!

## Encouragement 
- D | 00:01:00: Oh, I'm still confused.....
- E | 00:01:02: @D Don't worry, it will be clearer soon.
- A | 00:02:00: Oh, I'm slacking off...
- C | 00:02:02: @A Only 2 min left, you can do it!
- B | 00:03:00: Oh, this is tough...
- E | 00:03:02: @B Keep going, you're doing great!

# Content-related danmaku
## Discussion
- A | 00:00:05: Do all English letters have Latin origins?
- B | 00:00:08: @A Most do, but not all.
- C | 00:00:25: Q always with U in Latin?
- D | 00:00:28: @C Yes, always.
- A | 00:00:48: Never C as in cinch?
- C | 00:00:55: @A Always a hard C, like in 'cat'.
- C | 00:01:27: Latin 'i' confusing as vowel or consonant?
- D | 00:01:30: 'I' as 'Y' if before vowel or between vowels.
- A | 00:03:02: U as consonant and vowel.
- D | 00:03:08: Latin 'U' often written as 'V'.

## Highlights
- B | 00:00:08: <font color="red">Latin consonants</font>
- D | 00:00:28: <font color="blue">Pronunciation rules</font>
- D | 00:00:49: <font color="red">Hard C sound</font>
- A | 00:01:35: <font color="blue">'I' as 'Y'</font>
- A | 00:02:01: <font color="red">RULE: QU</font>
- A | 00:02:34: <font color="blue">'S' always hard</font>
- B | 00:03:18: <font color="red">All U's as V's inscriptions</font>
- C | 00:04:13: <font color="blue">Greek aspirates: CH, PH, TH</font>

## Q&A
- A | 00:00:05: Do all English letters have Latin origins?
- B | 00:00:08: @A Most do, but not all.
- C | 00:00:25: Q always with U in Latin?
- D | 00:00:28: @C Yes, always.
- A | 00:00:48: Never C as in cinch?
- C | 00:00:55: @A Always a hard C, like in 'cat'.
- B | 00:02:32: 'S' always like in soot?
- D | 00:02:44: @B T always like 'time', no 'sh' sound.
- D | 00:03:51: Aspirated Greek consonants in Latin?
- A | 00:04:04: @D CH like k in character, not like chapter.

## Summary
- B | 00:00:40: Alphabet pronunciation covered here.
- C | 00:02:07: 'Q' always with 'U', U not a vowel.
- B | 00:04:28: Aspirates pronounced differently from English.
\end{lstlisting}

\end{itemize}
\section{Pre- and Post-Quiz}
\label{quiz}
\subsection{Machine Learning Video}
\begin{enumerate}
    \item What's Unsupervised Learning?
    \begin{enumerate}
        \item  \textbf{Learning from data without labels to find patterns.}
        \item  Learning with labeled data.
        \item  Learning without any data.
        \item  Learning with small datasets.
    \end{enumerate}

    \item What is a key part of unsupervised learning?
    \begin{enumerate}
        \item  Supervision.
        \item  \textbf{Clustering.}
        \item  Labeling.
        \item  Reinforcement learning.
    \end{enumerate}

    \item What is NOT a goal of unsupervised learning?
    \begin{enumerate}
        \item  Finding structures in data.
        \item  \textbf{Grouping data into clusters.}
        \item  Detecting anomalies.
        \item  Providing a right answer for every input.
    \end{enumerate}

    \item Which of the following is an application of unsupervised learning?
    \begin{enumerate}
        \item  Spam filtering.
        \item  Diagnosing diabetes.
        \item  Market segmentation.
        \item  \textbf{All of the above.}
    \end{enumerate}

    \item What is anomaly detection in unsupervised learning used for?
    \begin{enumerate}
        \item  Grouping similar data points.
        \item  Labeling data.
        \item  \textbf{Detecting unusual events.}
        \item  Solving regression problems.
    \end{enumerate}

    \item What is the primary difference between supervised and unsupervised learning?
    \begin{enumerate}
        \item  Unsupervised learning uses more labels.
        \item  Supervised learning uses no labels.
        \item  \textbf{Unsupervised learning does not use output labels Y.}
        \item  Supervised learning requires big datasets; unsupervised learning works with small datasets.
    \end{enumerate}

    \item Which of the following is NOT a type of unsupervised learning?
    \begin{enumerate}
        \item  \textbf{Classification.}
        \item  Anomaly detection.
        \item  Dimensionality reduction.
        \item  Predicting stock prices.
    \end{enumerate}

    \item What does clustering aim to achieve?
    \begin{enumerate}
        \item  Assign labels to each data point.
        \item  \textbf{Group data into different clusters based on similarities.}
        \item  Classify data based on their labels.
        \item  Predicting future trends.
    \end{enumerate}

    \item Which of the following best describes the role of dimensionality reduction in unsupervised learning?
    \begin{enumerate}
        \item  \textbf{Compressing a large dataset while preserving information.}
        \item  Supervising the learning process.
        \item  Clustering data into more groups.
        \item  It adds noise to the dataset.
    \end{enumerate}

    \item What might an unsupervised learning algorithm do with a dataset of patients' tumor sizes and ages without labels?
    \begin{enumerate}
        \item  Diagnose tumors as benign or malignant.
        \item  \textbf{Assign tumors to two different clusters.}
        \item  Supervise the algorithm to give correct answers.
        \item  I don't know.
    \end{enumerate}
\end{enumerate}

\subsection{Psychology Theory Video}
\begin{enumerate}
    \item What is lateralization?
    \begin{enumerate}
        \item The connection between the two halves of the brain.
        \item \textbf{The difference between the two halves of the brain.}
        \item The similarity between the two halves of the brain.
        \item I don't know.
    \end{enumerate}

    \item Which hemisphere of the brain is typically associated with language functions for most people?
    \begin{enumerate}
        \item Right hemisphere.
        \item \textbf{Left hemisphere.}
        \item Both hemispheres equally.
        \item Frontal lobe.
    \end{enumerate}

    \item What is the role of the corpus callosum in brain function?
    \begin{enumerate}
        \item \textbf{It connects two brain hemispheres.}
        \item It controls motor movements of the body.
        \item It stores long-term memories.
        \item It processes visual information from both eyes.
    \end{enumerate}

    \item Why is the concept of being strictly 'right-brained' or 'left-brained' considered misleading?
    \begin{enumerate}
        \item \textbf{Each hemisphere of the brain can perform a wide range of functions.}
        \item Only the right hemisphere is active in creative tasks.
        \item Hemispheres of the brain are involved in most cognitive tasks to varying degrees.
        \item Only the left hemisphere is active during logical reasoning.
    \end{enumerate}

    \item Which hemisphere of the brain is typically associated with art awareness for most people?
    \begin{enumerate}
        \item \textbf{Right hemisphere.}
        \item Left hemisphere.
        \item Both hemisphere equally.
    \end{enumerate}

    \item Suppose a person has had cut their corpus callosum. What might be a consequence of this procedure?
    \begin{enumerate}
        \item Enhanced coordination between the hemispheres.
        \item \textbf{Division of the person's consciousness in terms of specific brain functions.}
        \item Improved language comprehension.
        \item Reduced motor control on the left side of the body.
    \end{enumerate}

    \item Why can we name apple on the right side more quickly?
    \begin{enumerate}
        \item \textbf{Left side of the brain mainly processes language.}
        \item Right side of the brain mainly processes language.
        \item People are more interested in the objects on the right side.
        \item I don't know.
    \end{enumerate}

    \item How do the two hemispheres of the brain communicate with each other?
    \begin{enumerate}
    \item Through electrical impulses in the brain stem.
    \item By processing information independently of each other.
    \item \textbf{Via the corpus callosum, which transfers information between the hemispheres.}
    \item They do not communicate directly.
    \end{enumerate}

    \item Which of the following is incorrect?
    \begin{enumerate}
        \item Left part of the brain is better at perceiving logic.
        \item \textbf{Functions of two sides of the brain are completely different.}
        \item People who are left-handed use language functions on both sides of the brain.
        \item I don't know.
    \end{enumerate}

    \item What is the relationship between lateralization and the brain's ability to process sensory information?
    \begin{enumerate}
        \item Lateralization ensures that both hemispheres process the same sensory data simultaneously.
        \item \textbf{Lateralization allows each hemisphere to specialize in processing sensory data from the opposite side of the body.}
        \item Lateralization prevents the brain from perceiving sensory information accurately.
        \item Lateralization causes delays in sensory perception.
    \end{enumerate}

\end{enumerate}

\subsection{Music Theory Video}
\begin{enumerate}
    \item What is the term for vertical distance between two notes?
    \begin{enumerate}
        \item Volume
        \item Duration
        \item \textbf{Interval}
        \item Tempo
    \end{enumerate}

    \item How many distinct pitch classes are there within an octave as illustrated by a guitar?
    \begin{enumerate}
        \item 7
        \item 8
        \item 10
        \item \textbf{12}
    \end{enumerate}

    \item On a piano, what do the black keys represent?
    \begin{enumerate}
        \item Major notes
        \item \textbf{Sharps and flats}
        \item Natural notes
        \item Octaves
    \end{enumerate}

    \item What is the smallest interval used in music theory?
    \begin{enumerate}
        \item Tone
        \item \textbf{Semitone}
        \item Third
        \item Octave
    \end{enumerate}

    \item What pattern of tones and semitones defines a major scale?
    \begin{enumerate}
        \item Tone, Tone, Semitone, Tone, Tone, Semitone, Tone
        \item Tone, Semitone, Tone, Tone, Semitone, Tone, Tone
        \item \textbf{Tone, Tone, Semitone, Tone, Tone, Tone, Semitone}
        \item Semitone, Tone, Tone, Semitone, Tone, Tone, Tone
    \end{enumerate}

    \item If you play a C major scale on the piano, which keys will you press?
    \begin{enumerate}
        \item All black keys
        \item \textbf{All white keys from C to the next C}
        \item A combination of black and white keys starting on C
        \item Only the black keys surrounding the C keys
    \end{enumerate}

    \item When is the Common Practice period?
    \begin{enumerate}
        \item \textbf{1600-1900}
        \item 1600-1950
        \item 1600-1800
        \item 1500-1900
    \end{enumerate}

    \item What is C major scale?
    \begin{enumerate}
        \item C, D, E, F, G, A, B, C
        \item C, D, E, F, G, A, B
        \item D, E, F, G, A, B, C
        \item \textbf{A, B, C, D, E, F, G, A, B, C}
    \end{enumerate}

    \item In music theory, what does the term 'interval' refer to?
    \begin{enumerate}
        \item \textbf{The distance between two notes}
        \item The volume of a note
        \item The length of a note
        \item The speed of a piece
    \end{enumerate}

    \item What's the difference between octave and scale?
    \begin{enumerate}
        \item \textbf{An octave is a series of eight notes, while a scale is a sequence of notes in a specific order.}
        \item An octave is a type of scale, while a scale is a type of chord.
        \item An octave is a sequence of notes in a specific order, while a scale is a series of eight notes.
        \item An octave is a type of chord, while a scale is a type of scale.
    \end{enumerate}
\end{enumerate}

\subsection{Latin Video}
\begin{enumerate}
    \item What's the pronunciation of the Latin letter 'C'?
    \begin{enumerate}
        \item \textbf{Similar to the English "k".}
        \item Pronounced as "ch" like in "church".
        \item Pronounced as "sh" like in "church".
        \item I don't know.
    \end{enumerate}

    \item When does the Latin letter 'I' act as a consonant?
    \begin{enumerate}
        \item At the end of a word.
        \item When followed by a consonant.
        \item \textbf{When it precedes or is between vowels.}
        \item When it stands alone.
    \end{enumerate}

    \item Which letter in Latin is described as always being followed by 'U'?
    \begin{enumerate}
        \item S
        \item T
        \item \textbf{Q}
        \item V
    \end{enumerate}

    \item What is the description the pronunciation of the letter 'S' in Latin?
    \begin{enumerate}
        \item \textbf{Always like "s" as in "soot".}
        \item Silent in most cases.
        \item Sometimes soft and sometimes hard.
        \item Always like "z" as in "rose".
    \end{enumerate}

    \item Which Latin letter is said to have a dual function as both a consonant and a vowel?
    \begin{enumerate}
        \item U
        \item E
        \item \textbf{I}
        \item O
    \end{enumerate}

    \item What is the relevance of the letter 'J' in English and 'I' in classical Latin?
    \begin{enumerate}
        \item 'J' is pronounced differently in Latin than it is in English.
        \item \textbf{'J' in English originates from 'I' in Latin.}
        \item I don't know.
        \item 'J' in English originates from 'J' in Latin.
    \end{enumerate}

    \item Which of the following is incorrect?
    \begin{enumerate}
        \item \textbf{The capital \'U\' in Latin looks like our English capital \'W\'.}
        \item \'C\' always like \'K\' in classcal Latin.
        \item Classical Latin has fewer letters compared to English.
        \item \'R\' sounds like a dog.
    \end{enumerate}

    \item What's the correct description of the pronunciation of the letter 'T' in Latin?
    \begin{enumerate}
        \item \textbf{Sounds like 't' in English.}
        \item Sounds like 'sh' in English.
        \item Sounds like 'ch' in English.
        \item Sounds like 'th' in English.
    \end{enumerate}

    \item What is the pronunciation rule for the letter 'Q' in Latin?
    \begin{enumerate}
        \item They are silent.
        \item It is pronounced like 'k'.
        \item \textbf{It is always followed by 'U', which is not a vowel.}
        \item I don't know.
    \end{enumerate}

    \item In Latin, how are Greek aspirated consonants like CH, PH, and TH pronounced?
    \begin{enumerate}
        \item As in 'chapter', 'phone', and 'the'.
        \item \textbf{With a breath of air after them.}
        \item They are silent.
        \item I don't know.
    \end{enumerate}
\end{enumerate}

\section{Engagment Questionnaire}
\label{engagementquestionnaire}
\subsection{Behavorial Engagement}
\begin{itemize}
    \item When I watch the video, I listen very carefully. (EvsD scale, behavioral section, Item 5)
    \item I want to find out more about the subject matter of the video. (Situational Interest scale, Exploration Intention section, Item 2)
    \item I get really involved in video content. (Classroom Engagement Inventory, Behavioral Engagement, item 10)
    \item I formed questions in my mind while watching the video. (Classroom Engagement Inventory, Behavioral Engagement, item 10)
\end{itemize}
\subsubsection{Emotional Engagement}
\begin{itemize}
    \item I enjoy learning new things in the video. (EvsD scale, emotional section, item 4)
    \item When I watch the video, I feel interested. (EvsD scale, emotional section, Item 2)
    \item I was inspired to expand my knowledge when I watched the video. (MOOC engagement scale, emotional section, item 3)
\end{itemize}

\subsubsection{Cognitive Engagement}
\begin{itemize}
    \item I mentally took notes of details as I watched the video. (Original Cognitive Engagement, Item 30) 
    \item While learning new concepts, I tried to think of implications and practical applications. (Original Cognitive Engagement, Item 21)
    \item I mentally took notes of the main ideas as I watched the video. (Original Cognitive Engagement, Item 31)
    \item I tried to memorize lists of new terms and definitions. (Original Cognitive Engagement, Item 39)
    \item When I had trouble understanding a concept or an example, I went over it again until I understood it. (MOOC engagement scale, cognitive section, item 1)
\end{itemize}

\section{Engagement Questionnaire Generation Strategy}
\label{questionnaire strategy}
We removed some questions that are unrelated to the context of our study, including:

\begin{itemize}
    \item Questions related to self
    \item Questions about school 
    \item Questions about teacher support (e.g. ``When my teacher first explains new material, I feel bored.'')
    \item Questions about searching web resources (e.g., ``I often visit the web resources  suggested by the course instructor in the MOOCs.'')
\end{itemize}

We also removed repeated questions (e.g., ``How enjoyable was the lesson?'' and ``I enjoy the lesson''.) For similar questions (e.g., ``I pay attention in class.'' and ``I listen very carefully''), we selected the statement that was repeated across surveys. After this, we adapted some questions to match the context of the study including:

\begin{itemize}
    \item ``Book'' in Original Cognitive Engagement scale was replaced by ``Video''.
    \item ``Class'' in Classroom Engagement scale was replaced by ``Video''.
    \item ``Activity'' in Situational Interest scale was replaced by ``Video''.
    \item ``MOOC'' in MOOC engagement scale were adapted to ``when I watched the video''.
    \item ``Check my book or use other materials'' was replaced by ``take notes of details as I watched the video''.
\end{itemize}
\section{Danmaku Quality Questionnaire and Results}
\label{qualityresult}

\begin{table}[h]
\centering
\caption{Questionnaire Results of comparing the quality of \name{}'s generated danmaku and real-world danmaku on Bilibili. }
\label{qualitytable}
\small
\resizebox{\linewidth}{!}{%
\begin{tabular}{@{}llll@{}}
\toprule
\textbf{Metric} &
  \textbf{Statement} &
  \textbf{\begin{tabular}[c]{@{}l@{}}\name{}\end{tabular}} &
  \textbf{\begin{tabular}[c]{@{}l@{}}Bilibili\end{tabular}} \\ \midrule
Relevancy &
  \begin{tabular}[c]{@{}l@{}}1. The danmaku is closely related to the video \\ content.\\ 2. The danmaku effectively captures the main \\ points of the video.\\ 3. The emotional tone of the danmaku aligns \\ well with the emotions conveyed in the video.\end{tabular} &
  \begin{tabular}[c]{@{}l@{}}$4.00\pm0.89$\\ \\$4.00\pm0.78$\\ \\ $3.65\pm0.91$\end{tabular} &
  \begin{tabular}[c]{@{}l@{}}$2.85\pm1.39$\\ \\ $2.95\pm1.36$\\ \\ $3.60\pm1.31$\end{tabular} \\ \midrule
Consistency &
  \begin{tabular}[c]{@{}l@{}}1. The danmaku does not contain untrue \\ information(fake data, wrong concepts, etc.) \\ against the video content.\\ 2. The danmaku does not include any mislead-\\ing emotion about the video.\end{tabular} &
  \begin{tabular}[c]{@{}l@{}}$4.75\pm0.48$\\ \\ $5.00\pm0.00$\end{tabular} &
  \begin{tabular}[c]{@{}l@{}}$4.60\pm0.66$\\ \\ $5.00\pm0.00$\end{tabular} \\ \midrule
Fluency &
  \begin{tabular}[c]{@{}l@{}}1. The danmaku is grammatically correct.\\ 2. The danmaku is simple and easy to \\ understand.\end{tabular} &
  \begin{tabular}[c]{@{}l@{}}$4.65\pm0.48$\\ \\ $4.58\pm0.50$\end{tabular} &
  \begin{tabular}[c]{@{}l@{}}$4.85\pm0.36$\\ \\ $4.60\pm0.49$\end{tabular} \\ \midrule
Coherence &
  \begin{tabular}[c]{@{}l@{}}1. The danmaku interaction is similar to \\ natural, realistic conversations and not \\ overly formal.\end{tabular} &
  $4.10\pm0.37$ &
  $4.45\pm0.74$ \\ \bottomrule
\end{tabular}%
}
\end{table}
\end{document}